\documentclass[abstracton, 12pt]{scrartcl}
\usepackage[ngerman]{babel}
\pdfoutput=1
\usepackage{url, subfigure}
\usepackage[T1]{fontenc} \usepackage{lmodern}
\usepackage[utf8]{inputenc}
\usepackage{lscape}
\usepackage[flushleft,alwaysadjust]{paralist}
\usepackage[hyperfootnotes=false, pagebackref=true, citebordercolor={1 1 1}, linkbordercolor={0 0 1}, urlbordercolor={1 1 1}]{hyperref}
\usepackage[]{epsfig}
\usepackage[sectionbib]{natbib}
\bibpunct[:]{(}{)}{;}{a}{}{,}
 
\usepackage{graphics,latexsym}
\usepackage[inner=2cm,outer=2cm,top=1.75cm,bottom=3.25cm]{geometry}
\usepackage[margin=0pt,font=normalsize,labelfont=bf, format=hang]{caption}
\usepackage{color}
\definecolor{dkgreen}{rgb}{0,0.6,0}
\definecolor{gray}{rgb}{0.5,0.5,0.5}
\definecolor{mauve}{rgb}{0.58,0,0.82}
\usepackage{listings}
\deffootnote{1.5em}{1em}{%
	\makebox[1.5em][l]{\thefootnotemark}}
\lstdefinelanguage{Scala}{
  morekeywords={abstract,case,catch,class,def,%
    do,else,extends,false,final,finally,%
    for,if,implicit,import,match,mixin,%
    new,null,object,override,package,%
    private,protected,requires,return,sealed,%
    super,this,throw,trait,true,try,%
    type,val,var,while,with,yield},
  otherkeywords={=>,<-,->,<\%,<:,>:,\#,@},
  sensitive=true,
  morecomment=[l]{//},
  morecomment=[n]{/*}{*/},
  morestring=[b]",
  morestring=[b]',
  morestring=[b]"""
}
\title{Typsichere Modellierung im Text-Mining}
\author{Fabian Steeg\footnote{Sprachliche Informationsverarbeitung, Universität zu Köln}}

\begin{document}
\maketitle
\thispagestyle{empty}

\begin{abstract}
Auf Basis einer Darstellung von annotationsbasierten Agenten in Kapitel \ref{chapter-theory} beschreibt Kapitel \ref{chapter-tools} Werkzeuge und eine formale Notation zur Definition und Durchführung von Experimenten im Text-Mining, in Form einer statisch typisierten, eingebetteten DSL. Am Beispiel von maschinellem Lernen zur Klassifikation zeigt Kapitel \ref{chapter-appl}, wie mit diesem Framework Text-Mining-Experimente entwickelt und automatisch dokumentiert werden können. Kapitel \ref{chapter-ausblick} stellt zusammenfassend dar, wie das Konzept der generischen, typsicheren Annotation dabei zur Modellierung von Prozessen der Informationsverarbeitung genutzt werden kann, und inwiefern es einem allgemeinen Informationsmodell entspricht, das über die Textprozessierung hinausgeht.
\end{abstract}

\setcounter{tocdepth}{1}
\tableofcontents
\newpage
\pagenumbering{arabic}
\section{Text-Mining mit annotationsbasierten Agenten} \label{chapter-theory}

Die ständig zunehmende Menge digital verfügbarer Daten erschwert zunehmend das Auffinden gesuchter Informationen. Text-Mining, die Analyse von unstrukturierten textuellen Daten, verspricht hier eine Verbesserung, etwa durch die Ermöglichung intelligenter Suche in Form eines semantischen Information-Re\-triev\-al. Es fehlt aber an Methoden und Werkzeugen zur Erstellung von modularen Verfahren, die mittel- oder langfristig Text-Mining auf menschlichem Niveau ermöglichen. Bestehende computerlinguistische Werkzeuge für komponentenbasiertes Text-Mining erfordern einen Mehraufwand bei Implementierung und Dokumentation gegenüber isolierten Lösungen und werden deshalb bisher wenig genutzt.

Ziel dieser Untersuchung ist daher die Entwicklung von Konzepten und Werkzeugen, welche die computerlinguistische Arbeit durch eine möglichst weitgehende Unterstützung bei der Modellierung, Dokumentation und Evaluierung zugleich erleichtern und verbessern. Um sowohl computerlinguistisch fundiert als auch für neue Bereiche ausbaufähig zu sein, werden die Werkzeuge auf Basis einer Verbindung von Konzepten aus Computerlinguistik und allgemeiner Künstlicher Intelligenz beschrieben und entwickelt, nämlich als annotationsbasierte Multiagentensysteme. Die Werkzeuge sollen soweit wie möglich mit bestehenden Softwarekomponenten integriert werden, um eine möglichst hohe praktische Nützlichkeit zu erreichen. Sie werden dabei von einfachen Beispielen ausgehend entwickelt und für komplexe Beispiele aus der lexikalischen Semantik ausgebaut, die verschiedene Verfahren des maschinellen Lernens zur Klassifikation einsetzen. Hierbei soll sich zeigen, inwiefern die möglichst einfachen Konzepte und Werkzeuge sowohl für grundlegende wie für komplexe Aufgaben eingesetzt werden können.

\subsection{Text-Mining und Modularität} \label{ml-mod}

\subsubsection{Computerlinguistische Komponentensysteme} \label{sale-theorie}

Dieser Abschnitt (\ref{sale-theorie}) gibt basierend auf eigenen Vorarbeiten \citep{Steeg2007} einen kurzen Überblick über komponentenbasierte Softwareentwicklung und computerlinguistische Komponentensysteme, der in den folgenden Abschnitten um eine Beschreibung der Rolle der Dokumentation (\ref{theorie-doku-bedeutung}, S. \pageref{theorie-doku-bedeutung}) und eine Betrachtung der Urspünge solcher Systeme (\ref{anno-grundlagen-technisch}, S. \pageref{anno-grundlagen-technisch}) ergänzt wird.

\paragraph{Komponenten in der Softwareentwicklung} \index{Service} \index{Komponenten}

Komponentenbasierte Software\-entwickl\-ung ist eine Wei\-ter\-ent\-wick\-l\-ung der objektorientierten Softwareentwicklung mit dem Ziel einer verbesserten Wiederverwertbarkeit größerer Softwarebausteine und einer daraus resultierenden Fokussierung auf die Geschäfts\-logik der zu erstellenden Anwendung anstelle der dazu nötigen Infrastruktur \citep[1]{GruhnAndThiel2000}. Eine Komponente ist ein gekapseltes Stück Software, das eine bestimmte Funktionalität, welche auch als \emph{Service} bezeichnet wird, anbietet. Die Anforderungen an eine Komponente werden durch das \emph{Komponentenmodell} definiert, in dem die Komponente verwendet werden soll \citep[15-6]{GruhnAndThiel2000}. 

\paragraph{Komponentensysteme in der Computerlinguistik} \label{theorie-sale}

\index{Komponenten!SALE}

Verfahren zur maschinellen Sprachverarbeitung werden heute z.T. nicht mehr nur isoliert, sondern integriert entwickelt, in einer \emph{Software Architecture for Language Engineering} (SALE), einer Infrastruktur für die maschinelle Sprachverarbeitung. Eine solche Infrastruktur besteht aus Frameworks, Referenzarchitekturen und einer Entwicklungsumgebung und bildet damit eine Art Werkzeugkasten für die computerlinguistische Arbeit (s. \citealt{Koehler2005, CunninghamAndBontcheva2006}, und \citealt{Ziegler2007}). Im Sinne der oben skizzierten Konzepte komponentenbasierter Entwicklung definiert eine SALE so unter anderem das Komponentenmodell, mit dem die sprachverarbeitenden Komponenten verwendet werden sollen.

Bekannte SALEs sind etwa GATE\footnote{General Architecture for Text Engineering, \url{http://gate.ac.uk/}}, LingPipe\footnote{LingPipe, \url{http://alias-i.com/lingpipe/}}, UIMA\footnote{Unstructured Information Management Architecture, \url{http://uima.apache.org/}} oder SMILA\footnote{Semantic Information Logistics Architecture, \url{http://www.eclipse.org/smila/}}. Die Ausgangsfragen dieser Untersuchung basieren auf eigenen Erfahrungen \citep{Steeg2007} bei der Entwicklung und Anwendung von Tesla (Text Engineering Software Laboratory)\footnote{Tesla, \url{http://www.spinfo.phil-fak.uni-koeln.de/spinfo-forschung-tesla.html}}, einer an der Abteilung für Sprachliche Informationsverarbeitung der Universität zu Köln entwickelten SALE, die konsequent dynamische Annotation (s. \citealt{BendenAndHermes2004, HermesAndBenden2005}) einsetzt.

Die Implementierung von Text-Mining-Verfahren in einer SALE ermöglicht eine Integration der einzelnen Komponenten (z.B. Parsing, POS-Tagging, Disambiguierung, etc.) mit ihrer Anwendung (z.B. Informationsextraktion, maschinelle Übersetzung, etc.). Der Einsatz einer SALE verbessert zudem Modularität und Wiederverwertbarkeit, sowohl fachlich durch \emph{dynamische Annotation}, bei der die ursprünglichen Daten stets verfügbar bleiben (vgl. \citealt{HermesAndBenden2005} und Abschnitt \ref{ml-base-corpus}, S. \pageref{ml-base-corpus}) als auch technisch durch \emph{dynamische Konfiguration} \citep[135]{HuntAndThomas2003} von wiederverwertbaren Komponenten.

\subsubsection{Dokumentation, Nachvollziehbarkeit und Modularität} \label{theorie-doku-bedeutung}

Das Ziel von computerlinguistischen Komponentensystemen ist zum Einen eine Vereinfachung von gewohnten Arbeitsabläufen -- etwa die Entwicklung eines Parsers -- und zum Anderen die Ermöglichung neuer Tätigkeiten, etwa die Erstellung von komplexen \emph{Workflows} (vgl. \citealt{BirdAndLiberman1999}) -- etwa die Integration des Parsers in einen Anwendungsfall zur maschinellen Übersetzung. Diese zwei Ziele sind aber nicht ohne Weiteres miteinander in Einklang zu bringen: In der Praxis ermöglichen gängige Systeme eher neue Dinge als dass sie gewohnte Arbeitsabläufe erleichtern -- sie verkomplizieren diese eher durch die nötige Einarbeitung in die Frameworks. Dies ist vermutlich ein Grund, weshalb keine Lösung bisher so weitgehende Akzeptanz gefunden hat, dass eine Anwendung solcher Systeme allgemeine Praxis wäre. Der Mangel an Werkzeugen für die Entwicklung von nachvollziehbaren Softwarexperimenten im Text-Mining trägt damit auch zu dem generellen Problem bei, dass die Veröffentlichung von forschungsrelevanter Software entgegen offensichtlich scheinender methodischer Grundsätze der Wissenschaft nicht die Regel ist\footnote{Zur grundlegenden Debatte um Empirie in der Computerlinguistik vgl. \citet{Pedersen2008b}, zur medialen Diskussion siehe etwa \emph{If you're going to do good science, release the computer code too},  The Guardian, 5. 2. 2010, \url{http://www.guardian.co.uk/technology/2010/feb/05/science-climate-emails-code-release} und \emph{Computational science: ...Error -- …why scientific programming does not compute},  Nature News, 13. 10. 2010, \url{http://www.nature.com/news/2010/101013/full/467775a.html}}.

\paragraph{Dokumentation als Bedingung für Wiederverwendung}

Ein zentrales Problem komplexer Softwaresysteme ist die Dokumentation. Gute Dokumentation ist eine Voraussetzung für wirklich modulare Software, denn nur sehr gut dokumentierte Komponenten werden wiederverwendet, da kein Programmierer gerne schlecht (oder gar nicht) dokumentierte Software analysiert und integriert, selbst wenn sie gut entworfen ist. So zitiert etwa Brooks (\citealt{Brooks1975}, \citealt[224]{Brooks1995}) David Parnas: 

\begin{quote}
``Reuse is something that is far easier to say than to do. Doing it requires both good design and very good documentation. Even when we see good design, which ist still infrequently, we won't see the components reused without good documentation.''
\end{quote}

\paragraph{Das Grundproblem von Softwaredokumentation}

Das Grundproblem der Softwaredokumentation ist die Trennung von Quelltext und Dokumentation (etwa in Form von Word-Dokumenten, PDF-Dateien, Websites, Wikis etc.). Da Code und Dokumentation laufend synchronisiert werden müssen, führt diese Trennung zu ständig veralteter Dokumentation \citep[169]{Brooks1995}. Trotz vorhandener Ansätze, vor allem über Dokumentationsgeneratoren (wie Javadoc, Doxygen oder DocBook), ist diese Trennung ein zwar lange erkanntes \citep{Brooks1975, Knuth1992} aber im Wesentlichen ungelöstes Problem der Softwaretechnik. Werkzeuge spielen für die Lösung dieses Problems eine wichtige Rolle, denn ohne Werkzeugunterstüzung setzten sich neuartige Konzepte nicht weitflächig durch. So beschreibt etwa schon \citet[165f.]{Brooks1975,Brooks1995} die Notwendigkeit von Unit-Tests zur Programmdokumentation -- ein Konzept das sich erst jetzt langsam durch Werkzeuge wie JUnit\footnote{JUnit, \url{http://junit.org}} durchsetzt.

Die etablierte Form der Bewahrung und Weitergabe von wissenschaftlichen Erkenntnissen ist die schriftliche wissenschaftliche Arbeit. Entsprechend kann auch Software aufbereitet werden: als illustriertes, an einen menschlichen Leser gerichtetes Dokument (\citealt{Knuth1992}, \citealt[164]{Brooks1995}). Eine solche Aufbereitung von Software nennt Knuth \emph{Literate Programming}\footnote{Beispiele für \emph{literate programs} finden sich z.B. unter \url{http://en.literateprograms.org/}}. Softwarebasierte Erkenntnisse sind nur dann vollständig dokumentiert, wenn die Veröffentlichung die entwickelte und verwendete Software enthält. In diesem Sinn kann Literate-Programming als eine Aufbereitung von Software für nachvollziehbare wissenschaftliche Veröffentlichungen betrachtet werden.

Es existieren Ansätze zur automatischen Dokumentation in bestimmten Do\-mänen (z.B. \citealt{AhlichEtAl2008} für die Dokumentation von Modellen, vgl. neuere Entwicklungen wie das Mylyn-Intent-Projekt\footnote{Mylyn Intent, \url{http://wiki.eclipse.org/Intent}}), doch es fehlt bisher -- insbesondere für die computerlinguistische Arbeit -- an modernen, integrierten Werkzeugen für automatische Dokumentation, Literate-Programming und vergleichbare Ansätze.

Eng verknüpft sind im Bereich des Text-Mining Dokumentation und Evaluation: Eine Interpretation von Ergebnissen erfordert eine vollständige Dokumentation der verwendeten Verfahren und Daten. Es existieren Ansätze einer generischen, automatischen Evaluierung von computerlinguistischen Verfahren \citep{BigertEtAl2003,Halpin2006}. Andere Ansätze sind meist für bestimmte Anwendungen spezialisiert, etwa für Textzusammenfassung \citep{Lin2004a,Lin2004b} oder maschinelle Übersetzung \citep{PapineniEtAl2002,VanZaanenAndZwarts2006}. Die existierenden Ansätze sind meist nicht mit anderen Werkzeugen integriert -- eine Ausnahme bildet hier das AnnotationDiff-Tool von Gate \citep{CunninghamEtAl2002}.

\subsubsection{Technische Ursprünge} \label{anno-grundlagen-technisch}

\paragraph{Blackboard-Architektur} \label{anno-blackboard}

Die Blackboard-Architektur gilt als Ursprung von anno\-tations\-ba\-siert\-en Systemen zur Sprachverarbeitung \citep{CunninghamAndBontcheva2006}. Die Tafel (engl. \emph{Blackboard}) ist eine alte Metapher der Softwareentwicklung: \citet{Newell1962} beschrieb als Erster explizit die Tafel als gemeinsame Datenstruktur für den Datenaustausch zwischen Methoden. In diesem Sinn ist die Tafel eine Metapher für das Prinzip von explizitem Zustand, einem Konzept das ein zentraler (und durchaus problematischer) Aspekt von Programmiersprachen ist (vgl. \citealt[405,569]{VanRoyAndHaridi2004}).

Die Kriterien einer Blackboard-Architektur (vgl. \citealt{Corkill2003}) umfassen: \begin{inparaenum}
\item mehrere kooperierende Quellen,
\item mehrere konkurrierende Hypothesen,
\item mehrere Abstraktionsebenen,
\item Feedback für die Quellen,
\item das Blackboard, ein assoziativer Speicher.
\end{inparaenum} Der Blackboard-Ansatz ist in verschiedenen Projekten angewendet worden, etwa im Hearsay-Spracherkennungsprojekt der 1970er Jahre \citep{ErmanEtAl1980} oder in der Cryptoanalyse. Hier sind die Quellen etwa Akustik, Phonetik, Artikulation, Syntax, Semantik und Pragmatik (Hearsay) bzw. Häufigkeitsanalyse, Schlüsselwörter, Codebooks und Übersetzer (Cryptoanalyse). Auf höherer Abstraktionsebene kommen Blackboard-Architekturen auch für die Entwicklung komplexer Systeme in der Industrie zum Einsatz, oder für die Entwicklung von Open-Source-Software (wo alle Programmierer die Quellen sind). Diese sehr unterschiedlichen Beispiele machen deutlich, dass die Blackboard-Architektur ein natürliches Prinzip zur Lösung komplexer Probleme darstellt.

\paragraph{Pandämonium-Modell}

\citet{Newell1962} nennt als Vorbild für seine Blackboard-Metapher das Pan\-dämon\-ium-Modell von \citet{Selfridge1959}, eine Architektur für ein Mustererkennungssystem. Als solches ist das Pan\-dämon\-ium selbst viel näher am Thema der annotationsbasierten Sprachverarbeitung als die in diesem Zusammenhang häufig zitierte Blackboard-Architektur, die ursprünglich ein sehr allgemeines Konzept des Programmierens beschreibt (expliziten Zustand, s.o.). Das von Selfridge beschriebene Pandämonium besteht aus vier Ebenen von Dämonen, die unter bestimmten Umständen die Dämonen der nächsthöheren Ebene informieren (s. Abb. \ref{tab-pan}).

\begin{figure}
\begin{center}
\begin{tabular}{c}
Entscheidungsdämonen \\ \hline
Kognitive Dämonen \\ \hline
Berechnungsdämonen\\ \hline
Datendämonen\\
\end{tabular}
\end{center}
\caption[Struktur des Pandämoniums]{Struktur des Pandämoniums von \citet{Selfridge1959}} \label{tab-pan}
\end{figure}

Schon diese Darstellung der Mustererkennung bei Selfridge enthält die zentralen Elemente des korpusbasierten, modularen maschinellen Lernens durch Klassifikation: die Datengrundlage, die Merkmalsberechnung auf den Daten, eine Verarbeitung der Merkmale, und eine Entscheidung über die Klassenzugehörigkeit. Bei der Verarbeitung im Pandämonium werden zunächst überwacht die Gewichte zwischen Berechnungsdämonen und kognitiven Dämonen ermittelt. Unnütze Dämonen werden elimiert und neue, zueinander in Konkurrenz stehende Dämonen werden erzeugt. Eine solche Eliminierung unnützer Dämonen kann im maschinellen Lernen etwa als Merkmalsauswahl modelliert werden, bei der Merkmale eleminiert werden, die schlechte Ergebnisse liefern. Die Konkurrenz der Dämonen entspricht auch einem zentralen Kernkonzept von Multiagentensystemen. Das Pandämonium bildet so die konzeptuelle Grundlage einer ganzen Reihe von Metaphern der modularen Softwarentwicklung, wie der dargestellten Blackboard-Architektur und Multiagentensystemen (s.u.), aber auch etwa dem Open-Spaces-Ansatz\footnote{Open-Spaces, vgl. \emph{Loosely Coupled Communication and Coordination in Next-Generation Java Middleware}, \url{http://today.java.net/pub/a/today/2005/06/03/loose.html}} oder Service-orientierten Architekturen.

\paragraph{Die Agentenmetapher als gemeinsames Grundkonzept} \index{Agenten!als Metapher in der Softwareentwicklung}

Ein etabliertes Konzept der Künst\-lich\-en Intelligenz (KI) sind Agenten \citep{RussellAndNorvig2003}. Diese Agenten interagieren über Sensoren und Aktuatoren in Multiagentensystemen \citep{Vlassis2003,Lin2007}. Wenn auch bisher selten so formuliert (eine Ausnahme bildet hier etwa \citealt{YuEtAl2008}), entsprechen sprachverarbeitende Komponenten solchen Agenten, die in einer Welt aus annotierten Korpora agieren.

Letztlich ist eine modulare Konzeption von Software der Grundgedanke aller Lösungsansätze von strukurierter \citep{Dijkstra1968,Wirth1971,DahlEtAl1972} und objektorientierter Programmierung, über das Konzept von Multiagentensystemen bis zu serviceorientierten Architekturen (wo \emph{Services} die Schnittstellen der Module sind, vgl. \citealt{GruhnAndThiel2000}). All diese Konzepte sind unterschiedliche Metaphern für das Zusammenspiel der Einzelteile in Softwarekomponentensystemen (vgl. \citealt{Corkill2003}).

Der Kern des Konzepts der Multiagentensysteme, das mit dem Pandämonium und der Black\-board-Architektur übereinstimmt ist so die ursprüngliche Bedeutung von \emph{Agent} als 'der handelnde Teil, die wirksame Substanz, das Tuende'. Aus dem Zusammenspiel dieser wirksamen Substanzen ergibt sich die Lösung komplexer Probleme, hier etwa für Aufgaben des Text-Mining. Eine Anwendung der allgemeinen Agentenmetapher im Bereich des Text-Mining verspricht dabei eine  Generalisierung sprachspezifischer Konzepte.

\subsection{Modularität durch Annotation} \label{anno-theorie} \index{Annotation!für Modularität}

Annotationen spielen eine zentrale Rolle in computerlinguistischen Softwaresystemen. Der Begriff der \emph{Annotation} kann  für die Anreicherung von Daten mit Metadaten oder für diese Metadaten selbst verwendet werden. Annotierte Texte dienen etwa als Input für Suchwerkzeuge, für linguistische Untersuchungen, oder zur Erzeugung neuer Annotationen. \citet[448]{McEnery2003} bezeichnet Annotationen als Treibstoff der maschinellen Sprachverarbeitung (``the raw fuel of NLP''), da sie es maschinellen Verfahren ermöglichen, die Intuition von menschlichen Experten, deren Analysergebnisse in Form von Annotationen festgehalten wurden, mithilfe maschinellen Lernens zu reproduzieren (s. Beispiele unten). Da außerdem über den Vergleich von Annotationen (von verschiedenen Verfahren, Goldstandard, etc.) ermittelt werden kann, welches Verfahren die Expertenmeinung am treffendsten reproduziert, bilden Annotationen den konzeptionellen und technischen Kern des Text-Mining, gerade in Hinblick auf die beschriebenen Ziele modularer, optimierter Komponenten.

\subsubsection{Fachliche Ursprünge}\label{ml-base-corpus} \label{anno-grundlagen-fachlich}

Dieser Abschnitt beschreibt die Grundlagen der Korpuslinguistik, basierend auf Darstellungen in eigenen Vorarbeiten \citep{Steeg2007}. Im anschließenden Abschnitt \ref{dynamische-annotation}, S. \pageref{dynamische-annotation} wird darauf aufbauend die Entwicklung der Annotation als logische Zwischenschicht in Systemen zur maschinellen Sprachverarbeitung beschrieben, einer zentralen theoretischen Grundlage der in den folgenden Kapiteln entwickelten Konzepte und Werkzeuge.

\paragraph{Korpuslinguistik}

Grundlegender fachlicher Ursprung des Annotationskonzepts ist die Korpuslinguistik, deren Gegenstand die Erstellung und Auswertung von Textkorpora \citep{Koehler2005}, sowie die Entwicklung von Werkzeugen für die Erstellung und Auswertung von Textkorpora ist. Die Korpora dienen dabei zur Beantwortung linguistischer Fragestellungen. Dementsprechend definiert etwa \citet{McEnery2003} ein Textkorpus als große Menge linguistischer Evidenz. Eine solche, auf empirischen Daten basierende Linguistik kann als datengetrieben beschrieben werden. Dieser Ansatz ist wissenschaftstheoretisch fundiert, da echte Daten Annahmen des Linguisten stützen, aber auch widerlegen können (vgl. \citealt{Labov1975,Labov1996}). Damit sind Korpora eine sehr wertvolle Quelle für die linguistische Arbeit. Korpuslinguistische Methoden werden nicht nur in der maschinellen Sprachverarbeitung angewendet, sondern auch in anderen sprachwissenschaftlichen Bereichen, etwa in der allgemeinen Sprachwissenschaft oder den Philologien, sowie auf verschiedenen Ebenen der Sprache, etwa Morphologie, Syntax, Semantik oder Pragmatik \citep[2]{McEneryAndWilson1996}. Da ein gro"ser Teil der Auswertung von Korpora mithilfe von statistischen Verfahren erfolgt, wird diese Art der Arbeit mit Korpora auch als \emph{Quantitative Linguistik} bezeichnet (s. \citealt{ManningAndSchuetze1999, Koehler2005}). 

\paragraph{Korpusannotation} \label{korpusannotation}

Annotierte Korpora sind Korpora, die mit verschiedenen linguistischen Informationen angereichert wurden \citep[24]{McEneryAndWilson1996}. Bei diesen Informationen handelt es sich linguistisch betrachtet nicht um die Hinzufügung neuer Informationen, sondern um die Explizitmachung bereits vorhandener linguistischer Struktur im Sprachmaterial \citep[453]{McEnery2003}. Ein Beispiel für Korpusannotation ist etwa die Kennzeichnung von Wortarten durch POS-Tags. Andere Annotationen enthalten etwa Informationen über Stammformen (als Ergebnis einer Stammformenreduktion oder Lemmatisierung), über die syntaktische Struktur (solche Korpora werden auch \emph{Baumbanken} genannt), sowie semantische, stilistische oder Informationen zur Diskursstruktur. Dabei sind morphologische Annotationen verbreiteter als syntaktische Annotationen und diese wiederum verbreiteter als semantische oder pragmatische Annotationen. Eine Anreicherung durch Annotation basiert immer auf einer bestimmten Interpretation der Daten. Beim Annotieren ist daher zu beachten, dass der ursprüngliche Text auch nach der Annotierung noch in seiner Rohform verfügbar sein sollte, etwa durch \emph{dynamische Annotation} (s. \citealt{BendenAndHermes2004}, \citealt{HermesAndBenden2005}, vgl. Abschnitt \ref{dynamische-annotation}, S. \pageref{dynamische-annotation}).

Über die rein linguistische Arbeit hinaus ermöglichen annotierte Korpora es Computerprogrammen, die Intuitionen von Experten (die Autoren der Annotationen) in Bezug zu den annotierten Texten zu setzen und so menschliche Intuitionen zu reproduzieren. Grundgedanke ist hierbei, dass von Menschen annotierte Korpora vom Computer verarbeitet werden, wobei der Computer die Muster im Kontext der Annotationen lernt. Konkret könnte etwa gelernt werden, dass im Englischen \emph{can} als Nomen oft nach \emph{a} oder \emph{the}, als Verb dagegen nach \emph{to} vorkommt (wenn zuvor Wortarten annotiert wurden). Ein Beispiel hierfür sind etwa auf Hidden-Markov-Modellen (HMM) basierende POS-Tagger, die aus annotierten Korpora lernen und dadurch in die Lage versetzt werden, neue Texte zu annotieren (s. etwa \citealt{ManningAndSchuetze1999}). Daher bilden annotierte Korpora eine etablierte Grundlage für das maschinelle Lernen in der Sprachverarbeitung (\citealt[459]{McEnery2003}).

\subsubsection{Dynamische Annotation als logische Applikationsschicht} \label{dynamische-annotation}

\begin{figure}
\begin{center}
  \includegraphics[width=11.5cm]{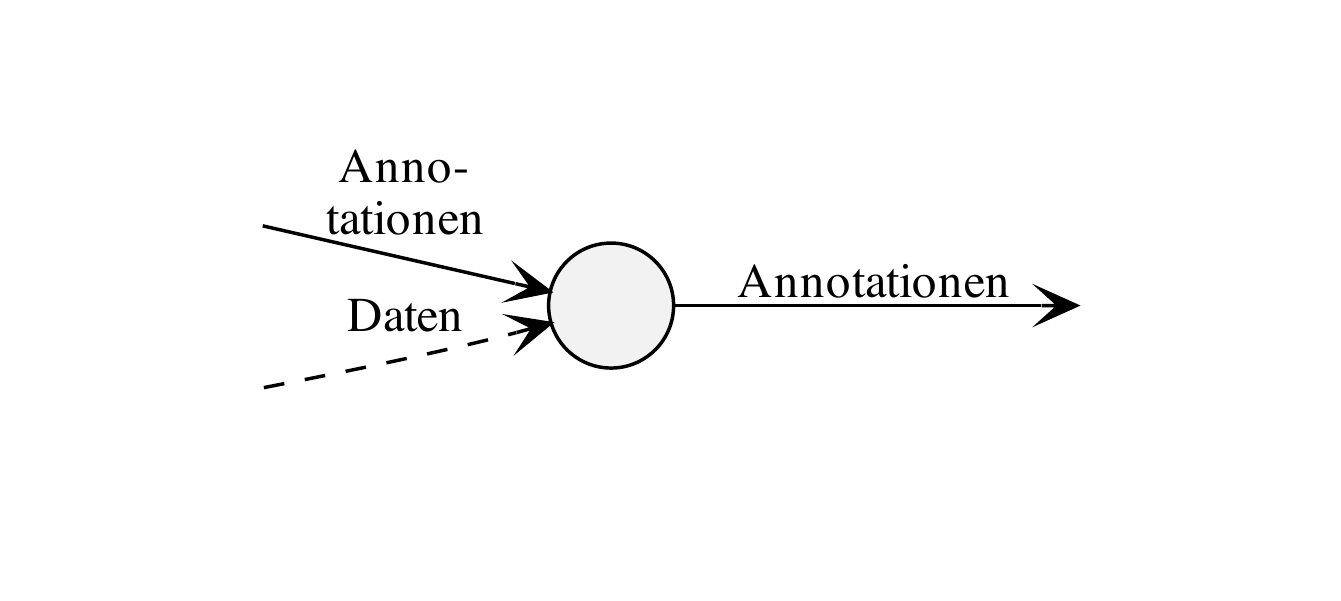}
  \caption{Input und Output einer Komponente im Text-Mining} \label{agent-io}
\end{center}
\end{figure}

Wie oben dargestellt spielen annotierte Korpora sowohl für die rein linguistische Arbeit wie für die maschinelle Sprachverarbeitung und damit für das Text-Mining eine herausragende Rolle. Aus diesem Grund stellen Annotationen das konzeptuelle Austauschformat der Komponenten einer Text-Mining-Architektur (oder SALE) dar: Einzelne Komponenten bekommen als Input (in Agenten-Terminologie über ihre Sensoren) vorhandene Annotationen (und optional die Daten, d.h. den Text), und produzieren auf dieser Basis neue Annotationen (in Agenten-Terminologie mit ihren Aktuatoren). Abbildung \ref{agent-io} veranschaulicht dieses Prinzip.

Die gängigen computerlinguistischen Komponentensysteme wie Gate oder UIMA basieren auf dem Annotationskonzept, das auf ATLAS (Architecture and Tools for Linguistic Analysis Systems, \citealt{BirdAndLiberman1999}) zurückgeht. Dabei ist es aus verschiedenen Gründen (s.u.) vorteilhaft, den Text und die Annotation zu trennen (\emph{Stand-off markup} oder dynamische Annotation, vgl. \citealt{BendenAndHermes2004}). Die Annotation bildet so eine logische Ebene zwischen den Daten und den darauf operierenden Verfahren, und bildet so einen zentralen Teil der maschinellen Sprachverarbeitung.

\paragraph{Ursprünge der physischen Trennung von Daten und Annotation}

Schon mit der aufkommenden Verbreitung von Markup-Sprachen wie HTML und XML wurde erkannt, dass eine Trennung von ausgezeichneten Daten und Metadaten statt einer gemeinsamen Speicherung wie in üblichen Markup-Dokumenten für die computerlinguistische Praxis sinnvoll ist. So nennen etwa \citet{ThompsonAndMcKelvie1997} drei Gründe für eine solche Trennung: \begin{inparaenum} 
    \item Kopieren der Daten in ein Markup-Format ist nicht immer möglich, z.B. für schreibgeschütztes Material oder zu große Datenmengen.
    \item Überlappendes, hierarchisches Markup.
    \item Rechtliche Probleme, z.B. kann der Text selbst urheberrechtlich geschützt sein, während die Auszeichnungen weitergegeben werden sollen (vgl. zur rechtlichen Situation \citealt{HermesAndBenden2005}).
\end{inparaenum} Darüber hinaus sollte aus methodischen Gründen der Rohtext immer verfügbar sein (vgl. Abschnitt \ref{korpusannotation}, S. \pageref{korpusannotation}).

Ihren Ursprung hat diese Trennung von Daten und Annotation im EU-geförderten MUL\-TEXT-Projekt, das den Schwerpunkt auf wiederverwertbare Korpora setzte, sowie im DARPA-Projekt TIPSTER \citep{Harman1992}, in dessen zweiter Projekthälfte Softwarearchitekturen für standardisierte Text\-ana\-lyse-Komponenten (und damit Text-Mining) im Mittelpunkt standen. Dies führte dazu, dass bereits im Corpus Encoding Standard (CES) von 1996 eine Form von SGML-basiertem Standoff-Markup enthalten war. Diese erste Form stellte allerdings noch keine vollständige Trennung von Daten und Annotation her: grundlgende Analysen wurden mit den Daten vermischt -- so wurden Token in der Hauptdatei annotiert, und diese von anderen Dateien referenziert und mit Zusatzinformationen angereichert, vgl. \citealt{BendenAndHermes2004} zu den Problemen einer solchen unvollständigen Trennung.

\paragraph{Weiterentwicklung des Annotationskonzepts in ATLAS} \label{atlas}

\begin{figure}
\begin{center}
  \includegraphics[width=9cm]{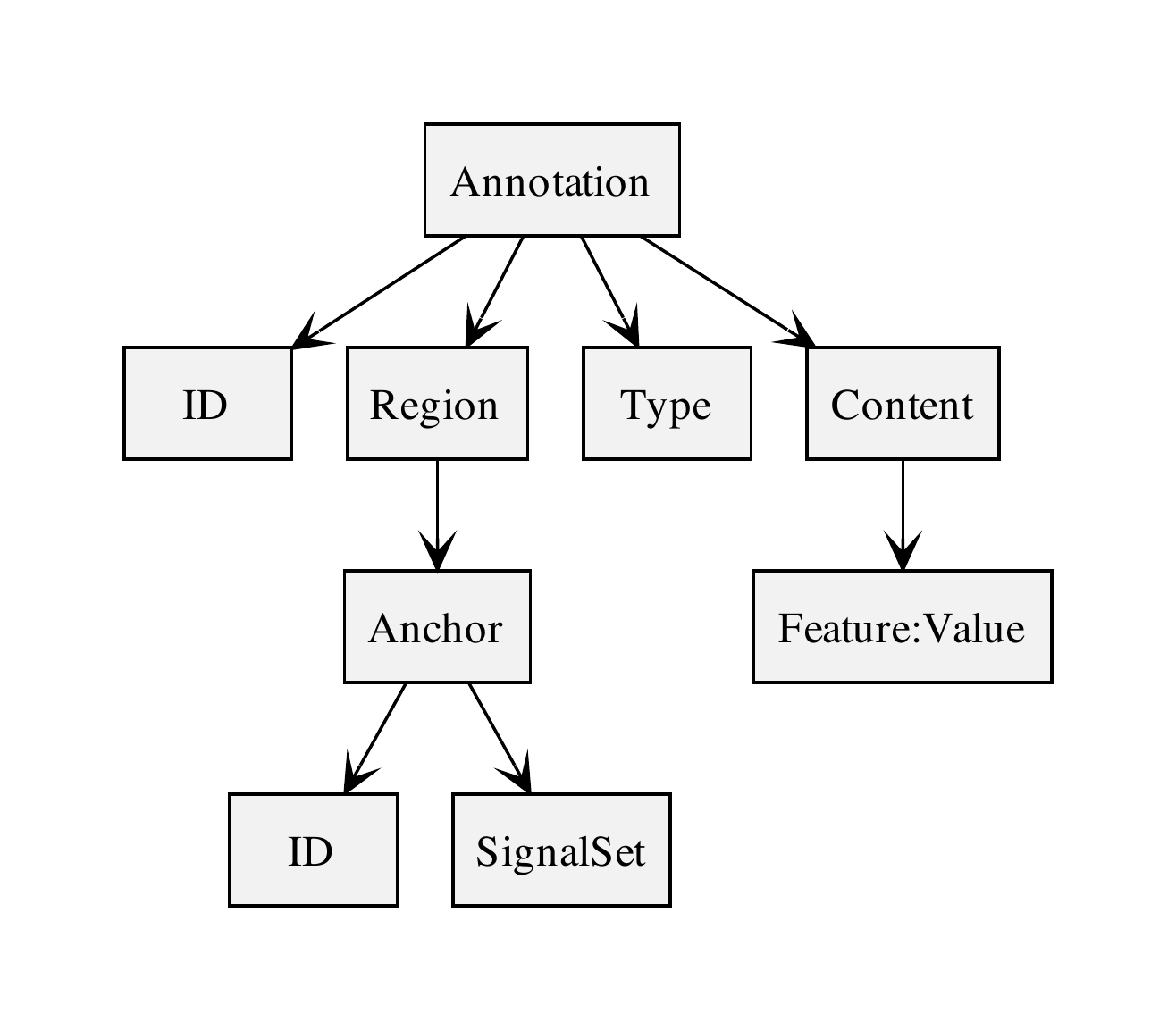}
  \caption[ATLAS-Objektmodell]{ATLAS-Objektmodell aus \citealt{BirdAndLiberman1999}}
  \label{atlas-model}
\end{center}
\end{figure}

Eine entscheidende Weiterentwicklung erfuhr das Annotationskonzept mit dem ATLAS-Projekt (\emph{Architecture and Tools for Linguistic Analysis Systems}, vgl. \citealt{BirdAndLiberman1999}), das Vorbild für spätere Entwicklungen wie Gate 2 \citep{CunninghamEtAl2002} und UIMA wurde. Vor ATLAS bestanden sprachverarbeitende Programme konzeptuell aus zwei Ebenen: der Anwendung und der Datenhaltung. Dazwischen etablierte ATLAS als logische Ebene die Annotation\footnote{In ersten Veröffentlichungen wird die logische Schicht in ATLAS als Annotationsgraph (\emph{annotation graph}) mit beschrifteten Kanten beschrieben, später vereinfacht zu einer Menge von Annotationen (\emph{annotation set}).}. Ziel einer solchen Schicht ist die Etablierung von flexiblen, erweiterbaren Werkzeugen und Frameworks durch die Entkoppelung der direkten Verbindung zwischen Daten und Algorithmen. Die Annotationsobjekte verweisen dabei auf Regionen in den Daten und enthalten die Metadaten (vgl. Struktur der Annotationsobjekte in Abb. \ref{atlas-model}).

Zum Speichern der Annotationen, und damit der Analyseergebnisse, gibt es verschiedene Standards (wie TEI, CES, XCES, TUSNELDA oder Tiger), die von den verschiedenen SALE-Systemen als Exportformat unterstützt werden und in diese importiert werden können. Intern werden meist eigene Formate verwendet. 

Als persistente Darstellung von Analyseergebnissen und als ``information aqueduct'' \citep{BirdEtAl2000} zwischen den Komponenten in ``pipelined apps'' (ebd.) bildet das Konzept der Annotation so den Kern von Text-Mining-Software. Da neben dieser logischen Schicht die Annotation eine wesentliche Rolle für die Evaluierung von sprachverarbeitenden Verfahren spielt, und annotierte Korpora die Datengrundlage linguistischer Untersuchungen bilden, kann die Annotation allgemein als das zentrale Konzept des Text-Mining betrachtet werden.

\section{Werkzeuge zur Modellierung im Text-Mining} \label{chapter-tools}

\subsection{Werkzeuge zur Modellierung, Dokumentation und Evaluierung} \label{ziele-tools}

Die Modellierung der Multiagentensysteme soll möglichst unkompliziert erfolgen. Die Agentenschnittstellen sollen als Teil der eigentlichen Funktionalität, d.h. direkt im Code und nicht in einer separaten Konfigurationsdatei definiert werden. Experimente sollen über eine einfache Fachsprache (domain-specific language, s. Abschnitt \ref{basis-dsl}, S. \pageref{basis-dsl}) definiert werden. Eine solche Modellierung von Experimenten über eine Fachsprache ermöglicht eine einfache Konfiguration der automatischen Evaluierung und Dokumentation: Die verschiedenen zu evaluierenden Aufbauten werden in der Fachsprache modelliert und können so automatisch ausgeführt und aufbereitet werden (vgl. Abschnitt \ref{api-doc}, S. \pageref{api-doc}). Die Dokumentation der Experimente soll die beteiligten Agenten, Visualisierungen ihrer Interaktionen, sowie tabellarische Aufbereitungen von Ergebnissen enthalten (vgl. Abschnitt \ref{appl-senseval-results}, S. \pageref{appl-senseval-results}). Im Unterschied zum klassischen Literate-Programming soll mit den Werkzeugen nicht die menschenlesbare Form editiert werden, sondern diese soll aus dem Code generiert werden (wie bei den Dokumentationsgeneratoren DocBook, Doxygen oder Javadoc). Ein Vorteil dieses Ansatzes ist, dass bestehender Code leicht integriert werden kann, da kein spezifisches, neues Format vorausgesetzt wird (wie bei bestehenden Literate-Programming-Lösungen, z.B. \emph{noweb}\footnote{\emph{A Simple, Extensible Tool for Literate Programming}, \url{http://www.eecs.harvard.edu/nr/noweb/}}).

Die im Folgenden beschriebenen Konzepte und Werkzeuge sollen die Entwicklung von modularen, optimierten und wohldokumentierten Text-Mining-Komponenten vereinfachen und so einen entscheidenden Teil zur Entwicklung von funktionierenden Text-Mining-Systemen beitragen\footnote{Eine Sicht auf die angestrebten Werkzeuge ist die eines Testframeworks für computerlinguistische Komponenten, konzeptuell vergleichbar mit den xUnit-Frameworks für testgetriebene Entwicklung \citep{Beck2003} von generellen Softwarekomponenten.}. Zur Evaluierung der in Kapitel \ref{chapter-tools} beschriebenen Werkzeuge werden diese in Kapitel \ref{chapter-appl} für die Umsetzung komplexer Text-Mining-Experimente eingesetzt. Eine zentrale Frage ist dabei, inwiefern die möglichst einfachen Konzepte und Werkzeuge sowohl für grundlegende wie auch für komplexe Anwendungen eingesetzt werden können.

\subsection{Basistechnologien}

Im Folgenden werden einige softwaretechnische Konzepte beschrieben, welche die Basis der Umsetzung der Werkzeuge bilden: statisch typisierte Programmiersprachen und Fachsprachen (domain-specific languages, DSLs).

\subsubsection{Statische Typisierung} \label{statische-typisierung} \index{Typisierung!statische, generell}

Schon im ATLAS-Objektmodell enthält eine Annotation ein \emph{Type}-Attribut (vgl. Abbildung \ref{atlas-model}, S. \pageref{atlas-model}). Dieses spezifiziert die Art der Annotation, z.B. \emph{Token}, \emph{POS}, \emph{Bedeutung}, etc. Das Konzept der Typisierung geht zurück auf die Aristotelische Kategorienlehre und entspricht generellen Prinzipien der Kognition, deren zentrales Grundkonzept die Kategorisierung ist (vgl. \citealt{Hawkins2004,Harnad2005}). Die Typisierung der Annotation ist damit ein wesentlicher Aspekt der Organisation von Information beim Text-Mining.

In der Softwaretechnik kann das Konzept der Typisierung nicht nur als zusätzliche Information verwendet werden (diese Daten sind eine Zahl, eine Zeichenkette, etc.), sondern zusätzlich zur "Uberprüfung, ob die Daten nur in sinnvoller Weise verwendet werden (z.B. zur Vermeidung von mathematischen Operationen mit Zeichenketten). In statisch typisierten Programmiersprachen werden solche Kontrollen vom Compiler ausgeführt und verhindern so, dass Fehler erst zur Laufzeit auftreten. So führt statische Typisierung zu robusteren Programmen mit weniger Fehlern. In der Softwaretechnik hat sich diese "Uberprüfung bewährt und gilt als ein Grund für die industrielle Dominanz von Programmiersprachen wie C++, Java oder C\#. Die Typisierung stellt zudem eine zusätzliche Dokumentation der Quelltexte dar, die laufend und automatisch vom Compiler überprüft wird. Dabei wird für einen eingeschränkten Bereich das grundlegende Problem der Synchronisation von Quelltext und Dokumentation gelöst.

Aufgrund der beschriebene Vorteile typisierter Sprachen und der Tatsache, dass Annotationen die Grundlage von Text-Mining-Software sind und typisiert beschrieben werden können (s. Abschnitt \ref{anno-theorie}, S. \pageref{anno-theorie}) ist es sinnvoll, Text-Mining typsicher zu modellieren. Die entwickelten Werkzeuge stellen daher ein Framework zur typsicheren Modellierung im Text-Mining dar und werden im Folgenden abgekürzt als \emph{TM2} (Typsichere Modellierung im Text-Mining) bezeichnet.

\subsubsection{Domain-specific languages: textuelles MDD} \label{basis-dsl}

Fachsprachen (domain-specific languages, DSL) sind formale Hochsprachen, die auf eine bestimmte Anwendung spezialisiert sind und mit denen man nah am Problem programmiert (\citealt[53]{HuntAndThomas2003}, \citealt[3]{Parr2007}). Sie bilden eine Form von domänenspezifischer textueller Modellierung oder formaler Notation, etwa für spezifische Aufgaben oder für Fachbereiche mit eigenem Vokabular\footnote{vgl. eigene Vorarbeiten in \citet{SteegEtAl2008} zur Implementierung einer formalen Notation für \emph{Functional Grammar} und \emph{Functional Discourse Grammar}.}. Beispiele für DSLs sind etwa die Datenbankabfragesprache SQL oder die Graphenbeschreibungssprache DOT. Ziel von domänenspezifischen Sprachen ist allgemein die Ermöglichung einer Formalisierung, die den Problembereich einfach und elegant erfasst. So kann etwa ein Graph in DOT folgendermassen modelliert werden:

 \begin{minipage}{2cm}
\begin{lstlisting}
digraph{

  1->2
  2->3
  2->4

}
\end{lstlisting}
  \end{minipage}
  \begin{minipage}{0.48\textwidth}
  \includegraphics[width=3cm]{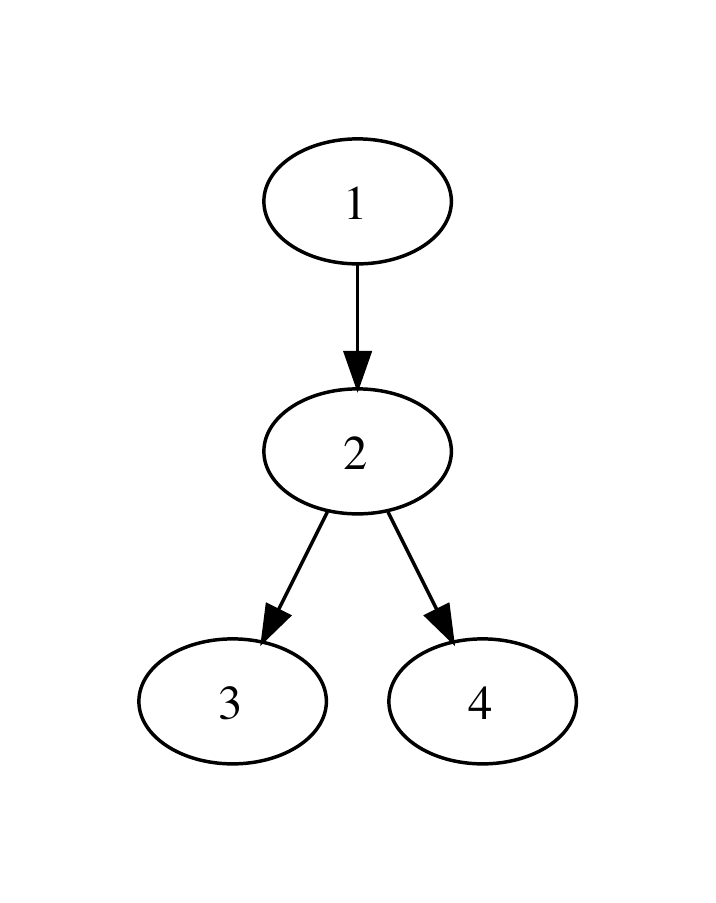}
  \end{minipage}

\paragraph{Implementierung von DSLs}

Zur Implementierung von DSLs lassen sich zwei Ansätze unterscheiden: \emph{eigenständige} (oder \emph{externe}) und \emph{eingebettete} (oder \emph{interne}) DSLs. Unter einer eigenständigen DSL versteht man eine komplett neu entwickelte Sprache, für die in der Regel eine Grammatik in einer eigenen Grammatikbeschreibungssprache geschrieben wird. Aus dieser wird mit Mitteln der modellgetriebenen Software-Entwicklung ein Parser generiert (z.B. mit ANTLR, vgl. \citealt{Parr2007}), der Quelltext in der durch die Grammatik beschriebenen Sprache verarbeiten kann. Aus der Grammatik können zudem Werkzeuge für die Sprache (z.B. Editoren) generiert werden (z.B. mit Xtext\footnote{Xtext Language Framework, \url{http://www.eclipse.org/xtext}}). Auf Basis der Objekte des Parse-Trees, der vom generierten Parser geliefert wird, kann dann Quelltext in der DSL verarbeitet werden.

Unter einer eingebetteten DSL versteht man die Verwendung einer existierende Sprache, die so flexibel ist, dass sie für die speziellen Zwecke einer Domäne angepasst werden kann. Frühestes Beispiel für eine flexible Sprache, die eine Definition neuer Konstrukte ermöglicht (und sich so für die Entwicklung eingebetteter DSLs eignet), ist Lisp. Eine weitere ältere Sprache mit eingebetteten DSLs ist Prolog, deren DCG-Bibliothek eine DSL zur Beschreibung von Grammatiken ist, und so als eingebettete DSL zur Entwicklung eingebetteter DSLs gesehen werden kann. Eine heute sehr beliebte Sprache zur Entwicklung eingebetteter DSLs ist Ruby -- eine Sprache, die durch das Framework \emph{Ruby on Rails} (das stark auf DSL-Konzepten basiert) weite Verbreitung fand. Diese Ursprünge erklären, warum eingebettete DSLs lange mit dynamisch typisierten Basissprachen (\emph{host languages}) assoziiert wurden. Mit neueren Sprachen wie Scala bleiben eingebettete DSLs jedoch nicht auf den Bereich dynamisch typisierter Sprachen beschränkt (s.u.).

\paragraph{Wahl der Implementierungssprachen} \label{impl-langs}

Die eingesetzten Sprachen sollen verbreitet sein und Unterstützung für weitgehende statische Typisierung in Form von parametrischer Polymorphie bieten. Diese ermöglicht, dass das Framework generische Elemente enthält, deren konkreter Typ erst durch die Implementierungen festgelegt wird, die das Framework verwenden (s.u.). Darüberhinaus soll die Möglichkeit für eine einfache Entwicklung von eingebetteten oder eigenständigen DSLs für die Modellierung von Experimenten gegeben sein.

Java ist eine moderne, weit verbreitete\footnote{Im Tiobe-Index, der die Häufigkeit der Suche nach bestimmten Programmiersprachen im Web erfasst, belegt Java im überwiegenden Teil der letzten 10 Jahre den ersten Platz, s. \url{http://www.tiobe.com/index.php/content/paperinfo/tpci/index.html}.}, hochperformante\footnote{Siehe etwa \url{http://en.wikipedia.org/wiki/Java_performance}.} objektorientierte Programmiersprache. Da Java zudem open-source unter der GPL2 zugänglich ist und die JVM zunehmend von weiteren Sprachen verwendet wird, ist zu erwarten, dass Java langfristig verfügbar sein wird. Java verfügt über parametrischen Polymorphismus (\emph{Java Generics}) und umfangreiche eigene und externe Programmbibliotheken, auch zur Entwicklung von eigenständigen DSLs, z.B. mit dem Xtext-Framework.

Alternative, Java-kompatible JVM-Sprachen eignen sich zur Implementierung von eingebetteten DSLs, z.B. Scala\footnote{Scala enthält eine Vielzahl von Eigenschaften, die zur Implementierung von eingebetteten Sprachen nützlich sind. Dazu zählen etwa die Umsetzung von Operatoren als Methoden und \emph{implicit conversions}, vgl. Abschnitt \ref{werk-dsl-intern}, S. \pageref{werk-dsl-intern}.}, Clojure (ein Lisp-Dialekt, dynamisch typisiert), und JRuby (eine Ruby-Im\-ple\-men\-tier\-ung in Java, dynamisch typisiert). All dies macht Java, oder allgemeiner die JVM, zu einer optimalen Wahl für ein Framework, das langfristig die Entwicklung von Text-Mining-Anwendungen auf Basis typsicherer Modellierung praktisch vereinfachen soll.

Eine Unterstützung für parametrische Polymorphie gibt es in Java seit der Einführung von \emph{Java Generics} in Version 1.5 der Sprache. Um die Vorteile der statischen Typisierung zu nutzen, wird daher bei der Umsetzung der Werkzeuge konsequent auf \emph{Generics} gesetzt. Dabei ergibt sich grundsätzlich folgendes Problem: Da Generics in Java über \emph{type erasure} implementiert sind, sind die Typinformationen zur Laufzeit nicht verfügbar (sie sind nicht \emph{reifiable}). Dabei gibt es eine Ausnahme: Bei einer Implementierung eines generischen Interface, dessen Typ oder Typen in der implementierenden Klasse definiert werden (und nicht erst bei der Instanziierung von Objekten der implementierenden Klasse) ist es möglich, die generischen Typen zu Laufzeit abzufragen. Im Kontext von TM2 sind dies etwa die Typen der Eingabe- und Ausgabeannotationen eines Agenten, die bei der Implementierung des Interface definiert werden (vgl. Abschnitt \ref{werk-api}, S. \pageref{werk-api}). So können die Typinformationen von Experimenten zur Laufzeit ermittelt und als Teil der generierten Dokumentation festgehalten werden (vgl. Abschnitt \ref{api-doc}, S. \pageref{api-doc})\footnote{Scala 2.8, das zur Implementierung einer DSL-Schicht in Abschnitt \ref{werk-dsl-intern}, S. \pageref{werk-dsl-intern} verwendet wird, bietet in diesem Zusammenhang besondere Unterstützung in Form von \emph{Manifests}, die einen einfacheren und umfassenderen Zugriff auf die Typ-Informationen zur Laufzeit als mit Java ermöglichen.}.

Aus den beschriebenen Basistechnologien ergibt sich der Rahmen für die Implementierung der Werkzeuge und des Frameworks im weiteren Verlauf dieses Kapitels: Java als typsichere, etablierte Sprache für das zentrale Framework und dessen API, Xtext für die Entwicklung einer eigenständigen DSL, sowie Scala für die Entwicklung einer typsicheren, eingebetteten DSL.

\subsection{Ein Framework für typsicheres Text-Mining} \label{werk-api} \label{section-api}

\subsubsection{Fachliche Konzepte} \label{api-fach}

\begin{figure}
\begin{center}
  \includegraphics[width=11.5cm]{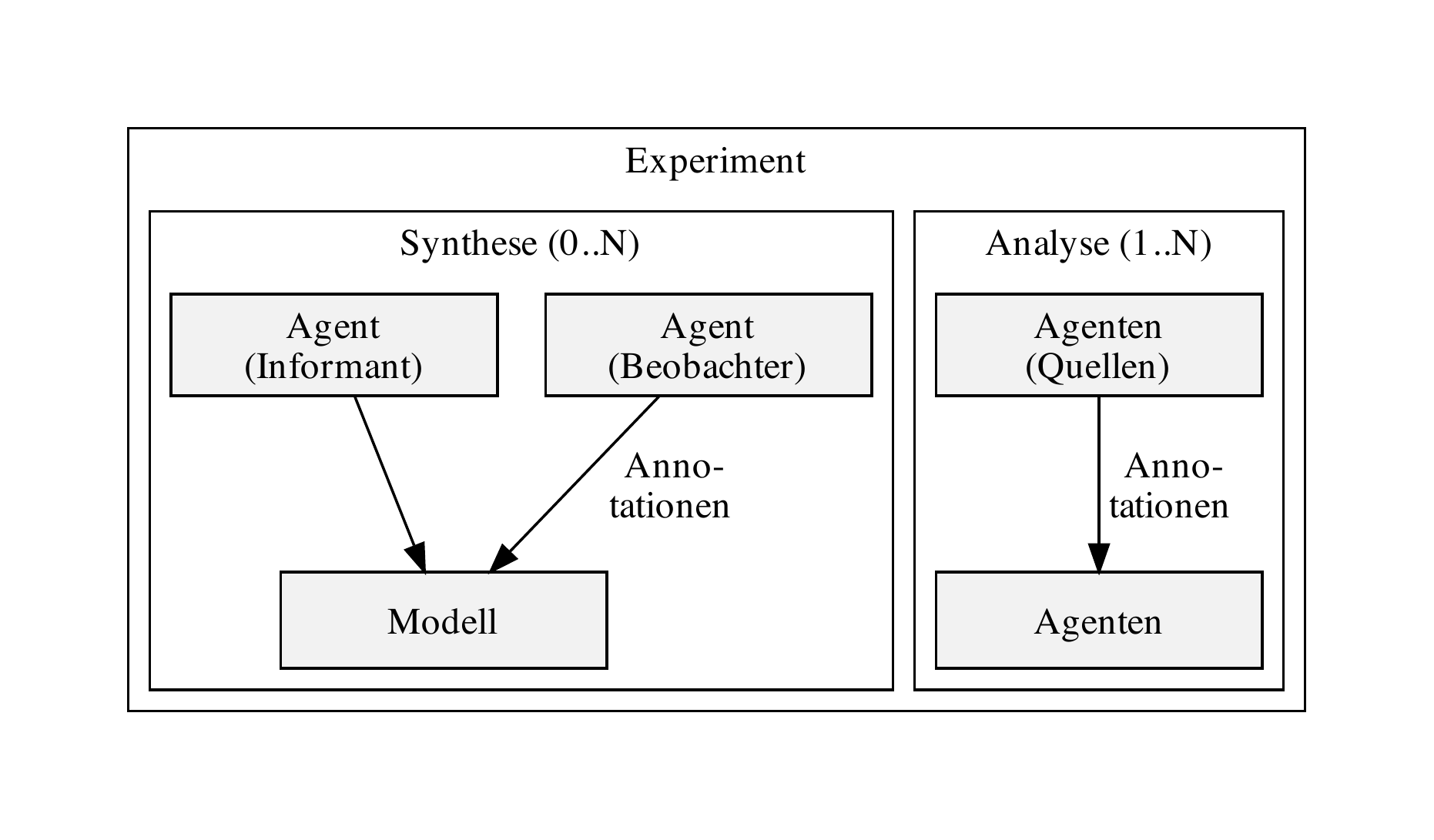}
  \caption{Zentrale API-Konzepte und Zusammenhang}
  \label{konzepte-zusammenhang}
\end{center}
\end{figure}

Zur Implementierung der in Kapitel \ref{chapter-theory} beschriebenen inhaltlichen Konzepte des Text-Mining (Annotationen, Agenten, Experimente, Evaluation) werden diese im Folgenden um Analysen, Synthesen und Modelle ergänzt. Die grundlegenden neuen Konzepte sind dabei die lineare (Analyse) und die zusammenführende (Synthese) Interaktion zwischen Agenten. Analysen sind Prozesse, bei denen ein Zielagent die Ergebnisse seiner Quellen verarbeitet, und daraus neue Ergebnisse erzeugt. Synthesen beschreiben den Zusammenfluss der Ergebnisse zweier Agenten zur Bildung eines Modells. Experimente bestehen aus solchen Agenteninteraktionen, d.h. aus Analysen und Synthesen. Das Zusammenspiel dieser Begriffe beschreibt Abbildung \ref{konzepte-zusammenhang}.

\paragraph{Typsichere Modellierung} \index{Modellierung!typsicher} \index{Typisierung!bei der Modellierung} \index{Annotation!typsichere A.}

Eine im Projekt Tesla \citep{HermesAndSchwiebert2007} entwicklete Implementierung der Annotationen ist die \emph{typsichere} Annotation, die in Java mit \emph{Generics} formuliert werden kann: \lstinline!Annotation<T>!, d.h. eine Annotation vom Typ T. Wird nun in einem Programm eine solche Annotation verwendet, die etwa auf den Ganzzahltyp \emph{Integer} typisiert ist, kann der Java-Compiler garantieren, dass mit den Daten der Annotation nur gültige Operationen ausgeführt werden.

Im Kontext dieser Arbeit wird dieser Gedanke auf die Interaktion der Agenten ausgeweitet: Annotationsbasierte Agenten interagieren über typsichere Annotationen, und zwar in typsicheren Interaktionen, z.B.: \lstinline!Analysis<T>!, d.h. eine Analyse vom Typ T, wobei T der analysierte Annotationstyp ist. Dies erlaubt eine sehr weitgehende "Uberprüfung bei der Modellierung von Experimenten: Für eine solche Analyse vom Typ T etwa werden als Quellen Agenten benötigt, die Annotationen vom Typ T produzieren, und als Ziele Agenten, die Annotationen vom Typ T verarbeiten können. So kann schon vor der Ausführung die Struktur und Semantik von Experimenten validiert werden.

\subsubsection{API der zentralen Elemente} \label{api}

Im Folgenden wird die API der zentralen Elemente zur Modellierung von Text-Mining-Ex\-pe\-ri\-ment\-en mit TM2 beschrieben, speziell von Annotationen, Agenten und Interaktionen zwischen Agenten (Analysen und Synthesen), sowie zur Durchführung und Evaluation von Experimenten.

\lstset{language=Java}

\paragraph{Annotationen}

Annotationen bilden die Grundlage des Text-Mining in der hier beschriebenen Ausprägung. Dem ATLAS-Objektkmodell folgend, haben Annotationen im Wesentlichen einen Wert, eine Position und einen Typ (vgl. Abschnitt \ref{atlas}, S. \pageref{atlas}). Dieser Typ der Annotation wird in der TM2-API mit Generics typsicher modelliert.

Eine normale Instanziierung einer generischen Klasse in Java erfordert eine doppelte Angabe der generischen Typen\footnote{Für Java 7 ist ein neues Sprachfeature angekündigt, das diese Verdopplung vermeidet, s. \url{http://openjdk.java.net/projects/coin/}}, etwa für eine Map, die als Schlüssel \emph{Strings} und als Werte \emph{Integers} enthält: 

\begin{lstlisting}
Map<String,Integer> map = new HashMap<String,Integer>();
\end{lstlisting}

Statische Factory-Methoden bieten hier eine Alternative, da sie eine Form von Typ-Inferenz er\-mög\-lich\-en (vgl. \citealt[9]{Bloch2008}). So kann etwa eine Annotation vom Typ \emph{String} auf folgende Weise instanziiert werden (unter Verwendung einer Klasse \emph{Annotations}, die statische Funktionalität für Annotationen bereitstellt): 

\begin{lstlisting}
Annotation<String> annotation = Annotations.create("Firma", 0, 1, data);
\end{lstlisting}

Neben dem Typ enthalten Annotationen ihren Wert (oben etwa \emph{Firma}), Start- und Endposition, sowie eine Referenz auf die annotierten Daten (vgl. Abschnitt \ref{impl-daten}, S. \pageref{impl-daten}). Die Implementierung der oben verwendeten Factory-Methode sieht dabei folgendermaßen aus:

\begin{lstlisting}
public static <T> Annotation<T> create(T value, int start, int end, URL data) {
  return new Annotation<T>(value, start, end, data);
}
\end{lstlisting}

Auf diese Weise wird die Verdopplung des Typs in eine Methode ausgelagert. Der Typ wird dabei automatisch durch den Typ der Variable gefüllt, auf den die zurückgegebene Instanz zugewiesen wird, die redundante Typdeklaration bei der Erzeugung wird dadurch überflüssig. 

\paragraph{Agenten}

Konzeptuell erzeugt ein annotationsbasierter Agent aus Eingabeannotationen Ausgabeannotationen. Dies lässt sich durch ein generisches Interface modellieren, dessen Implementierungen die konkreten, zu verarbeitenden Typen spezifizieren, etwa für die Informationsextraktion mit einem \emph{Gazetteer}, der Wörter mit ihrer Bedeutung auszeichnet:

\begin{lstlisting}
class Gazetteer implements Agent<Word, Sense> { ... }
\end{lstlisting}

Ein solcher Agent erstellt aus \emph{Word}-Annotationen \emph{Sense}-Annotationen, er annotiert also Wörtern mit ihrer Bedeutung. Die Methode, die diese Funktionalität für den Gazetteer umsetzt sähe in Java etwa folgendermaßen aus:

\begin{lstlisting}
List<Annotation<Sense>> process(List<Annotation<Word>> words) { ... }
\end{lstlisting}

Über das vom Gazetteer implementierte Agent-Interface und dessen generische Typen kann ge\-währ\-leist\-et werden, dass eine Klasse, die einen \lstinline!Agent<Word,Sense>! implementiert, auch die entsprechende Methode definiert, und eine falsch implementierte Methode (der etwa ein falscher Typ von Annotationen übergeben wird) bereits vom Compiler erkannt wird. Die entsprechende Definition des gemeinsamen Interface aller Agenten sieht folgendermaßen aus:

\begin{lstlisting}
public interface Agent<I, O> {
  public List<Annotation<O>> process(List<Annotation<I>> input);
}
\end{lstlisting}

Dies ist zu lesen als: Ein Agent mit den Typen I und O definiert eine Methode \lstinline!process!, die als Parameter eine Menge von Annotationen des Typs I bekommt und als Rückgabewert eine Menge von Annotationen des Typs O liefert. Ein  \lstinline!Agent<I,O>! kann also Elemente vom Typ I zu Elementen vom Typ O verarbeiten.

\paragraph{Agenteninteraktionen} \label{agent-interaction}

Es können zwei Arten der Interaktion von Agenten unterschieden werden: Analysen, bei denen Annotationen eines Typs linear von Agenten verarbeitet werden, und Synthesen, bei denen potentiell unterschiedliche Typen von Annotationen von zwei Agenten zu einem Modell zusammengeführt werden.

\paragraph{Analysen}

Bei einer Analyse erstellt ein Agent aus Eingabeannotationen eines bestimmten Typs Ausgabeannotationen eines bestimmten Typs. Daher besteht eine Interaktion von Agenten hier aus dem Austausch von Annotationen eines bestimmten Typs, und zwar des Ausgabetyps des einen Agenten (der Quelle) und des Eingabetyps des anderen Agenten (des Ziels). Entsprechend wird eine Analyse bei ihrer Instanziierung auf einen bestimmten Typ festgelegt, z.B. \emph{Word} (unter Verwendung einer Klasse \emph{Analyses}, die statische Funktionalität für Analysen bereitstellt):

\begin{lstlisting}
Analysis<Word> a = Analyses.create();
\end{lstlisting}

Eine Analyse kann aus mehreren Quell- und mehreren Zielagenten bestehen. Einer Analyse sollen als Quellen nur Agenten zugefügt werden können, die Annotationen des entsprechenden Typs erzeugen (\emph{Word} an zweiter Position, als \lstinline!O! eines \lstinline!Agent<I,O>!), etwa: 

\begin{lstlisting}
Agent<?, Word> tokenizer = new Tokenizer(); a = a.withSource(tokenizer);
\end{lstlisting}

Analog sollen als Ziele nur Agenten hinzugefügt werden können, welche die entsprechenden Annotationen als Eingabe akzeptieren (\emph{Word} an erster Position, als \lstinline!I! eines \lstinline!Agent<I,O>!): 

\begin{lstlisting}
Agent<Word, ?> gazetteer = new Gazetteer(); a = a.withTarget(gazetteer);
\end{lstlisting}

Durch die Verwendung von Generics kann sichergestellt werden, dass nur passende Agenten der Analyse hinzugefügt werden. Zur Implementierung wird dazu zunächst das Interface generisch typisiert:

\begin{lstlisting}
interface Analysis<T> { ... }
\end{lstlisting}

Als Quelle werden nur Agenten akzeptiert, welche die entsprechend typisierten Annotationen erzeugen (\emph{T} an zweiter Position, als \lstinline!O! eines \lstinline!Agent<I,O>!):

\begin{lstlisting}
Analysis<T> withSource(Agent<?, T> s) { sources.add(s); return this; }
\end{lstlisting}

Als Ziel werden nur Agenten akzeptiert, welche die entsprechend typisierten Annotationen verlangen (\emph{T} an erster Position, als \lstinline!I! eines \lstinline!Agent<I,O>!):

\begin{lstlisting}
Analysis<T> withTarget(Agent<T, ?> t) { targets.add(t); return this; }
\end{lstlisting}

Die Listen, in denen intern die Agenten einer Analyse verwaltet werden sind ebenfalls entsprechend typisiert:

\begin{lstlisting}
List<Agent<?, T>> sources;
List<Agent<T, ?>> targets;
\end{lstlisting}

\paragraph{Synthesen}

In entsprechender Weise werden Synthesen modelliert. Wir erzeugen eine Synthese, die auf die zu synthetisierenden Elemente typisiert ist, z.B. zur Ermittlung von Worthäufigkeiten (unter Verwendung einer Klasse \emph{Syntheses}, die statische Funktionalität für Synthesen bereitstellt):

\begin{lstlisting}
Synthesis<Word, Frequency> s = Syntheses.create();
\end{lstlisting}

Wie bei Analysen können nun kompatible Agenten als beobachtete Daten (z.B. Wörter) oder Information (z.B. Häufigkeiten) der Synthese definiert werden. Dabei sind bei Synthesen immer die Ausgabetypen der beteiligten Agenten (an zweiter Position eines \lstinline!Agent<I,O>!) relevant, da diese synthetisiert werden:

\begin{lstlisting}
Agent<?, Word> tokenizer; s = s.withData(tokenizer);
Agent<?, Frequency> indexer; s = s.withInfo(indexer);
\end{lstlisting}

Das Ergebnis einer durchgeführten Synthese ist ein Modell, das Elemente der synthetisierten Typen einander zuordnen kann, hier etwa Wörter zu ihren Häufigkeiten:

\begin{lstlisting}
Model<Word, Frequency> frequencies = synthesis.model();
\end{lstlisting}

Die Modellbildung erfolgt über ein Training, bei dem die zu synthetisierenden Elemente als Eingabe dienen:

\begin{lstlisting}
Model<Word, Frequency> train(
  List<Annotation<Word>> data, List<Annotation<Frequency>> info) { ... }
\end{lstlisting}

So gewährleistet die API durch statische Typisierung von Agenteninteraktionen in Form von Analysen und Synthesen schon zur Kompilierzeit, dass Experimente nur mit kompatiblen Agenten und sinnvollen Interaktionen definiert werden.

\paragraph{Experimente}

Mehrere Interaktionen werden zu Experimenten zusammengefasst. So kann ein Experiment folgendermassen aus Analysen und Synthesen zusammengestellt und ausgeführt werden (unter Verwendung einer Klasse \emph{Experiments}, die statische Funktionalität für Experimente bereitstellt):

\begin{lstlisting}
Experiment x = Experiments.create(...).withAnalysis(a).withSynthesis(s).run();
\end{lstlisting}

Ein vollständiges Experiment kann über die Java-API von TM2 wie in Listing \ref{api-full} erfolgen, vgl. die kompaktere Modellierung über DSLs in Abschnitt \ref{werk-dsl}, S. \pageref{werk-dsl}.

\lstset{language=Java}
\begin{lstlisting}[float, label=api-full, caption={Prinzipielle Modellierung eines einfachen Experiments über die Java-API von TM2, vgl. Abschnitt \ref{werk-dsl}, S. \pageref{werk-dsl} für die alternative Modellierung über eine DSL}]
public class NE implements Runnable {
    public static void main(String[] args) { new NE().run(); }
    public void run() {
        String output = "output/Result";
        Experiment x = Experiments.create("NE", "data/Corpus.txt", output);
        Agent<String,Token> tokenizer = new Tokenizer();
        List<Analysis<?>> interactions = new ArrayList<Analysis<?>>();
        /* [Corpus] via String zu [Tokenizer]: */
        Analysis<String> interaction0 = Analyses.create()
          .withSource(new Corpus())
          .withTarget(tokenizer);
        interactions.add(interaction0);
        /* [Tokenizer] via Token zu [Gazetteer, Counter]: */
        Analysis<Token> interaction1 = Analyses.create()
          .withSource(tokenizer)
          .withTarget(new Gazetteer())
          .withTarget(new Counter());
        interactions.add(interaction1);
        /* Ausfuehrung und Evaluierung des Experiments: */
        for (Analysis<?> i : interactions) { x = x.withAnalysis(i); }
        x.run();
        Evaluation evaluation = new Evaluation(output + ".xml")
          .evaluate("data/Gold_Gazetteer.xml", Gazetteer.class)
          .evaluate("data/Gold_Counter.xml", Counter.class);
        Documentation.of(experiment, evaluation);
    }
}
\end{lstlisting}

\paragraph{Evaluierung}

Für die Evaluierung von Analyseergebnissen bietet das TM2-Framework generische Evaluierungskomponenten, die auf Basis der beschriebenen API-Elemente erzeugte Annotationen mit einem Goldstandard vergleichen (als Synthese von Ergebnissen und Goldstandard). Ein Beispielexperiment, das die generische Evaluierung verwendet, findet sich in Abschnitt \ref{appl-pseudoeval}, S. \pageref{appl-pseudoeval}, vgl. die alternative Senseval-Evaluierung in Abschnitt \ref{appl-senseval}, S. \pageref{appl-senseval}.

\subsubsection{Interne Datenverwaltung} \label{impl-daten}

\paragraph{Blackboard} \label{impl-blackboard}

Bei der Ausführung von Experimenten schreiben die einzelnen Agenten ihre Annotationen in die gemeinsame Datenstruktur, das Blackboard (vgl. Abschnitt \ref{anno-blackboard}, S. \pageref{anno-blackboard}). Dieses ist als typsicherer, heterogener Container \citep[142]{Bloch2008} umgesetzt und komplett von der Anwender-API isoliert, d.h. das Framework kümmert sich um die Verwaltung des Blackboards, das Abrufen von benötigten Annotationen, etc. Vom Anwender müssen lediglich die einzelnen Agenten implementiert und wie oben dargestellt in Experimenten miteinander kombiniert werden.

\paragraph{Ressourcen}

In der gesamten TM2-API werden Daten (das Korpus, Agentenressourcen, etc.) als URLs angegeben. Auf diese Weise können über einen einheitlichen Mechanismus sowohl lokale Dateien, als auch Daten auf einem Webserver (z.B. des Gutenberg-Projekts) verwendet werden. Auf diese Weise können etwa vergleichende Experimente mit lokalen und Serverdateien in einer einheitlichen Notation definiert werden, z.B. für das Korpus:

\begin{lstlisting}
Corpus local  = new Corpus("file:///Users/fsteeg/Documents/faust.txt");
Corpus remote = new Corpus("http://www.gutenberg.org/files/14591/14591-8.txt");
// Verwendung von 'local' und 'remote' in vergleichendem Experiment
\end{lstlisting}

\subsubsection{Nebenläufigkeit} \label{impl-concurrency} \index{Parallele Programmierung}

Die Entwicklung von Mikroprozessoren in den vergangenen Jahren ist von einer zunehmenden Anzahl von Kernen gekennzeichnet\footnote{So haben einzelne Standardprozessoren inzwischen bis zu 32 Kerne; der angekündigte Intel \emph{Knights Ferry} etwa unterstützt dabei pro Kern hardwaremäßig 4 Threads und damit auf Hardwarebene bis zu 128 parallele Prozesse in einem einzelnen Prozessor (\url{http://www.tomshardware.com/news/knights-ferry-corner-mic-xeon,11036.html}), auf dem Standardanwendungen laufen (etwa auf Basis von Java, wie die hier beschriebene Implementierung).}. Es gibt zwar Ansätze zu einer alternativen Nutzung der wachsenden Transistordichte in Form von rekonfigurierbaren Chip-Architekturen (vgl. \citealt{BerekovicAndHochberger2008}), doch durch mehrere Prozessorkerne werden aktuell und auch auf absehbare Zeit die fortschreitenden Möglichkeiten der Miniaturisierung am Besten ausgenutzt. Um solche Hardware auszunutzen, muss dabei die Software die Mehrkernarchitektur unterstützen. Bei einer rein sequenziellen Verarbeitung bieten mehrere Prozessorkerne keinerlei Vorteile. Die softwareseitige Unterstützung mehrerer Prozessorkerne ist daher heute eines der wichtigsten Elemente zur Steigerung der Laufzeiteffizienz von Computerprogrammen.

\paragraph{Grundlegende Herausforderungen durch Nebenläufigkeit}

Eine Unterstützung mehrerer Prozessorkerne auf Seiten der Softwarentwickler ist nicht einfach, da paralleles Programmieren grundsätzlich fehleranfälliger ist als sequenzielles \citep[1]{GoetzEtAl2006}.

Eine grundlegende Unterscheidung paralleler Algorithmen kann danach erfolgen, ob von den parallel laufenden Subprozessen gemeinsame Daten verändert werden oder nicht. Wird auf gemeinsame veränderliche Daten (\emph{mutable data}) zugegriffen, muss der parallele Zugriff koordiniert\footnote{Gängig ist der \emph{shared memory} Ansatz mit Sperren und synchronisiertem Datenzugriff (zu Details s. \citealt[405,569]{VanRoyAndHaridi2004}). Andere Ansätze bilden etwa Actors oder \emph{software transactional memory} (STM).} werden, sonst wird der Prozess nondeterministisch, d.h. das gleiche Programm kann bei unterschiedlichen Durchläufen unterschiedliche Ergebnisse liefern \citep[20]{VanRoyAndHaridi2004}. Verfahren ohne gemeinsamen Zustand (\emph{shared state}) sind dagegen deutlich einfacher zu realisieren, da Nebenläufigkeit ohne die Notwendigkeit, gemeinsamen Datenzugriff zu koordinieren, ein vergleichsweise simples Konzept ist \citep[14-5]{VanRoyAndHaridi2004}. Daher ist eine Formulierung von Algorithmen ohne gemeinsamen Zustand erstrebenswert. Zugleich ist aber wie in Abschnitt \ref{anno-blackboard}, S. \pageref{anno-blackboard} beschrieben die Natur des zu lösenden Problems \emph{stateful}. Dies führt zu einem Paradoxon bei der Implementierung von Text-Mining-Systemen: Probleme, deren Natur auf Kontext und gemeinsamem Wissen basieren, sollen möglichst atomar und funktional formuliert werden.

Der Ansatz in TM2 ist hier, die veränderlichen Daten (das Blackboard) vom Anwender zu isolieren (vgl. Abschnitt \ref{impl-daten}, S. \pageref{impl-daten}). Auf dieser Grundlage unterstützt das TM2-Framework soweit möglich Nebenläufigkeit automatisch -- mit dem Ziel, Vorgänge, die logisch nebenläufig sind, automatisch parallel auszuführen\footnote{Diese Verbindung von domänenspezifischer Modellierung (im späteren Verlauf dieses Kapitels auch in Form von DSLs) und automatischer Nebenläufigkeit wird auch in anderen Bereichen und in größerem Maßstab (in Bezug auf die Parallelität) verfolgt, vgl. \citet{ChafiEtAl2010}.}. So werden die Möglichkeiten augeschöpft, die ein Framework gegenüber einer reinen Softwarebibliothek bietet: die Nutzer des Frameworks verwenden Architektur, und klinken sich so in die nebenläufige Verarbeitung ein, ohne dass die Komplexität der Problemdomäne durch die parallele Verarbeitung akzidentell gesteigert wird.

\paragraph{Parallele Verarbeitung von Experimenten}

Einzelne Experimente sind in sich geschlossen und konzeptuell voneinander unabhängig. Mehrere Experimente können so grundsätzlich parallel verarbeitet werden, ohne dass auf gemeinsame Daten zugegriffen werden muss. Das TM2-Framework unterstützt daher eine automatische parallele Ausführung mehrerer Experimente. Die Ausführung erfolgt über eine statische Methode, der die Experimente übergeben werden, z.B.:

\begin{lstlisting}
Batch.run(experiment1, experiment2, experiment3);
\end{lstlisting}

\paragraph{Parallele Verarbeitung von Interaktionen}

Auch innerhalb eines Experiments können bestimmte Interaktionen zwischen Agenten parallel ausgeführt werden: Eine Analyse hat nach der Verarbeitung aller ihrer Quellen alle Annotationen gesammelt, die von den Zielen benötigt werden. So können alle Ziele einer Analyse ihre Daten ohne gemeinsamen Zustand parallel verarbeiten. Daher verarbeiten im TM2-Framework die Zielagenten einer Analyse ihre Eingaben automatisch nebenläufig (z.B. unterschiedliche Tagger, die auf Basis der Tokens eines einzigen Tokenizers arbeiten). 

\subsubsection{Retrieval} \label{api-retrieval}

\paragraph{Ziele}

Die in Experimenten generierten Annotationen sollen neben der automatischen Auswertung bei einer Evaluierung auch etwa für eine erweiterte Suche in den analysierten Daten verwendet werden können. Für das in Kapitel \ref{chapter-theory} beschriebene exemplarische Gesamtziel eines semantischen Information-Retrieval ist eine solche intelligente Suche sogar der eigentliche Zweck des Text-Mining. Dazu unterstützt die TM2-API eine Suche in den Annotationen. Dabei kann nach einem Wert in den Annotationen eines bestimmten Agenten gesucht werden, während die korrespondierenden Annotationen eines anderen Agenten als Ergebnis geliefert werden können.

Mithilfe dieses generischen Mechanismus sind unterschiedliche Anwendungsszenarien denkbar. So kann etwa in den Annotationen eines POS-Taggers nach Verben gesucht werden und die korrespondierenden Annotationen des Tokenizers ausgegeben werden (wir erhalten die als Verben getaggten Wortformen). Es können aber auch die korrespondierenden Annotationen eines Indexers ausgegeben werden (und wir erhalten die Häufigkeit von Wortformen, die Verben sind), etc. Beim Retrieval können so die von den verschiedenen Agenten gesammelten Informationen flexibel genutzt und neu kombiniert werden: jeder Agent hat nur seine spezifische Aufgabe erfüllt (der Tokenizer hat zerlegt, der Tagger getaggt, der Indexer gezählt), aber die Ergebnisse lassen sich zu neuen, so zuvor nicht definierten Aufgaben verbinden. Annotationen stellen so eine Strukturebene in den Daten dar, die quasi klassisches Data-Mining (im Sinne einer Analyse strukturierter Daten) auf Texten ermöglicht.

\paragraph{API}

Die Retrieval-API unterstützt eine typsichere Suche. So muss beim Suchen in den Annotationen eines bestimmten Agenten ein Objekt zum Suchen verwendet werden, das den Typ hat, den der Agent produziert und nur ein solches -- was wie oben bei Experimenten zur Folge hat, dass der Compiler sicherstellt, dass nur sinnvolle Suchen möglich sind.

Über die API kann etwa nach den Bedeutungen einer Wortform gesucht werden (Suche nach Tokenizer-Aus\-ga\-be\-typ \emph{String}, Ergebnis ist Gazetteer-Aus\-ga\-be\-typ \emph{Sense}):

\begin{lstlisting}
List<Sense> senses = retrieval.find("Rhine", Tokenizer.class).by(Gazetteer.class);
\end{lstlisting}
Oder umgekehrt, eine Suche nach den Wortformen einer Bedeutung (Suche nach Gazetteer-Aus\-ga\-be\-typ \emph{Sense}, Ergebnis ist Tokenizer-Aus\-ga\-be\-typ \emph{String}):

\begin{lstlisting}
List<String> tokens = retrieval.find(Sense.NAME, Gazetteer.class).by(Tokenizer.class);
\end{lstlisting}

Oder etwa eine Suche nach Bedeutungen von Wortformen, die zweimal vorkommen (Suche nach Indexer-Aus\-ga\-be\-typ \emph{Integer}, Ergebnis ist Gazetteer-Aus\-ga\-be\-typ \emph{Sense}):

\begin{lstlisting}
List<Sense> senses = retrieval.find(2, Indexer.class).by(Gazetteer.class);
\end{lstlisting}

Als Anwendungsbeispiele dieser API enthält die TM2-Software eine einfache Konsolenapplikation, die eine textuelle Nutzung dieser Retrievalfunktionalität ermöglicht. Als Brücke zwischen einer textuellen Eingabe und der typsicheren Such-API kann für eine solche Anwendung die optionale String-Repräsentation von Annotationswerten dienen (vgl. Abschnitt \ref{impl-xml}, S. \pageref{impl-xml}).

\subsubsection{Dokumentationsgenerierung} \label{api-doc}

Die generierte Dokumentation von Experimenten im TM2-Framework besteht aus einer Beschreibung des Versuchsaufbaus, den verwendeten Daten, den Verfahren, die auf die Daten angewendet wurden (d.h. den Agenten), sowie den so ermittelten Resultaten in Form von Annotationen und Evaluationsergebnissen.

\paragraph{Abbildungen}

Auf Basis des Objektmodells eines Experiments, d.h. auf Basis seines Aufbaus aus Interaktionen und Agenten, kann eine Visualisierung des Experiments für die Dokumentation generiert werden. In TM2 erfolgt dies auf Basis des von AT\&T entwickelten Graphviz-Pakets\footnote{Graphviz, \url{http://www.graphviz.org/}}. Die Beschreibungen der zu generierenden Abbildungen in der Graphenbeschreibungssprache DOT\footnote{DOT, \url{http://en.wikipedia.org/wiki/DOT_language}} werden mit EMF und der Template-Sprache JET\footnote{JET, \url{http://www.eclipse.org/modeling/m2t/}} generiert. Die Abbildungen zu den verschiedenen Experimenten in Kapitel \ref{chapter-appl} (z.B. Abbildung \ref{anwendung-wsd-senseval}, S. \pageref{anwendung-wsd-senseval}) wurden auf diese Weise erstellt.

\subsubsection{HTML-Ausgabe über WikiText}

Die Gesamtdokumentation wird in Fom eines HTML-Dokuments generiert, welches mit JET (s.o.) und Mylyn WikiText\footnote{Mylyn WikiText, \url{http://wiki.eclipse.org/Mylyn/WikiText}} erstellt wird. Dabei können die verschiedenen Ressourcen (wie Input-Texte, XML-Export von produzierten Annotationen oder Goldstandard-Annotationen) mittels Hyperlinks verbunden und Evaluationsergebnisse tabellarisch aufbereitet werden. So erhält man automatisch eine sehr kompakte und nützliche Dokumentation zu einem Experiment oder einer Serie von Experimenten (vgl. Experimente in Kapitel \ref{chapter-appl} und Anhang \ref{ref-wsd}, speziell Abb. \ref{generated-table}, S. \pageref{generated-table}).

Eine erste Implementierung der Dokumentationsgenerierung mit LaTeX stellte sich als schwergewichtige Lösung heraus: Die LaTeX-Abhängigkeit ist nicht trivial (eine vollständige TeX-Installation umfasst meist über 1 GB) und der Aufruf des nativen \emph{pdflatex}-Programms ist fehleranfällig und plattformabhängig. Eine flexiblere, rein Java-basierte Alternative stellt die Generierung von Wiki-Markup dar, welches mithilfe der WikiText-Komponente von Mylyn in HTML exportiert wird. WikiText kann die generierte Zwischenrepräsentation alternativ in andere Formate wie DocBook\footnote{DocBook, \url{http://www.docbook.org/}}, DITA\footnote{DITA, \url{http://dita.xml.org/}} oder PDF (über XSL-FO)\footnote{XSL-FO, \url{http://en.wikipedia.org/wiki/XSL_Formatting_Objects}} transformieren.

\paragraph{XML-Export} \label{impl-xml}

Annotationen können in TM2 in ein simples XML-Format exportiert werden. Aus diesem lassen sich die Annotationen wieder als Objekte instanziieren. Als ein XML-Attribut eines Annotationselements wird dafür die serialisierte und als Base64-String codierte\footnote{Die Base64-Kodierung ermöglicht eine Abbildung von Binärdaten auf den ASCII-Zeichensatz.} Version des Wertes der Annotation gespeichert.

Dieses Vorgehen hat zwei Vorteile: \begin{inparaenum}
  \item Aus den textuellen XML-Dateien können so beliebig komplexe Objekte instanziiert werden.
  \item Bei der Entwicklung von Komponenten wird durch die Forderung einer Implementierung des Serializable-Interface bei der Typisierung der Agenten-Daten vom Java-Kompiler schon zur Kompilierzeit sichergestellt, dass die verwendeten Daten mit Java-Mitteln persistiert werden können.
\end{inparaenum} 

"Uber die API können auf diese Weise Annotationen persistiert und typsicher geladen werden:

\begin{lstlisting}
Agent<?, Sense> agent = new Gazetteer(); 
/* ...Experiment, 'agent' schreibt Annotationen die wir speichern und laden koennen: */
AnnotationWriter w = new AnnotationWriter(blackboard, data).writeAnnotations(location);
List<Annotation<Sense>> senses = new AnnotationReader(location).readAnnotations(agent);
\end{lstlisting}

Für die Verwendung von Annotationenen, die in einer anderen Sprache als Java oder manuell erstellt wurden (z.B. für einen Goldstandard), und die daher keine serialisierten Objekte enthalten, können die Objekt auch aus einer menschenlesbaren Darstellung instanziiert werden. Dazu muss der zu instanziierende Datentyp einen Konstruktor mit einem einzelnen String-Parameter haben, der beim Laden der Annotationen von TM2 über die \emph{Java Reflection API} aufgerufen wird. Für die gängigen Datentypen wie Strings, Ganzzahlen und Fließkommazahlen ist dies automatisch gewährleistet. Für eigene Datentypen muss ein entsprechender Konstruktor erstellt werden, wenn eine Instanziierung über menschenlesbare Darstellungen gewünscht ist.

\paragraph{Exportformat} \label{impl-xsd}

\begin{figure}
\begin{center}
  \includegraphics[width=3.3cm]{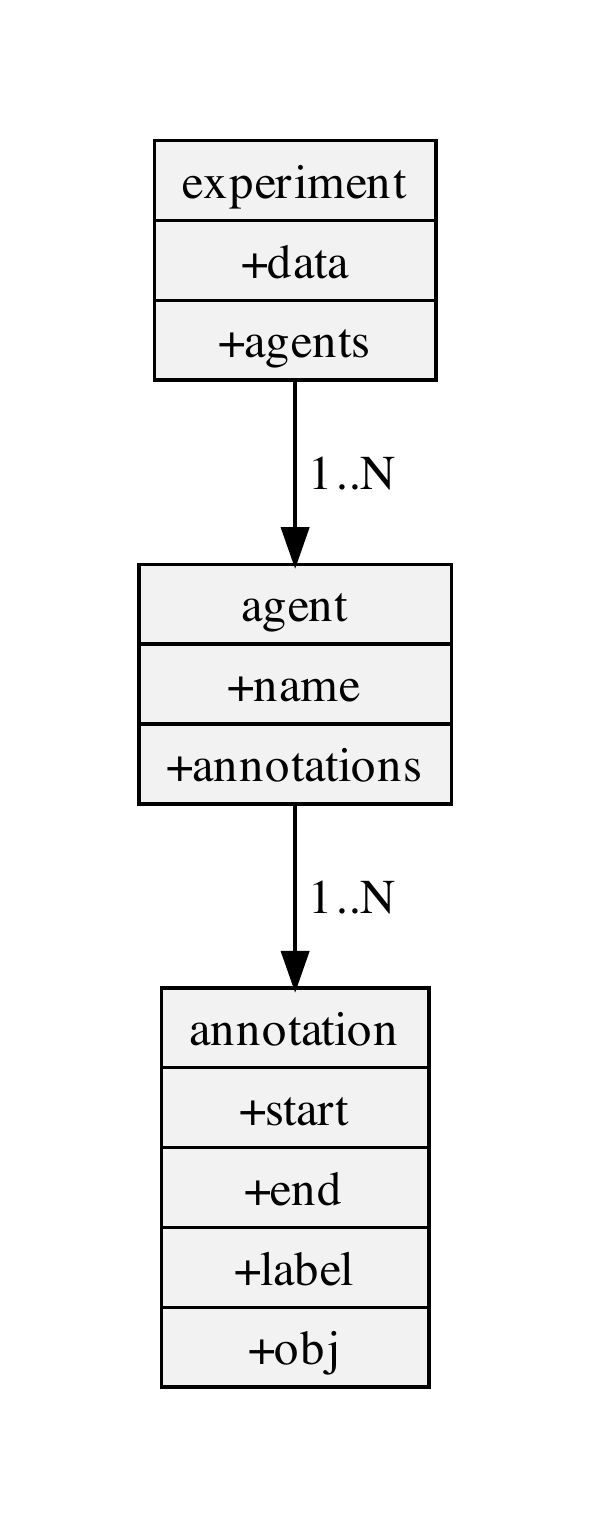}
  \caption{Logische Struktur der XSD des TM2-Exportformats}
  \label{xsd-struktur}
\end{center}
\end{figure}

Die Struktur des Exportformats wird mit einer XSD\footnote{XML Schema Definition, \url{http://www.w3.org/XML/Schema}} definiert. Die logische Struktur des XML-Formats beschreibt Abb. \ref{xsd-struktur}, die vollständige XSD findet sich in Anhang \ref{anhang-xml}, S. \pageref{anhang-xml}. Durch die Validierung gegen eine XSD können fehlerhafte Eingabedateien früh erkannt werden, was eine Form des Fail-Fast-Konzepts \citep{Shore2004} darstellt, d.h. ein möglichst frühes Erkennen von Fehlern zur Vereinfachung der Fehlersuche.

Im Zusammenspiel mit Werkzeugen bietet eine XSD zudem die Möglichkeit einer Validierung beim manuellen Erstellen von XML-Dateien (z.B. für einen Goldstandard) in einem validierenden Editor, etwa im XML-Editor von Eclipse. So stellt die Verwendung einer XSD die Grundlage für eine vielfältige Verwendung der mit TM2 erzeugten XML-Annotationsdateien dar (vgl. Abschnitt \ref{xcl}, S. \pageref{xcl}).

\subsection{Textuelle Modellierung im Text-Mining} \label{werk-dsl}  \index{Modellierung!textuell} \index{DSL} \index{Agenten!Modellierung}

Experimentserien sollen in einer \emph{domain-specific language} (DSL) beschrieben werden, welche die Funktionalität der API in einer kompakteren Form zugänglich macht. Zur Implementierung von DSLs lassen sich zwei Ansätze unterscheiden: \emph{eigenständige} und \emph{eingebettete} DSLs (vgl. \ref{basis-dsl}, S. \pageref{basis-dsl}). In dieser Arbeit werden beide Ansätze verfolgt und verglichen: zum Einen eine eigenständige DSL mit dem auf ANTLR \citep{Parr2007} basierenden Xtext-Framework\footnote{Xtext Language Development Framework, \url{http://eclipse.org/xtext}} des \emph{Eclipse Modeling Project}, zum Anderen eine eingebettete DSL in der JVM-basierten, statisch typisierten Sprache Scala. Beide Ansätze werden in Abschnitt \ref{dsl-fazit}, S. \pageref{dsl-fazit} hinsichtlich der Komplexität ihrer Umsetzung und der praktischen Anwendbarkeit verglichen.

\subsubsection{Eigenständige DSL} \label{werk-dsl-extern} \index{DSL!extern} \index{DSL!eigenständig}

\paragraph{Modellierung}

Eine Modellierung der Interaktionen wie in Abschnitt \ref{section-api}, S. \pageref{section-api} für die API beschrieben (vgl. Listing \ref{api-full}, S. \pageref{api-full}) ist aus Sicht eines Java-Entwicklers nicht ungewöhnlich, würde man jedoch unvorbelastet\footnote{Eine solche unvorbelastete Herangehensweise ist ein Grundgedanke des DSL-Ansatzes -- ganz im Sinne Wittgensteins: \emph{Die Grenzen meiner Sprache bedeuten die Grenzen meiner Welt} \citep{Wittgenstein1922}, vgl. \emph{Notation And Thinking}, \url{http://rjlipton.wordpress.com/2010/11/30/notation-and-thinking/} zur Bedeutung von Notation für den Fortschritt in der Mathematik.} anfangen, ist eine typisierte Interaktion von Agenten in der einfachsten textuellen Form etwa so modellierbar: 

\begin{lstlisting}
Tokenizer -> Token -> Gazetteer.
\end{lstlisting}

D.h. der Tokenizer produziert Token-Instanzen, die vom Gazetteer verarbeitet werden. Dem\-ent\-sprech\-end kann mit Xtext eine Sprache implementiert werden, in der Experimente wie in Listing \ref{code-dsl-xtext}, S. \pageref{code-dsl-xtext} modelliert werden können. Jede Zeile nach der Definition der Metadaten in der ersten Zeile entspricht dabei einer Interaktion. Der Operator \lstinline!->! zwischen den Agenten beschreibt den Informationsfluss in der Interaktion: von den Agenten, die als Quellen fungieren (oben z.B. \lstinline!Tokenizer!), über den Typ der ausgetauschten Daten (oben z.B. \lstinline!Token!), zu den Zielagenten (oben z.B. \lstinline!Gazetteer!). Jede Interaktion schließt mit einem Punkt ab. Die Angabe der Daten, die zwischen den Agenten ausgetauscht werden, stellt eine Form von Typisierung dar, die bei der Weiterverarbeitung automatisch verifiziert werden könnte (vgl. Abschnitte \ref{statische-typisierung}, S. \pageref{statische-typisierung} und \ref{werk-dsl-intern}, S. \pageref{werk-dsl-intern}). Eine solche Sprache ist deklarativ, da sie den Aufbau eines Experiments beschreibt, das dann gestartet werden kann - im Gegensatz zu einer prozeduralen DSL, welche die Ausführung eines Programms steuert\footnote{Eine solche prozedurale DSL wird etwa in \citet{Bilagher2006} für Tesla beschrieben.}.

\begin{lstlisting}[float, label=code-dsl-xtext, caption={Experiment in der eigenständigen Xtext-DSL}]
Experiment "NE" Data "input/Corpus" Out "output/Result" Import "tm2.agents"
/* Als Erstes wird das Korpus tokenisiert: */
Corpus -> String -> Tokenizer.
/* Dann werden die Token mit Eigennamen versehen und indexiert: */
Tokenizer -> Token -> Gazetteer Indexer.
/* Schliesslich koennen wir z.B. die Eigennamen evaluieren: */
Evaluate Gazetteer Against "input/Gold"
\end{lstlisting}

\paragraph{Implementierung mit Xtext}

Xtext\footnote{Xtext Language Framework, \url{http://www.eclipse.org/xtext}} ist ein auf dem Eclipse Modeling Framework (EMF\footnote{Eclipse Modeling Framework, \url{http://eclipse.org/emf}}) basierendes Framework zur Entwicklung von DSLs. Ausgangspunkt bei der Entwicklung einer DSL mit Xtext stellt die Grammatik der zur entwickelnden Sprache dar, die in einer EBNF-artigen\footnote{Extended Backus–Naur Form, eine Syntax zur Spezifikation von Grammatiken} Syntax beschrieben wird. Ausgehend von dieser Grammatik generiert Xtext für die Sprache ein EMF-Metamodell\footnote{Das Metamodell bildet das Schema für die EMF-interne Repräsentation der geparsten Modelle}, einen Parser und graphische Elemente wie einen Editor mit Fehlerüberprüfung, Syntaxfärbung, kontextsensitiver Hilfe, Strukturübersicht, etc. Dies macht sowohl die Entwicklung von Sprachen (inklusive passenden Werkzeugen), als auch die Integration mit bestehenden EMF-Werkzeugen sehr einfach. Die Definition der Grammatik für die oben dargestellte Sprache in Xtext ist dabei sehr übersichtlich (s. Listing \ref{code-dsl-xtext-grammar}, S. \pageref{code-dsl-xtext-grammar}). Mithilfe des generierten Parsers kann ein Experiment wie in Listing \ref{code-dsl-xtext} eingelesen und anschließend ausgeführt werden. In dem generierten Editor können solche Experimente komfortabel erstellt werden.

\lstset{}
\begin{lstlisting}[float, label=code-dsl-xtext-grammar, caption={Grammatik der eigenständigen Xtext-DSL}]
Experiment : /* Attribute eines Experiments: Name, Daten, Ausgabeort */
"Experiment" name=STRING "Data" corpus=STRING "Out" output=STRING
/* Es verwendet Agenten und Datentypen aus einem spezifizierten Package: */
"Import" (imports+=STRING)+
 /* Es besteht aus einer Menge von Interaktionen: */
(interactions+=Interaction)+
/* Es enthaelt optional eine Evaluierung: */
("Evaluate" (evalAgents+=ID)+ "Against" evalLocation=STRING)?;
/* Jede Interaktion hat Quellen, einen Typ und Ziele... */
Interaction : source=Source  "->" type=ID "->" target=Target "."; 
/* ...potentiell mehrere Quellen... */
Source : (sourceAgents+=Agent)+;
/* ... und potentiell mehrere Ziele...:*/
Target : (targetAgents+=Agent)+;
/* ...alle dargestellt durch Agenten: */
Agent : name=ID;
\end{lstlisting}

\paragraph{"Ubersetzung} \label{section-xpand}

Ein Nachteil der beschriebenen Xtext-DSL gegenüber der direkten Nutzung der Java-API ist, dass der Typ der Interaktionen lediglich als String in der Grammatik modelliert wurde und der generierte Parser so nicht automatisch eine "Uberprüfung der Korrektheit der Typen durchführt. Um die Vorteile der problemnahen Modellierung in einer DSL mit der statischen Typisierung und den Überprüfungen zur Laufzeit in der beschriebenen Java-API zu verbinden, kann die DSL z.B. in die Java-Version übersetzt werden\footnote{Alternativen bilden etwa Xtext \emph{constraint checks} oder das neuere \emph{Xtext Typesystem Framework} (\url{http://voelterblog.blogspot.com/2010/08/xtext-typesystem-framework.html}).}. Diese generierte Form nutzt die Java-API von TM2. Sie kann vom Java-Compiler überprüft werden und entsprechende Fehlermeldungen zurückgeben. Auf diese Weise wird eine Duplikation der Regelmodellierung verhindert (statt diese -- neben der Java-API -- noch einmal mit anderen Mitteln für die DSL zu implementieren). Durch die Integration des generierten Xtext-Editors in Eclipse ist eine rudimentäre Integration mit der Java-Entwicklungsumgebung in Eclipse bei diesem Ansatz automatisch vorhanden.

Eine Transformation des textuellen Modells in Java ist mithilfe von Xpand-Templates sehr unkompliziert möglich: Xpand ermöglicht die typsichere Konvertierung des geparsten Modells (hier aus der DSL) in eine alternative textuelle Form (hier in Java-Code). In analoger Weise kann aus dem Modell auch eine graphische Aufbereitung mit Graphviz erfolgen, die zur Generierung der Übersichtsdokumentation verwendet werden kann. In Anhang \ref{anhang-impl-xpand}, S. \pageref{anhang-impl-xpand} finden sich die verwendeten Xpand-Templates und ein komplettes Beispiel aus Eingabe-DSL, Templates und Java- sowie DOT-Ergebnissen. 

\paragraph{Fazit externe DSL}

\begin{figure}
\begin{center}
  \includegraphics[width=11cm]{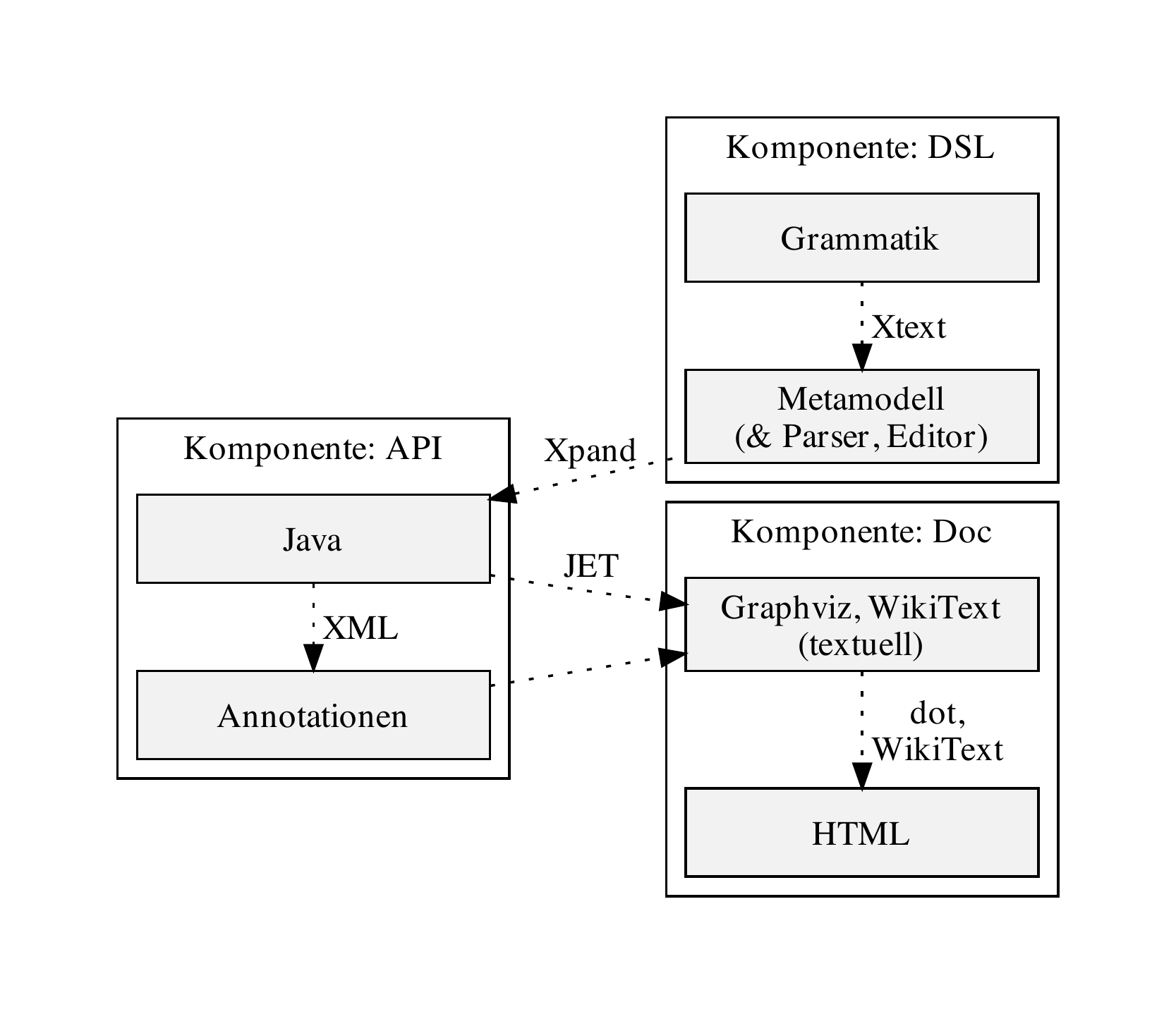}
  \caption[Gesamtübersicht]{Gesamtübersicht und Zusammenspiel der TM2-Komponenten bei einer Lösung mit der eigenständigen Xtext-basierten DSL}
  \label{modelle}
\end{center}
\end{figure}

Abbbildung \ref{modelle} zeigt die Komponenten der beschriebenen externen DSL mit Xtext und Xpand im Gesamtzusammenhang des TM2-Frameworks. Vorteile einer solchen Lösung sind die Integration in EMF-basierte Werkzeuge und die hervorragende Werkzeugunterstützung -- sowohl beim Erstellen von Grammatik und Code-Generatoren bei der Implementierung der DSL, wie auch zur Erstellung textueller Modelle in der Zielsprache. Die enge EMF- bzw. allgemeine Eclipse-In\-te\-gra\-tion kann je nach Anwendungsfall jedoch auch einen Nachteil darstellen -- etwa wenn für eine zu entwickelnde DSL weder Java- noch Eclipse-Integration erforderlich sind. Zudem ist der Aspekt der typsicheren Modellierung hier nur indirekt umgesetzt, nämlich durch die Nutzung der typsicheren Java-API in dem aus der DSL generierten Code.

\subsubsection{Eingebettete DSL} \label{werk-dsl-intern} \index{DSL!intern} \index{DSL!eingebettet}

Zum Vergleich mit der beschriebenen Implementierung einer eigenständigen DSL mit Xtext wird dieser im Folgenden eine eingebettete, typsichere DSL in Scala gegenübergestellt.

\paragraph{Scala für eingebettete DSLs}

Scala (\emph{A scalable language}, s. \citealt{OderskyEtAl2008}) ist eine statisch typisierte, funktional-objektorientierte Hybridsprache, die auf der JVM läuft. Operatoren sind in Scala als Methoden implementiert -- so wird etwa ein Aufruf wie \lstinline!1 + 2! vom Scala-Compiler in \lstinline!(1).+(2)! umgewandelt, d.h.  \lstinline!+! ist der Name einer Methode der Klasse von  \lstinline!1!. Methoden können aus beliebigen symbolischen oder alphanumerischen Unicode-Zeichen bestehen (d.h. auch etwa '$\heartsuit$'), sowie in Infix- oder Präfix-Notation verwendet werden. Dies ist ein Grund warum sich Scala hervorragend zur Definition eingebetteter DSLs eignet\footnote{Scala bietet über die hier dargestellten Merkmale hinaus unterschiedliche weitergehende Unterstützung für DSLs, sowohl durch grundlegende Sprachmerkmale (etwa Möglichkeiten zur Definition von kontrollstrukturartigen Bibliotheken, vgl. \citealt[161ff.]{OderskyEtAl2008}), als auch durch spezialisierte Bibliotheken (etwa in Form von \emph{combinator parsing}, vgl. \citealt[619ff.]{OderskyEtAl2008}).}.

\paragraph{Implizite Typumwandlung und Nutzung der typsicheren API}

Scala ermöglicht darüber hinaus eine implizite Typumwandlung. Durch solche \emph{implicit conversions} werden Objekte eines Typs implizit in Objekte eines anderen Typs konvertiert. Dies ermöglicht eine Erweiterung der Funktionalität von Klassen, ohne die Klassen selbst zu verändern. Scala verwendet diese Funktionalität etwa, um Java-Strings um zusätzliche Methoden zu erweitern. Die in Scala implementierte DSL-Schicht kann daher auf die Java-Klassen der TM2-API zugreifen und über \emph{implicit conversions} die Java-Klassen mit der Scala-DSL anreichern. So ermöglicht die implizite Typumwandlung in TM2 die Verwendung von in Java implementierten TM2-Agenten mit der Syntax der Scala-DSL. Scala ist eine optimale Sprache für diesen Anwendungsfall, da sie neben den dargestellten generellen Möglichkeiten bei der Implementierung einer DSL zudem wie Java statisch typisiert ist und so automatisch die Vorteile der typsicheren Java-API nutzen kann. Diese Regeln der Java-API müssten bei der Verwendung einer externen DSL komplett neu definiert werden, oder könnten wie oben beschrieben nur indirekt genutzt werden (vgl. Abschnitt \ref{section-xpand}, S. \pageref{section-xpand}, vgl. Abschnitt \ref{impl-langs}, S. \pageref{impl-langs}).

\paragraph{Enge Scala-Integration}

Eine eingebettete DSL ermöglicht zudem grundsätzlich die Verwendung von allen Möglichkeiten der Basissprache und ihrer Werkzeuge, wie die Definition von Variablen, die Verwendung (und Entwicklung, vgl. \citealt[161ff.]{OderskyEtAl2008}) von Kontrollstrukturen, oder interaktives Debugging. Im Kontext von TM2 ist etwa ein sehr nützliches Scala-Sprachfeature die vielseitige \emph{for}-Syntax (vgl. \citealt[486]{OderskyEtAl2008}), mit der kombinatorische Probleme sehr kompakt beschrieben werden können -- etwa komplexe Experimente mit variablen Teilen zur einfachen Definition von Experimentserien (s. Abb. \ref{dsl-scala-form}, S. \pageref{dsl-scala-form}, vgl. Abschnitt \ref{appl-wsd-model}, S. \pageref{appl-wsd-model}).

Diese enge Integration in die Basissprache kann auf anderer Ebene auch als Nachteil gesehen werden: Fehlermeldungen beziehen sich semantisch immer auf die Basissprache, nicht die entwickelte DSL. Eine eigenständige DSL bietet hier mehr Möglichkeiten für angepasste Fehlermeldungen, die allerdings immer eine separate Modellierung der Regeln voraussetzen (und damit hier eine Duplikation der Regeln aus der Java-API, s.o.).

\paragraph{Modellierung von Interaktionen}

In Anlehnung an die Modellierung von Interaktionen in der eigenständigen DSL werden auch in der eingebetteten DSL Interaktionen mit  \lstinline!->! dargestellt. Im Unterschied zur eigenständigen DSL verwenden wir hier die eingebaute Typisierung von Scala, bzw. der zugrunde liegenden Java-API, z.B. bei der Modellierung einer Analyse:

\begin{lstlisting}
tokenizer -> gazetteer : Analysis[Token]
\end{lstlisting}

In vergleichbarer Weise können in der Scala-DSL Synthesen definiert werden, z.B. zur Ermittlung von Worthäufigkeiten:

\begin{lstlisting}
(tokenizer, indexer) -> model : Synthesis[Token, Frequency]
\end{lstlisting}

\paragraph{Modellierung von Experimenten} \label{dsl-model-experiments}

Durch eine Verknüpfung von Interaktionen mit dem selbst definierten Operator \lstinline!|! können wir mit der Scala-DSL z.B. einfache Experimente ohne Evaluierung definieren, die wir mit dem ebenfalls selbst definierten Operator \lstinline!!! ausführen können:

\begin{lstlisting}
val experiment = corpus -> tokenizer | tokenizer -> gazetteer !
\end{lstlisting}

Wir können über die Scala-DSL auch -- wie in Abschnitt \ref{section-api}, S. \pageref{section-api} für die Java-API beschrieben -- flexibel in den Annotationen eines Experiments suchen, etwa nach den Tokenizer-Entsprechungen von Gazetteer-Annotationen mit dem Wert \emph{NAME}\footnote{Die Verwendung der Java-API aus Scala kann in einem solchen Fall durch Weglassen der Punkte vereinfacht werden.}:

\begin{lstlisting}
experiment find (Sense.NAME, gazetteer) by tokenizer
\end{lstlisting}

Neben der manuellen Suche in den erzeugten Annotationen kann das Ergebnis eines Experiments auch automatisch evaluiert werden, z.B. zur Evaluierung einer Informationsextraktion mithilfe eines Goldstandards:

\begin{lstlisting}
val experiment = corpus -> tok | tok -> (ie, gold) | (ie, gold) -> evaluation !
\end{lstlisting}

\paragraph{Experimentserien} 

Zur Modellierung von Experimentserien unter Verwendung der Scala-DSL definieren wir zunächst die beteiligten Agenten mit den unterschiedlichen, zu vergleichenden Konfigurationen und definieren dann deren einheitliches Zusammenspiel in den Experimenten. Die generelle Form von Experimentserien in der Scala-DSL ist in Abb. \ref{dsl-scala-form} dargestellt. Dabei beschreibt der \emph{for}-Block die unterschiedlichen Konfigurationen der im \emph{yield}-Block definierten Experimente, die durch den optionalen \emph{run}-Block ausgeführt und automatisch dokumentiert werden.

\begin{figure}[hb]
\begin{center}
	 \lstinline! run { for { <configuration> } yield { <experiment> } } !
  \caption{Grundlegende Form von Experimentserien in der Scala-DSL}
  \label{dsl-scala-form}
\end{center}
\end{figure}

Die Beschreibung in Listing \ref{sample-exp}, S. \pageref{sample-exp} etwa verwendet zwei unterschiedliche Korpora und zwei unterschiedliche Präprozessoren. Die Permutationen der Konfigurationsparameter stellen dabei die unterschiedlich konfigurierten Experimente dar, die auf Basis einer solchen Beschreibung parallel ausgeführt und vergleichend dokumentiert werden (d.h. hier z.B. 4 Experimente: \emph{WorksOfShakespeare} und \emph{WorksOfGoethe} jeweils mit einem \emph{RuleBasedTokenizer} und einem \emph{TrainableTokenizer}).

Diese DSL wird in Kapitel \ref{chapter-appl} für die Durchführung von Beispielexperimenten zur Informationsextraktion (Abschnitt \ref{appl-ie}, S. \pageref{appl-ie}) sowie zur WSD mit Pseudoambiguität (Abschnitt \ref{appl-pseudoeval}, S. \pageref{appl-pseudoeval}) und Senseval-Daten (Abschnitt \ref{appl-senseval}, S. \pageref{appl-senseval}) verwendet. Die Implementierung der eingebetteten Scala-DSL ist in Anhang \ref{anhang-embedded-dsl}, S. \pageref{anhang-embedded-dsl} dargestellt.

\lstset{language={Scala}}
\begin{lstlisting}[float, label=sample-exp, caption={Beispiel für Experimentserie in der Scala-DSL}]
run { 
 for {
    corpus <- List(new WorksOfShakespeare, new WorksOfGoethe)
    tokenizer <- List(new RuleBasedTokenizer, new TrainableTokenizer)
    gazetteer = new Gazetteer
    gold = new GazetteerGoldStandard
    evaluation = new SimpleEvaluation
 } yield {
    corpus -> tokenizer |
    tokenizer -> (gazetteer, gold) |
    (gazetteer, gold) -> evaluation 
   }
}
\end{lstlisting}

\subsubsection{Editor}

Der Editor, ein zentraler Teil der Werkzeugunterstützung für die Modellierung mit einer DSL, ist bei einer eingebetteten Sprache in der Regel schon vorhanden, und im besten Fall ausgereift. So bedeutet die Entwicklung einer eingebetteten TM2-DSL in Scala im Grunde die Entwicklung einer dünnen Scala-API, die auf die bestehende Java-API aufbaut (indem sie diese aufruft und mit einer spezialisierten, domänenspezifischen Syntax zugänglich macht). Durch die sehr weitgehende Generierung von Werkzeugen beim Erstellen von eigenständigen DSLs mit Xtext ist der Unterschied in diesem Bereich nicht mehr so gro"s wie bis vor Kurzem, dennoch haben hier eingebettete DSLs einen generellen Vorteil, da die Werkzeugunterstützung nicht selbst erzeugt und angepasst werden muss, und sehr weitgehend sein kann (z.B. automatisches Refactoring, interaktives Debugging, etc.).

\subsubsection{Fazit DSL} \label{dsl-fazit}

Die Verwendung von Scala zur Implementierung einer DSL für TM2 bestätigt, dass mit Scala Anwendungsfälle mit einer eingebetteten DSL umsetzbar sind, für die in anderen Sprachen eigenständige DSLs eingesetzt werden müssen \citep[447]{OderskyEtAl2008}. Die Umsetzung als eingebettete DSL macht dabei die Eigenentwicklung vieler Komponenten überflüssig, etwa zur Unterstützung von statischer Typisierung oder von Werkzeugen, etwa in Form von Editoren und Debuggern. Mit diesen Vorteilen macht Scala eingebettete DSLs generell zu einer sehr attraktiven Alternative gegenüber eigenständigen DSLs -- selbst ohne die Nutzung weitergehender Möglichkeiten von Scala bei der Entwicklung von DSLs, etwa die Definition neuer Kontrollstrukturen (vgl. \citealt[161ff.]{OderskyEtAl2008}) oder \emph{combinator parsing} (vgl. \citealt[619ff.]{OderskyEtAl2008}).

Die beschriebenen Umsetzungen von DSLs zur typsicheren Modellierung im Text-Mining zeigen ein Spannungsfeld zwischen der Forderung nach statischer Typisierung zur Umsetzung der Konzepte aus Kapitel \ref{chapter-theory} und der häufig dynamisch typisierten Natur von DSLs. Hier umgesetzt wurden zwei Ansätze: \begin{inparaenum}
\item Eine eigenständige DSL, die mit Xtext implementiert wurde und die über Xpand zu Java-Code kompiliert wird. Dieser wird mit dem Java-Compiler verarbeitet, dessen Meldungen über fehlerhafte Typisierung den Benutzer der DSL bei der Modellierung unterstützen können.
\item Eine eingebettete Scala-DSL, welche die Typisierung von Scala nutzt und durch ihre Java-Kompatibilität die Java-Klassen und -Interfaces des TM2-Frameworks direkt nutzen kann.
\end{inparaenum} 

\begin{figure}
\begin{center}
  \includegraphics[width=8cm]{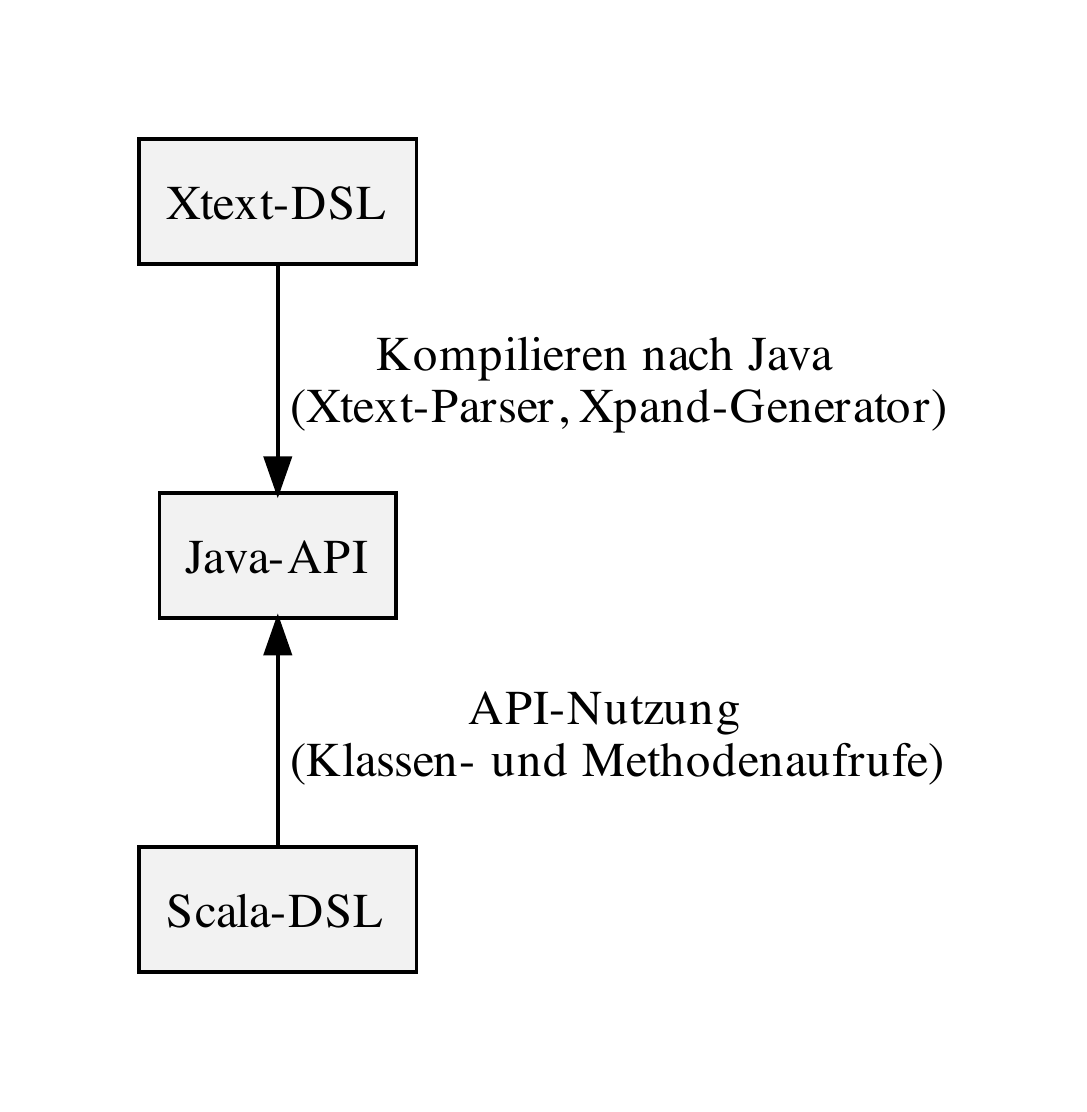}
  \caption{Unterschiedliche Nutzung der Java-API durch Xtext- und Scala-DSL}
  \label{dsl-dsl-api}
\end{center}
\end{figure}

Abbildung \ref{dsl-dsl-api} gibt einen "Uberblick über die Interaktion der alternativen Implementierungen der DSL mit der zugrundeliegenden Java-API. Die Experimente in Kapitel \ref{chapter-appl} wurden mit der leichtgewichtigen\footnote{Die Scala-DSL ist insgesamt leichtgewichtiger als die Xtext-Lösung, da neben Scala selbst keine zusätzlichen Bibliotheken oder Frameworks notwendig sind, während der Xtext-Ansatz gerade von seiner starken Integration in andere Bibliotheken und Frameworks (speziell EMF und Eclipse allgemein) profitiert.}, typsicheren Scala-DSL implementiert, die zu diesem Zweck ausgebaut wurde und so die Basis der TM2-Werkzeuge bildet. Für Details zur prototypischen Xtext-DSL s. Hinweise in Anhang \ref{anhang-archiv}, S. \pageref{anhang-archiv}.

\section{Typsichere Modellierung von Klassifikationsproblemen} \label{chapter-appl}  \index{Modellierung!typsicher} \index{Typisierung!bei der Modellierung}

\subsection{Informationsextraktion als einführendes Beispiel} \label{appl-ie}

Im folgenden Abschnitt (\ref{chapter-semantische-annotation}) wird aufbauend auf eigenen Vorarbeiten \citep{Steeg2007} eine einfache Form der Informationsextraktion dargestellt, die im anschließenden Abschnitt (\ref{tm2-ie}, S. \pageref{tm2-ie}) als einführendes Beispiel mit TM2 implementiert wird.

\subsubsection{Semantische Annotation und Informationsextraktion} \label{chapter-semantische-annotation} \label{eigennamen} \index{Annotation!semantische A.}

Das Ergebnis einer Informationsextraktion aus Texten kann als semantische Annotation festgehalten werden, bei der Textabschnitte mit ihrer Bedeutung annotiert werden, etwa \emph{Rhein} als `Fluss'. Eine solche semantische Annotation ist für verschiedene direkte Anwendungsfälle hilfreich (z.B. bei der Suche in den Texten), und für zahlreiche Probleme der maschinellen Sprachverarbeitung sogar notwendig (z.B. für die maschinelle Übersetzung von ambigen Wörtern in der Quellsprache, etwa engl. \emph{bank}).

Eine einfache Form der Informationsextraktion ist die Eigennamenerkennung (\emph{named entity recognition}), bei der in einem Text Eigennamen bestimmter Kategorien gesucht werden, z.B. in biologischen Texten Spezies, in medizinischen Medikamente oder bei Nachrichten Personen, Firmen oder Orte. \emph{Eigenname} ist dabei hier, wie bei \citet[27]{Frege1892}, die Bezeichnung eines bestimmten einzelnen Gegenstandes, wobei diese Bezeichnung aus einem einzigen oder auch aus mehreren Wörtern\footnote{In diesem Sinn ist die später modellierte Wortsinndisambiguierung (WSD) eigentlich eine Eigennamendisambiguierung, da abhängig davon, was als Sinneinheit annotiert wurde, einzelne Wörter, Wortgruppen oder Wortteile disambiguiert werden können.} bestehen kann.

\paragraph{Wortlisten zur Annotation} \index{Wortlisten} \index{Metaprogrammierung}

Eine einfache Umsetzung einer solchen Eigennamenerkennung kann durch Listen mit den Eigennamen\footnote{Auf Wortlisten basiert etwa ANNIE, die Informationsextraktionskomponente von GATE (\url{http://www.gate.ac.uk/ie/annie.html})} erfolgen. Hierbei wird pro Kategorie eine Liste angelegt, z.B. für alle Flüsse oder alle Städte, welche dann konkrete Einträge enthalten, wie \emph{Rhein} bzw. \emph{Köln}.

Wortlisten bilden eine gut wartbare, flexible Form eines maschinenlesbaren Lexikons. So ist etwa neben der Nutzung zur Informationsextraktion eine Integration von POS-Tags ebenso denkbar wie jede andere Form von Klassifikation: es müssen lediglich entsprechende Wortlisten (manuell oder automatisch) erstellt werden. Dabei ist keine Modifikation des Programms nötig. Es handelt sich hierbei um eine Form von Metaprogrammierung oder dynamischer Konfiguration, bei der das Programm die Abstraktionen enthält, während die Metadaten -- hier in Form der Listen -- die Details bereitstellen (vgl. \citealt[135]{HuntAndThomas2003}).

\paragraph{Ambiguität als praktisches Problem} \label{appl-ie-ambig}

Wäre in menschlicher Sprache jeder Wortform eindeutig eine Bedeutung zugeordnet, wäre eine semantische Annotation von Texten so ohne gro"se Schwierigkeiten automatisierbar -- einfach durch ein Abrufen der Bedeutung einer Wortform in einer Datenstruktur, die einem Wörterbuch entspricht und die auf Grundlage der Listen gefüllt wurde. Die Namen der Listen bilden dabei die semantische Kategorie oder Bedeutung der in der Liste enthaltenen Wörter. Beim Vorfinden einer Wortform (etwa \emph{Rhein} oder \emph{Köln}), lässt sich diese so mit ihrer Bedeutung auszeichnen (etwa als `Fluss' bzw. als `Stadt'). 

Mehrdeutige Wörter jedoch tauchen in mehreren Listen auf, so könnte etwa \emph{Bank} in den Listen `Möbel' (mit Einträgen wie \emph{Tisch} und \emph{Stuhl}) und `Ort' (mit Einträgen wie \emph{Kino} oder \emph{Supermarkt}) enthalten sein. Daraus ergibt sich folgendes Problem: Wie sollen Einträge, die in mehreren Listen enthalten sind, annotiert werden? Denn im Kontext des Vorkommens ist in der Regel nur eine der Lesarten richtig. Hier benötigen wir  einen Mechanismus zur Disambiguierung (WSD, Wortsinndisambiguierung, vgl. Abschnitt \ref{appl-wsd}, S. \pageref{appl-wsd}).

\paragraph{Konzeptuelle Umsetzung}

Bei einer Eigennamenerkennung ohne WSD würden bei Einträgen, die in mehreren Listen sind, alle Kategorien vergeben, etwa für alle Vorkommen von \emph{Bank} sowohl `Möbel' als auch `Ort'. Das Resultat dieses Aufbaus ist so ein semantisch annotierter, aber nicht disambiguierter Text.

Eine solche semantische Annotation, die nicht disambiguiert ist, ermöglicht bei der Extraktion einen Recall von 100\%, da die richtige Bedeutung in jedem Fall enthalten ist. Dass die Precision durch ambige Wörter reduziert ist, kann bei einer Verarbeitung der Annotationen durch Menschen unproblematisch sein. Für eine maschinelle Weiterverarbeitung allerdings ist eine Disambiguierung nötig (vgl. Abschnitt \ref{wsd-def}, S. \pageref{wsd-def}). Ein WSD-Verfahren müsste hier als nachgelagerte Komponente die ambigen Annotationen disambiguieren. In einem solchen Szenario sind die möglichen Klassen der WSD die tatsächlich vergebenen Kategorien der Informationsextraktion. Die WSD ist so ihrer Natur entsprechend komplett eingebettet in ihren Anwendungsfall der Informationsextraktion und verwendet keine für den konkreten Anwendungsfall arbiträren Einträge aus einem Lexikon.

\subsubsection{Umsetzung in TM2} \label{tm2-ie}

Als einführendes Beispiel zur Anwendung der in Kapitel \ref{chapter-tools} beschriebenen Werkzeuge wird in den folgenden Abschnitten die Umsetzung einer einfachen Informationsextraktion mit TM2 beschrieben.

\paragraph{Agentenmodellierung}

Basierend auf der bschriebenen konzeptuellen Darstellung der IE könnte ein entsprechender Aufbau in TM2 durch drei Agenten implementiert werden: einem Korpus, einem Tokenizer (als Präprozessor) und einem Gazetteer (zur Informationsextraktion). Der Tokenizer verarbeitet den Text des Korpus zu Token und kann in Scala etwa mit folgender Signatur implementiert werden:

\begin{lstlisting}
class Tokenizer extends Agent[String, Token]
\end{lstlisting}

Darauf aufbauend verarbeitet der Gazetteer die Token zu Bedeutungen:

\begin{lstlisting}
class Gazetteer extends Agent[Token, Sense]
\end{lstlisting}

\paragraph{Grundlegender Aufbau}

\begin{figure}
\begin{center}
  \includegraphics[width=5cm]{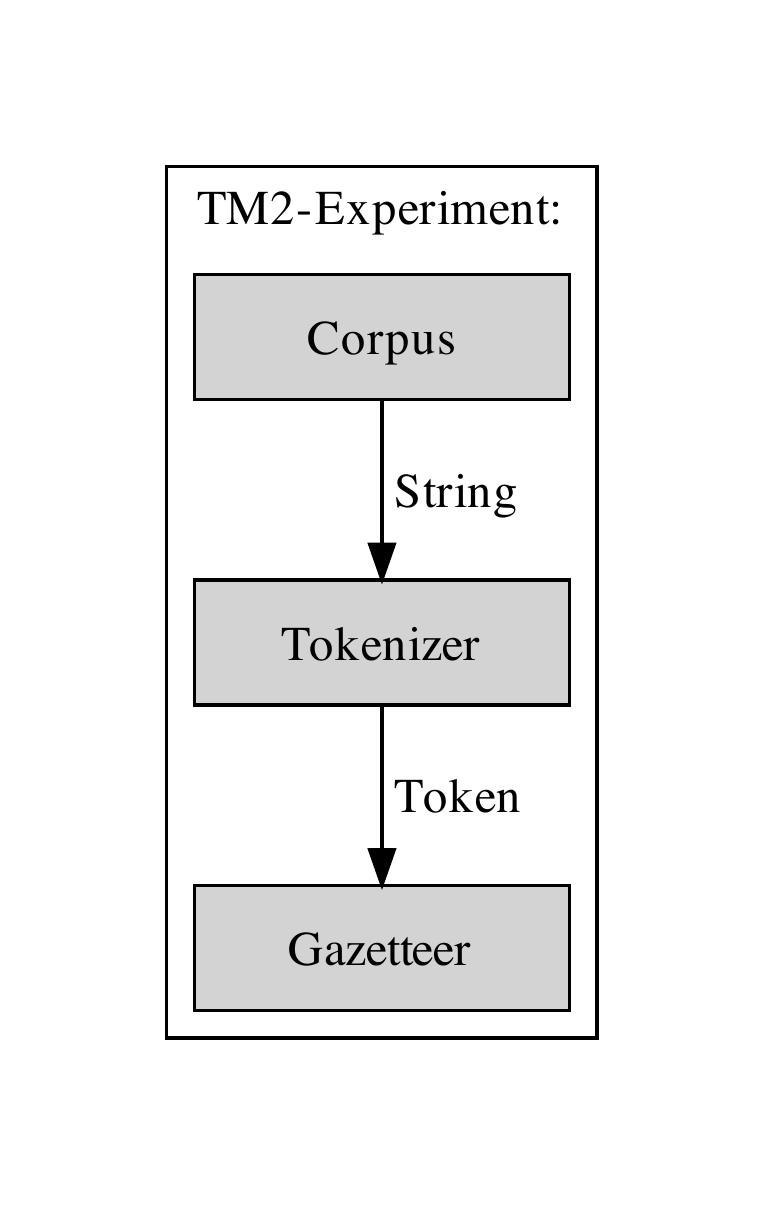}
  \caption{Generiertes Diagramm für Experiment in Listing \ref{exp-full-simple}, mit optionaler Ausgabe der ausgetauschten Typen}
  \label{full-simple}
\end{center}
\end{figure}

Unter der Verwendung solcher Agenten könnte eine einfache Informationsextraktion (inklusive Implementierung der Agenten) wie in Listing \ref{exp-full-simple}, S. \pageref{exp-full-simple} umgesetzt werden. Bei einem Export des Experiments enthält die generierte Dokumentation die Darstellung in Abbildung \ref{full-simple}, S. \pageref{full-simple}.

\lstset{language={Scala}}
\begin{lstlisting}[float, label=exp-full-simple, caption={Ein einfaches Experiment zur Informationsextraktion mit TM2 in Scala, inklusive Implementierung von Tokenizer und Gazetteer}]
object FullSimple extends Application {
  /* Tokenizer, ein Agent[String, Token]: */
  case class Token(form: String) extends Comparable[Token] with Serializable {
    def compareTo(that: Token) = this.form.compareTo(that.form)
  }
  class Tokenizer extends Agent[String, Token] {
    def process(t: java.util.List[Annotation[String]]): java.util.List[Annotation[Token]] =
      ("\\p{L}+".r findAllIn t.get(0).getValue()).matchData.map((m: Match) =>
        Annotations.create[Token](
          classOf[Tokenizer], Token(m.matched), m.start, m.end))
  }
  /* Gazetteer, ein Agent[Token, Sense]: */
  case class Sense(form: String) extends Comparable[Sense] with Serializable {
    def compareTo(that: Sense) = this.form.compareTo(that.form)
  }
  class Gazetteer extends AbstractAgent[Token, Sense] {
    def process(t: Token): Sense = {
      val prop = new Properties(); prop.load(new FileInputStream("files/dict0.properties"))
      Sense(prop.getOrElse(t.form, ""))
    }
  }
  /* Beteiligte Agenten und Experiment: */
  val (c, t, g): (Agent[String, String], Agent[String, Token], Agent[Token, Sense]) =
    (new Corpus, new Tokenizer, new Gazetteer)
  val x = c -> t | t -> g !
  /* Exeplarische Suche in den Ergebnissen: */
  println(x find t); println(x find g); println(x find (Sense("det"), g) by t))
}
\end{lstlisting}

\begin{figure}
\begin{center}
  \includegraphics[width=7cm]{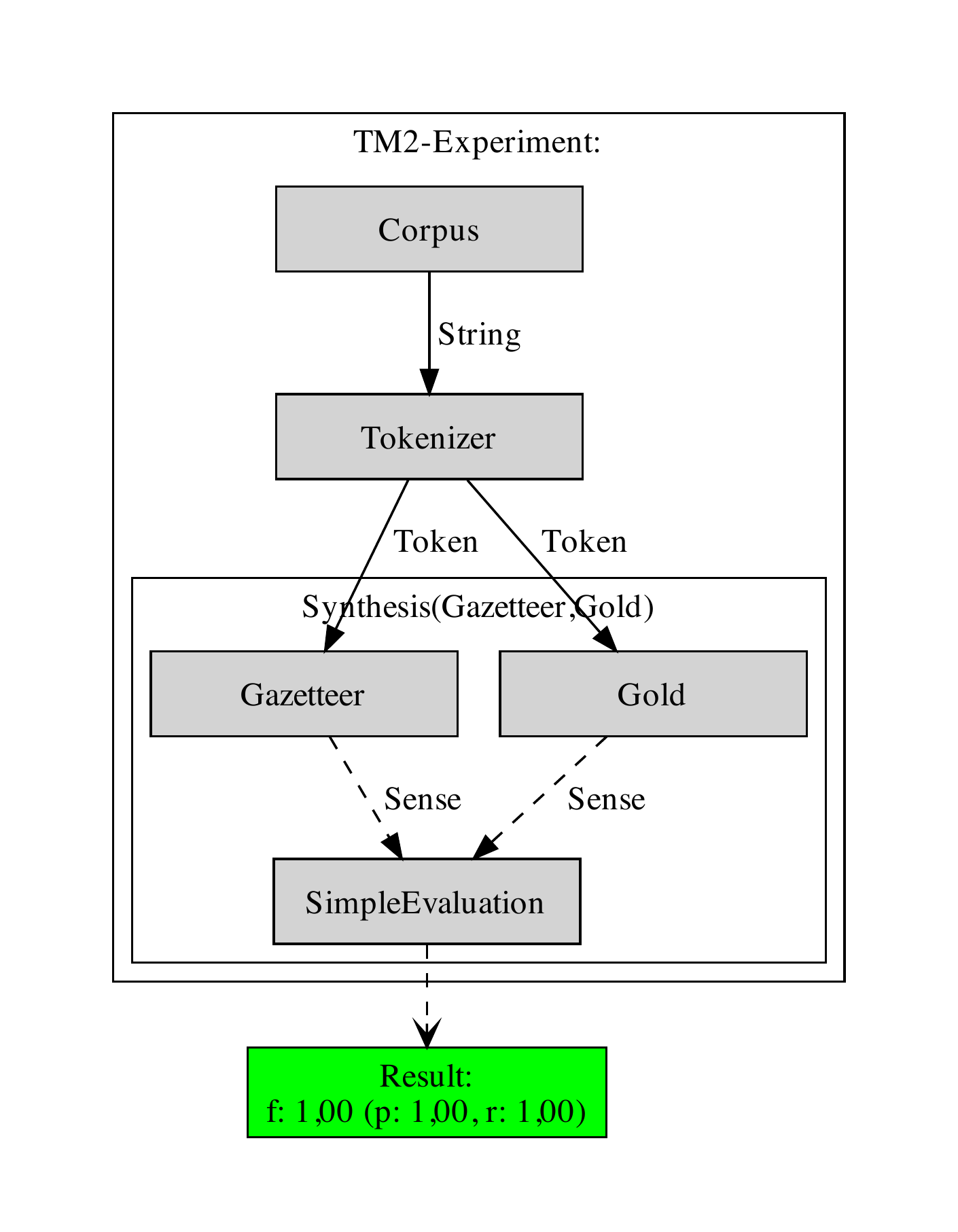}
  \caption{Generiertes Diagramm für Experiment in Listing \ref{exp-ie}, mit optionaler Ausgabe der ausgetauschten Typen}
  \label{anwendung-gazetteer-experiment}
\end{center}
\end{figure}

Ein entsprechendes Experiment inklusive Evaluierung könnte wie in Listing \ref{exp-ie}, S. \pageref{exp-ie} definiert und ausgeführt werden (hier ohne Implementierung der Agenten). Die Evaluierung erfolgt hier nicht gegen einen echten Goldstandard sondern gegen den Gazetteer selbst (\emph{Gold} ist eine Subklasse von \emph{Gazetteer}). Dies soll die Modellierung mit TM2 einführen und die generische Natur der Umsetzung unterstreichen: Der Goldstandard kann ein Agent wie jeder andere sein, er muss nur so typisiert sein, dass er die gleichen Ausgabedaten produziert wie der zu evaluierende Agent, vgl. Evaluierung echter Experimente in den Abschnitten \ref{appl-pseudoeval}, S. \pageref{appl-pseudoeval} und \ref{appl-senseval}, S. \pageref{appl-senseval}. Auch hier enthält die für ein solches Experiment generierte Dokumentation eine graphische "Ubersicht des Experiments (s. Abb. \ref{anwendung-gazetteer-experiment}, S. \pageref{anwendung-gazetteer-experiment}). Details zur Implementierung der beteiligten Agenten finden sich in Anhang \ref{anhang-ie}, S. \pageref{anhang-ie}.

\lstset{language={Scala}}
\begin{lstlisting}[float, label=exp-ie, caption={Ein einfaches Experiment zur Informationsextraktion, inklusive Evaluierung}]
val corpus = new Corpus
val tokenizer = new Tokenizer
val gazetteer = new Gazetteer
val gold = new Gold
val evaluation = new SimpleEvaluation[Sense]
val experiment = 
    corpus -> tokenizer |
    tokenizer -> (gazetteer, gold) |
    (gazetteer, gold) -> evaluation !
\end{lstlisting}

\paragraph{Suche in Ergebnissen}

In den Ergebnissen der Experimente kann über die API nach bestimmten Annotationen gesucht werden, etwa nach allen Annotationen des Gazetteer, der die IE durchgeführt hat:

\begin{lstlisting}
val result = experiment list gazetteer.getClass
\end{lstlisting}

Wie zuvor für API, GUI und DSL allgemein beschrieben, kann die Suche dabei in Annotationen eines bestimmten Agenten erfolgen, und Ergebnisse eines anderen Agenten für den gleichen Bereich zurückgeben, hier etwa die Token zu Elementen, die vom Gazetteer als \emph{NAME} annotiert wurden:

\begin{lstlisting}
val result = experiment find (Sense.NAME, gazetteer.getClass) by tokenizer.getClass
\end{lstlisting}

\paragraph{Experimentserien} \label{tdd}

Eine Definition mehrerer, unterschiedlich konfigurierter Experimente für einen solchen Aufbau ist wie in Abschnitt \ref{dsl-model-experiments}, S. \pageref{dsl-model-experiments} beschrieben möglich (s. speziell Listing \ref{sample-exp}, S. \pageref{sample-exp}, vgl. allgemeine Form der Experimentserien in Abb. \ref{dsl-scala-form}, S. \pageref{dsl-scala-form}). Die Ergebnisse der einzelnen Aufbauten werden in der generierten Dokumentation tabellarisch aufbereitet, um eine einfache Analyse der Ergebnisse zu ermöglichen (vgl. komplette Experimentserie zur WSD in Abschnitt \ref{appl-senseval}, S. \pageref{appl-senseval}). Auf diese Weise kann etwa beim Entwickeln neuer Verfahren zur IE laufend kontrolliert werden, ob und wie Veränderungen der Agentenkonfigurationen die Ergebnisse beeinflussen, etwa wie sich komplexe Tokenisierungsalgorithmen oder die Verwendung unterschiedlicher Korpora auswirken. Dies ermöglicht eine Art testgetriebener Entwicklung \citep{Beck2003} von Text-Mining-Komponenten und -Verfahren.

\subsection{Wortsinndisambiguierung durch Kontextabstraktion} \label{chapter-wsd-ml}

Das einführende Beispiel zur Informationsextraktion in Abschnitt \ref{appl-ie}, S. \pageref{appl-ie} zeigt die grundlegende Verwendung des TM2-Frameworks und stellt zugleich die Bedeutung der Disambiguierung für die Entwicklung von Verfahren der maschinellen Sprachverarbeitung dar: ohne Disambiguierung ist etwa eine präzise Informationsextraktion nicht möglich. Dies gilt ebenso für andere Anwendungsfälle, z.B. maschinelle Übersetzung. Zur Vorbereitung der vollständigen Evaluierung der TM2-Werkzeuge durch ihre Anwendung in Abschnitt \ref{appl-wsd}, S. \pageref{appl-wsd} führen die folgenden Abschnitte zunächst in die Domäne der Wortsinndisambiguierung durch maschinelles Lernen und die sich aus diesem Anwendungsbereich ergebenden Ziele und Anforderungen für die Experimente ein. Dazu werden eigene Vorarbeiten \citep{Steeg2007} aufgegriffen, erweitert und durch die Implementierung in TM2 konzeptuell verallgemeinert (vgl. abschließende Darstellung in Kapitel \ref{chapter-ausblick}, S. \pageref{chapter-ausblick}).

\subsubsection{Motivation und "Uberblick}

Mehrdeutige Wörter sind Teil der menschlichen Sprache und seit Beginn der Schriftkultur belegt \citep[153]{Haarmann1991}. Wortsinndisambiguierung (WSD, engl. \emph{word sense disambiguation}), der Prozess der Auflösung der Mehrdeutigkeit eines Wortes anhand seines Kontextes, fällt Menschen leicht; maschinell ist dieser Prozess jedoch bislang nicht in vergleichbarer Form durchführbar. Dies ist ein wesentlicher Grund dafür, dass Computer Sprache nicht verstehen können und macht so die WSD zu einem Kernproblem der Computerlingu\-istik.

Der Mensch abstrahiert beim kognitiven Prozess der WSD von den konkreten Kontexten der ambigen Wörter, vermutlich auf Grundlage eines ``einheitlichen Modus [...] der Informationsverarbeitung'' \citep[145]{Singer2002}, mit dem Daten unterschiedlicher Herkunft (d.h. die  verschiedenen Sinneswahrnehmungen) verarbeitet werden. Diese Verbindung aus domänenspezifischen Daten, die mit einem domänenübergreifenden Mechanismus verarbeitet werden, entspricht Prinzipien des maschinellen Lernens, dessen Datenbasis in der Sprachverarbeitung Korpora bilden (vgl. Abschnitte \ref{ml-base-corpus}, S. \pageref{ml-base-corpus}).

Diese Konzepte werden im Folgenden zur Evaluierung des TM2-Frameworks mit verschiedenen Klassifikationsverfahren und Daten des British National Corpus (BNC) zur WSD umgesetzt (s. Abschnitt \ref{appl-senseval}, S. \pageref{appl-senseval}). Die modulare Umsetzung der WSD im TM2-Framework macht das WSD-Verfahren für unterschiedliche Anwendungen in der maschinellen Sprachverarbeitung zugänglich. Sie eröffnet zudem zahlreiche Möglichkeiten zur Reproduktion und Weiterentwicklung des Verfahrens selbst sowie darüber hinaus, etwa durch die Nutzung einzelner Bestandteile des Verfahrens in anderen Zusammenhängen (vgl. Abschnitt \ref{tm2-fazit}, S. \pageref{tm2-fazit}).

\subsubsection{Wortsinndisambiguierung} \label{wsd-def}

\paragraph{Definition und Einführung} \index{Ambiguität} \index{Kognition!WSD als kognitives Problem}

Wortsinndisambiguierung ist eine kognitive Leistung bei der men\-sch\-lich\-en Sprach\-ver\-arbeitung und ein zentrales Problem der Computerlinguistik. Gegenstand der WSD ist die Auswahl der im Kontext passenden Lesart mehrdeutiger Wörter (\citealt{AgirreAndEdmonds2006a}), etwa in \emph{Hast du die neue Bank gesehen?} die zu aktivierende Lesart von \emph{Bank}, z.B. als `Möbel' oder als `Ort'.

WSD ist eine Voraussetzung für verschiedene Aufgaben der maschinellen Sprachverarbeitung\footnote{z.B. bei der maschinellen "Ubersetzung: wenn etwa ein MÜ-System in einem englischen Text auf das Wort \emph{bank} trifft, so muss dies je nach Lesart etwa im Deutschen als \emph{Bank} oder als \emph{Ufer} übersetzt werden -- oder bei der Informationsextraktion: werden etwa in englischen Texten Begriffe des Finanzwesens gesucht, sollte eine Fundstelle von \emph{bank} je nach Lesart ausgewählt oder übergangen werden.}. WSD ist für Menschen in der Regel\footnote{Eine Vielzahl von Witzen basiert auf falsch oder nicht aufgelöster Ambiguität, z.B.: \emph{Treffen sich zwei Jäger. -- Beide tot.}} einfach, für Computer aber bislang, insbesondere für Polysemie, nicht vollständig möglich. Aufgrund seiner Abhängigkeit von verschiedenen kognitiven Leistungen wie logischem Denken und Weltwissen (\citealt[2]{IdeAndVeronis1998}, \citealt[1]{AgirreAndEdmonds2006a}) kann WSD als kognitives Problem charakterisiert werden. Auf Grundlage einer Vorstellung von Wortsinn als Kontextabstraktion soll in den folgenden Abschnitten das TM2-Framework zur Umsetzung eines Verfahrens zur WSD verwendet und so evaluiert werden.

\paragraph{Forschungsstand der WSD} \label{chapter-wsd-stand}

Durch die Arbeit von Lexikographen und durch die Digitalisierung von Lexika (vgl. \citealt[vii]{WilksEtAl1996}) stehen vielfältige Ressourcen für die maschinelle Sprachverarbeitung zur Verfügung, in denen die Bedeutung eines Wortes auf Grundlage der Entscheidungen von Lexikographen enthalten ist. Die Arbeitshypothese in der maschinellen Sprachverarbeitung und speziell der WSD, lautete daher für den Bereich der Bedeutung lange: Bedeutung ist das, was im Lexikon steht. Dies führt jedoch zu verschiedenen Problemen, die in der WSD-Community aus pragmatischen Gründen\footnote{Die Natur des WSD-Problems erfordert ein Bedeutungsinventar, aus dem die passende Bedeutung ausgewählt wird (vgl. Abschnitt \ref{appl-ie}, S. \pageref{appl-ie}).} lange ignoriert wurden (\citealt[9]{AgirreAndEdmonds2006b}, vgl. \citealt{Kilgarriff2006}). Die allgemeine Gültigkeit von Verfahren und der Wert von Erkenntnissen auf Basis bestimmter Lexika ist aber fragwürdig, da z.B. unterschiedliche Anwendungen der WSD eine unterschiedliche Granularität der möglichen Bedeutungen erfordern \citep[58]{IdeAndWilks2006}, vgl. Abschnitt \ref{appl-ie}, S. \pageref{appl-ie}.

Unterschiedliche Verfahren zur isolierten WSD sind inzwischen ausführlich beschrieben \citep{WilksEtAl1996,IdeAndVeronis1998,ManningAndSchuetze1999,Stevenson2003,AgirreAndEdmonds2006b}. Forschungsbedarf besteht dagegen allgemein im Bereich der Integration mit der eigentlichen Anwendung: entsprechend seiner Natur als Mittel sollte WSD im Kontext einer konkreten Anwendung evaluiert werden, was bisher jedoch kaum geschieht (\citealt{Stevenson2003, AgirreAndEdmonds2006b}). 

Die maschinelle Disambiguierung von Homonymie ist erheblich einfacher als die von Polysemie. So wird automatische WSD für Homonymie mit Ergebnissen über 95\% schon bei kleinen Trainingsinputs als gelöstes Problem angesehen\footnote{Eine solche Aussage, dass WSD für Homonymie gelöst sei \citep[14]{AgirreAndEdmonds2006b}, heißt dabei im Grunde aber, dass WSD bisher nur für die einfachsten Fälle von Mehrdeutigkeit funktioniert.} \citep[14]{AgirreAndEdmonds2006b}. WSD für Polysemie ist dagegen deutlich schwieriger, da eine Abgrenzung der polysemen Lesarten häufig nicht einfach ist (je nach Korpus Leistungen von 60-70\%, vgl. \citealt[14]{AgirreAndEdmonds2006b}); dies gilt dabei nicht nur für maschinelle Verfahren, sondern auch für die menschliche Leistung: so beträgt etwa die Übereinstimmung der menschlichen Annotatoren (\emph{inter-annotator agreement}, ITA) für die Daten des \emph{English Lexical Sample Task} im Rahmen von Senseval-3 (s. Abschnitt \ref{appl-senseval}, S. \pageref{appl-senseval}) lediglich 67,3\% (\citealt{MihalceaEtAl2004}; \citealt[14]{AgirreAndEdmonds2006b}).

\subsection{Typsichere Modellierung von maschinellem Lernen} \label{appl-wsd} 

Im Folgenden wird eine Umsetzung von maschinellem Lernen zur WSD mit TM2 beschrieben.

\subsubsection{Merkmalsberechnung} \label{appl-merkmale} \index{Merkmale!Merkmalsberechnung} \index{Abstraktion!von sprachlichen Symbolen}

Als Beispielagenten zur Evaluierung des TM2-Frameworks werden Verfahren zur Kontextrepräsentation aus eigenen Vorarbeiten (\citealt{Steeg2007}) gegeneinander evaluiert: die Wörter, Wort\-läng\-en, sowie buchstabenbasierten Tri- und Heptagramme (7-Gramme) im Kontext der Ziel\-wörter. Die drei dargestellten Verfahren zur Merkmalsberechnung haben nicht den Anspruch, praxistaugliche Algorithmen zur numerischen Merkmalsrepräsentation zu sein, sondern stellen im Sinne eines \emph{Tutorials} einfache exemplarische Überlegungen und Fragestellungen dar, mit denen die Verwendung des TM2-Frameworks demonstriert werden soll.

Konzeptuell wird bei all diesen Verfahren der Kontext eines Tokens mit einem einzigen Merkmalsvektor dargestellt, dessen Merkmale auf den Wörtern oder buchstabenbasierten N-Grammen der Wörter im Kontext des Zielwortes basieren. Alternative Repräsentationen wären etwa die Verwendung eigener Vektoren für jedes Wort im Kontext, die zusammengenommen die Merkmale des Zielwortes darstellen könnten. Solche Verfahren könnten als zusätzliche Agenten implementiert und in den dargestellen Experimenten evaluiert werden.

\subsubsection{Agentenmodellierung} \index{Agenten!Modellierung}

Die verschiedenen Verfahren zur Merkmalsberechnung können in TM2 (unter Verwendung der beschriebenen standardisierten Darstellung von Merkmalen als numerische Vektoren) als einziger Agent implementiert werden, dem bei der Intanziierung ein Konfigurations\-string mit dem zu verwendenden Verfahren übergeben wird (\lstinline!"word", "length", "3-gram"!, etc.), z.B. \lstinline!new Features("3-gram")! für Trigramme. Ein solcher Agent erzeugt Merkmalsvektoren für Kontexte:

\begin{lstlisting}
class Features implements Agent<Context, FeatureVector>
\end{lstlisting}

Der \lstinline!Context! liefert dabei die darzustellenden Daten, ein \lstinline!FeatureVector! enthält die erzeugten numerischen Werte in einer \lstinline!List<Float>!. Details zur Implementierung der Agenten finden sich in Anhang \ref{anhang-wsd-agent-impl}, S. \pageref{anhang-wsd-agent-impl}.

\subsubsection{Integration bestehender Klassifikatoren: Weka} \label{appl-weka}

Weka\footnote{Weka Machine Learning Project, \url{http://www.cs.waikato.ac.nz/~ml/}, vgl. \citet{WittenAndFrank2000}} enthält eine gro"se Menge wohldokumentierter Umsetzungen von etablierten und experimentellen Klassifikationsalgorithmen in Java. Es bietet damit eine wertvolle Ressource für die Umsetzung von Text-Mining-Experimenten allgemein, und speziell zur Klassifikation in TM2.

\paragraph{Weka-Klassifikatoren}

Weka-Klassifikatoren (z.B. \lstinline!weka.classifiers.bayes.NaiveBayes!) erben von einer gemeinsamen Superklasse (\lstinline!weka.classifiers.Classifier!). Durch die Aggregation eines solchen Weka-Klass\-ifi\-ka\-tors lässt sich ein Weka-Wrapper generisch implementieren: der Wrapper wird mit der konkreten Instanz eines  Weka-Klassifikators erzeugt -- z.B.  \lstinline!new WekaWrapper(new NaiveBayes)! -- und delegiert die eigentliche Klassifikation an diesen, vgl. \citet[81]{Bloch2008}, \citet[20]{GammaEtAl1995}. Details zur Implementierung\footnote{Wie bestimmte Teile von Experimenten (vgl. Abschnitt \ref{impl-concurrency}, S. \pageref{impl-concurrency}) können auch Training und Klassifikation ohne gemeinsamen Zustand und so konzeptuell einfach nebenläufig implementiert werden, indem die Klassifikatoren für die unterschiedlichen Kategorien (bei der WSD etwa die ambigen Lemmata, z.B. \emph{Bank}) getrennt trainiert und verwendet werden. So können alle untersuchten Lemmata parallel verarbeitet werden. Das in Abschnitt \ref{appl-senseval} (S. \pageref{appl-senseval}) verwendete Korpus etwa enthält 57 ambige Lemmata, wodurch hier bei Training und Disambiguierung 57 separate Threads gestartet werden könnten. Im gleichen Maße, in dem bei einem Anstieg der verwendeten Lemmata der Aufwand für Training und Disambiguierung zunehmen würde, würde so automatisch zunehmend parallele Rechenleistung nutzbar gemacht.} der Weka-Integration finden sich in Anhang \ref{anhang-wsd}, S. \pageref{anhang-wsd}. Implementiert ein solcher Wrapper zudem das \emph{Agent}-Interface, ist das eigentliche Klassifikationsverfahren ein Konfigurationsdetail eines Agenten und kann ausgetauscht und gegen Alternativen evaluiert werden (vgl. Experimente in Abschnitt \ref{appl-senseval}, S. \pageref{appl-senseval}).

\paragraph{Agentenmodellierung}

In Begriffen der TM2-API kann maschinelles Lernen mit den beschriebenen Agenten für Merkmalsberechnung und Klassifikation als Kombination aus Analyse und Synthese modelliert werden: Beim Training wird in einer Synthese von Merkmalen und Bedeutung ein Modell gebildet, das die spätere Klassifikation ermöglicht (\lstinline!Model<FeatureVector, Sense>!). Bei der Klassifikation analysiert der Klassifikator Merkmale eines \lstinline!Agent<?, FeatureVector>! (vgl. beschriebene Agentenmodellierung für die Merkmalsberechnung) und ermittelt daraus Bedeutungen, d.h. er ist ein \lstinline!Agent<FeatureVector, Sense>!. Für den Wrapper ergibt sich damit in Java folgende Typsignatur:

\begin{lstlisting}
class WekaWrapper implements Agent<FeatureVector, Sense>, Model<FeatureVector, Sense>
\end{lstlisting}

Die entsprechende Signatur in Scala beschreibt die modellierte Funktion noch treffender:

\begin{lstlisting}
class WekaWrapper extends Agent[FeatureVector, Sense] with Model[FeatureVector, Sense]
\end{lstlisting}

So kann der Weka-Wrapper in Analysen und Synthesen (vgl. Abschnitt \ref{agent-interaction}, S. \pageref{agent-interaction}) mit anderen Agenten zur Durchführung von Experimenten verwendet werden (vgl. Experimente in Abschnitt \ref{appl-senseval}, S. \pageref{appl-senseval}).

\subsubsection{WSD mit Pseudoambiguität} \label{appl-pseudoeval}

Ein einfaches Mittel zur Evaluierung von WSD-Systemen ist automatisch generierte Ambiguität in Form sogenannter Pseudowörter. Dabei wird jedes Auftreten von zwei Wortformen der gleichen Wortart durch eine Zusammensetzung der Wörter ersetzt (etwa alle Vorkommen von \emph{banana} und \emph{door} durch \emph{banana-door}). Die korrekten Lesarten sind so durch die Ursprungsform gegeben (z.B. \emph{banana} oder \emph{door}), vgl. \citealt[233]{ManningAndSchuetze1999}.

Gegen eine Evaluierung auf dieser Grundlage spricht, dass es sich bei Pseudowörtern um kein echtes sprachliches Phänomen handelt und daher gute Ergebnisse nur bedingt Rückschlüsse auf Phänomene der lexikalischen Semantik erlauben. So entsprechen Pseudowörter etwa nie Polysemie, sondern stets eher Homonymie \citep[86]{PalmerEtAl2006}. Das heißt im Zusammenhang mit der in Abschnitt \ref{chapter-wsd-ml}, S. \pageref{chapter-wsd-ml} dargestellten Vorstellung von Wortsemantik als Kontextabstraktion, dass Pseudoambiguität im Vergleich zu natürlicher Mehrdeutigkeit zu einfach aufzulösen ist. Daher wird Pseudoambiguität hier nur als Beispiel zur Einführung der Modellierung von WSD mit TM2 verwendet. Vollständige Experimente finden sich in der Senseval-Evaluierung in Abschnitt \ref{appl-senseval}, S. \pageref{appl-senseval}.

Aufbauend auf die oben zur Agentenmodellierung verwendeten Elemente des TM2-Frame\-works kann ein einfaches, auf Pseudoambiguität basierendes WSD-Experiment etwa wie in Listing \ref{code-wsd-pseudo}, S. \pageref{code-wsd-pseudo} definiert und ausgeführt werden. Durch die Nutzung von Pseudoambiguität kann die gleiche Klasse für die Bereitstellung von Trainings-, Test- und Goldstandard-Annotationen verwendet werden (im Experiment in Form von Subklassen der Klasse \lstinline!PseudoAmbig!).

\begin{lstlisting}[float, label=code-wsd-pseudo, caption={WSD-Experiment mit Pseudoambiguit"at}]
class Ambig extends PseudoAmbig("and", "the")
class Gold extends PseudoAmbig("and", "the")

val (corpus, ambiguity, gold, tokenizer, evaluation) = 
  (new Corpus, new Ambig, new Gold, new Tokenizer, new SimpleEvaluation)
   
val (features, classifier) = 
  (new Features("3-gram"), new Classifier(new SMO))
   
val x =
  /* Vorverarbeitung: */
  corpus -> tokenizer | tokenizer -> ambiguity |
  /* Trainingsmerkmale: */
  (tokenizer, ambiguity) -> features | 
  /* Testmerkmale: */
  ambiguity -> features |
  /* Training des Klassifikators: */
  (features, ambiguity) -> classifier | 
  /* Klassifikation: */
  features -> classifier | 
  /* Evaluierung: */
  tokenizer -> gold | (classifier, gold) -> evaluation !
\end{lstlisting}

\begin{figure}
\begin{center}
  \includegraphics[width=6cm]{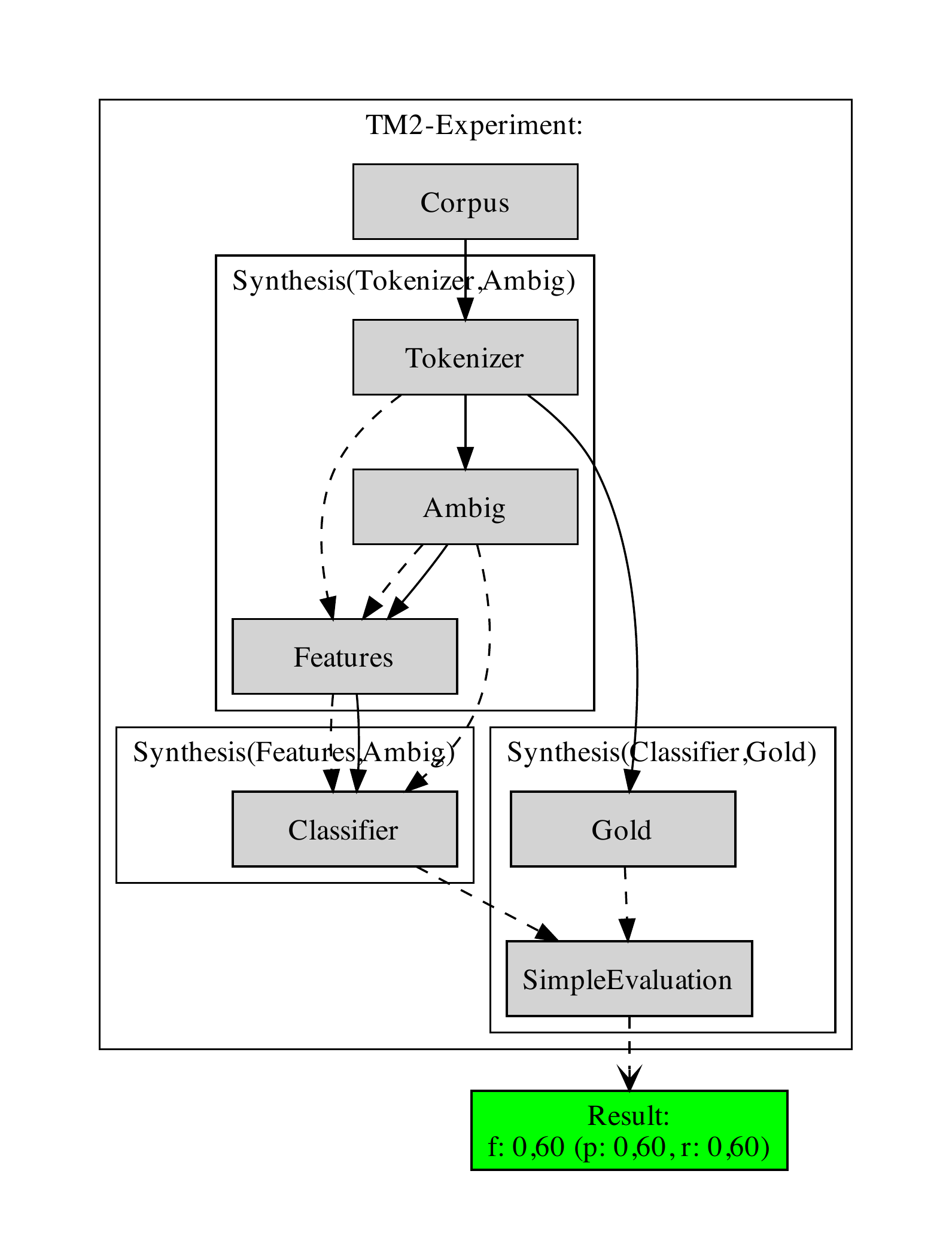}
  \caption[Für WSD-Experiment mit Pseudoambiguität generiertes Diagramm]{Für WSD-Experiment mit Pseudoambiguität generiertes Diagramm (durchgezogene Linien beschreiben Analysen, gestrichelte Linien beschreiben Synthesen)}
  \label{anwendung-wsd-pseudo}
\end{center}
\end{figure}

Die generierte Dokumentation enthält eine "Ubersicht über den Aufbau des Experiments, die in Abbildung \ref{anwendung-wsd-pseudo}, S. \pageref{anwendung-wsd-pseudo} wiedergegeben ist (vgl. Modellierung eines alternativen WSD-Experiments in Abschnitt \ref{appl-senseval}, S. \pageref{appl-senseval}). Der von diesem einführenden Experiment verwendete Text und Details zur Implementierung der beteiligten Agenten finden sich in Anhang \ref{anhang-wsd}, S. \pageref{anhang-wsd}.

\subsubsection{Mögliche Experimente und Fragestellungen} \label{appl-pseudoeval-results}

Auf Basis des beschriebenen Grundaufbaus und der dargestellten Implementierungen der Agenten kann nun durch spezielle Experimente und deren Evaluierung bestimmten Fragestellungen nachgegangen werden, etwa der Suche nach optimalen Verfahren zur Merkmalsberechnung oder Klassifikation. Diese Möglichkeiten werden im Rest dieses Abschnitts beschrieben.

\paragraph{Generische Evaluierung durch Annotationsvergleich}

Bei der dargestellten Evaluierung mit Pseudoambiguität liegen die Ergebnisse und der Goldstandard im gleichen Format vor, sie werden im Experiment von zwei Agenten mit einer gemeinsamen Superklasse erzeugt (\lstinline!PseudoAmbig!). Dadurch kann ein Vergleich auf Basis eines generischen Evaluations-Agenten erfolgen, der unabhängig von den konkreten Typen die Werte der Annotationen vergleicht\footnote{Die Typen von TM2-Annotationen müssen das Java-Interface \emph{Comparable} implementieren und haben so einen einheitlichen Vergleichsmechanismus, der etwa für eine generische Evaluierung (aber auch Sortierung) genutzt werden kann.} (vgl. im Gegensatz dazu die native Senseval-Evaluierung in Abschnitt \ref{appl-senseval}, S. \pageref{appl-senseval}). Die Evaluierung kann dabei wie die Klassifikation als Modell beschrieben werden, das hier in einer Synthese aus dem Ergebnis der WSD und dem Goldstandard gebildet wird (siehe Abb. \ref{anwendung-wsd-pseudo}, S. \pageref{anwendung-wsd-pseudo}).

\paragraph{Evaluierung von Verfahren zur Merkmalsberechnung}

Zur Evaluierung von verschiedenen Verfahren zur Merkmalsberechnung kann auf Basis des oben dargestellten Aufbaus eine Experimentserie formuliert werden. Dabei würden die unterschiedlich konfigurierten Agenten zur Merkmalsberechnung etwa folgendermaßen deklariert:

\begin{lstlisting}
for { // ...
  impl <- List("3-gram", "7-gram", "word", "length")
  feat = new Features(impl, 4)
} // ...
\end{lstlisting}

In diesem Beispiel würde das Experiment so viermal ausgeführt, jeweils mit den vier unterschiedlichen Verfahren zur Merkmalsberechnung (vgl. Grundform für Experimentserien in Abb. \ref{dsl-scala-form}, S. \pageref{dsl-scala-form}).

\paragraph{Evaluierung von Klassifikatoren}

Analog kann eine Evaluierung verschiedener Klassifikationsverfahren umgesetzt werden, etwa unter Verwendung unterschiedlicher Weka-Implementierun\-gen (vgl. Abschnitt \ref{appl-weka}, S. \pageref{appl-weka}).

Wie bei der Evaluierung der Merkmalsberechnung könnte hier das Experiment viermal mit unterschiedlichen Klassifikationsalgorithmen ausgeführt werden, die (wie oben die Strings zur Spezifikation der Merkmalsberechnung) als Konfigurationsdetail dem Agenten übergeben werden können:

\begin{lstlisting}
for { // ...
  algo <- List(new BayesNet, new NaiveBayes, new SMO, new HyperPipes)
  clas = new Classifier(algo)
} // ...
\end{lstlisting}

\paragraph{Experimentserien und Evaluierung durch Auswertung}

Ebenso können sämtliche Variablen in einer einzigen Experimentserie formuliert werden, und unterschiedliche Fragestellungen (z.B. nach Merkmalsberechnung oder Klassifikation) auf Basis der tabellarisch aufbereiteten Ergebnisse beantwortet werden. Bei einer solchen Reihe wären beide Agenten variabel, und sämtliche Permutationen der Konfigurationen würden automatisch ausgeführt (hier 4 Verfahren zur Merkmalsberechnung und 4 Klassifikatoren, also 4*4=16 Durchläufe):

\begin{lstlisting}
for { // ...
  impl <- List("3-gram", "7-gram", "word", "length")
  algo <- List(new BayesNet, new NaiveBayes, new SMO, new HyperPipes)
  feat = new Features(impl, 4)
  clas = new Classifier(algo)
} // ...
\end{lstlisting}

Details zur Implementierung dieser einführenden Experimente finden sich in Anhang \ref{ref-wsd}, S. \pageref{ref-wsd}. Eine vollständige Beschreibung solcher Experimentserien mit Senseval-Daten (und damit auf Basis echter Mehrdeutigkeit) findet sich im folgenden Abschnitt \ref{appl-senseval}.

\subsection{Senseval-Evaluierung} \label{appl-senseval}

Der folgende Abschnitt \ref{senseval-workshops} beschreibt basierend auf eigenen Vorarbeiten \citep{Steeg2007} die Senseval-Workshops und das verwendete Senseval-Korpus. Darauf aufbauend wird ab Abschnitt \ref{agentenmodellierung}, S. \pageref{agentenmodellierung} die Implementierung einer Senseval-Evaluierung mit TM2 dargestellt.

\subsubsection{Korpora zur Evaluierung} \label{senseval-workshops} \index{Korpora!zur Evaluierung} \index{Evaluierung!mit Korpora}

Mit den Senseval-Workshops\footnote{Evaluation Exercises for the Semantic Analysis of Text, \url{http://www.senseval.org/senseval3}} existiert eine Reihe von Veranstaltungen, die (nach dem Vorbild der \emph{Message Understanding Conferences} in der Informationsextraktion, s. etwa \citealt{GrishmanAndSundheim1996}) eine vergleichende Evaluierung von Systemen zur WSD auf der Basis gemeinsamer Korpora zum Gegenstand haben (vgl. \citealt[86]{PalmerEtAl2006}). Für diese Workshops wurden Trainings- und Testkorpora bereitgestellt, für die Vergleichswerte wie das \emph{inter-annotator agreement} (ITA) verfügbar sind. Im Folgenden wird die grundsätzliche, zuvor beschriebene Modellierung von WSD mit TM2 mit Korpora des \emph{english lexical sample task} im Rahmen von Senseval-3 (vgl. \citealt{MihalceaEtAl2004}) umgesetzt. So soll die praktische Anwendbarkeit des TM2-Frameworks für eine komplexe computerlinguistische Fragestellung dargestellt werden.

\begin{table}
\begin{center}\begin{tabular}{l|c|c|c}
Wortarten & Lemmata & Lesarten & Lesarten \\
& & (fein) & (grob)\\\hline
Nomen & 20 & 5,8 & 4,4 \\
Verben & 32 & 6,3 & 4,6 \\
Adjektive & 5 & 10,2 & 9,8 \\\hline
Gesamt & 57 & 6,5 & 5,0\end{tabular} \caption[Überblick über das Bedeutungsinventar]{Überblick über das Bedeutungsinventar im verwendeten Senseval-Korpus: Wortarten der ambigen Lemmata und durchschnittliche Anzahl von Lesarten pro Wort bei feiner und grober Körnung (aus: \citealt{MihalceaEtAl2004})}\label{inventar}
\end{center}
\label{defaulttable}
\end{table}

Die Senseval-Trainingskorpora enthalten ambige Zielwörter, die mit ihren korrekten Lesarten gekennzeichnet wurden. Die Umsetzung der Senseval-Annotation als Agent macht die verwendeten Senseval-Daten auch für andere Experimente zugänglich, die keine anderen der hier genutzten Agenten (etwa für Merkmalsberechnung und Klassifikation) verwenden, und erschließt so prinzipiell auch die Senseval-Korpora\footnote{Die Daten haben grundsätzlich ein gemeinsames Format und sind so prinzipiell wie hier implementiert nutzbar. Sie sind jedoch z.T. nicht in validem XML (z.B. keine maskierten Sonderzeichen), sowie manchmal in einer einzelnen Datei und manchmal für verschiedene Lemmata auf verschiedene Dateien verteilt. Hier müsste also ein Import der verschiedenen Varianten des Formats oder eine Normalisierung der Daten implementiert werden.} der anderen \emph{tasks}\footnote{Senseval 3 Tasks, \url{http://www.senseval.org/senseval3/tasks.html}} für solche Experimente. Auf der anderen Seite sind neben der WSD auch andere Aufgaben der maschinellen Sprachverarbeitung tokenbezogene Klassifikationsprobleme, etwa das Auszeichnen von Wortarten (\emph{part of speech tagging}, POS-Tagging). Hier würde etwa ein bloßer Austausch des Senseval-Agenten (der das Korpus bereitstellt) gegen ein mit POS-Tags ausgezeichnetes Korpus die Nutzung der eigentlich zur WSD implementierten Merkmalsberechnung und  Klassifikation zum POS-Tagging ermöglichen.

Das Senseval-3 Korpus des \emph{english lexical task} enthält Ausschnitte des BNC -- für das Training 8529 Vorkommen 57 ambiger Lemmata, zur Disambiguierung 5693 Vorkommen. Einen Überblick über das Bedeutungsinventar des Korpus gibt Tabelle \ref{inventar}, S. \pageref{inventar}. Die Bedeutungen stammen für Nomen und Adjektive aus WordNet 1.7.1, für Verben aus Wordsmyth (\citealt{MihalceaEtAl2004}).

\begin{figure}
    \begin{center}
 \includegraphics[height=4.5cm]{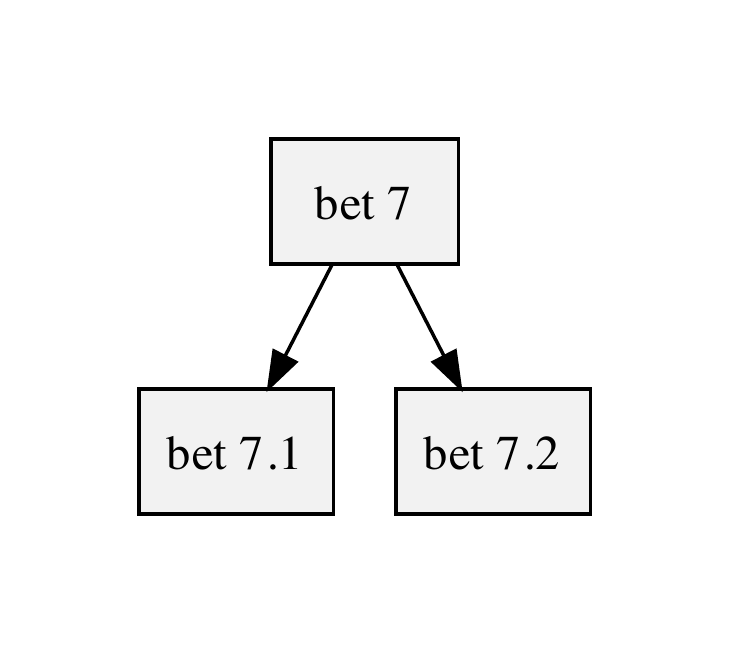}
\caption[Hierarchisch stukturierte Lesarten im Senseval-Korpus]{Hierarchisch stukturierte Lesarten im Senseval-Korpus (\citealt[79]{PalmerEtAl2006})}
\label{grained}
    \end{center}
   \end{figure}

Wie die Bedeutungen nicht-ambiger Wörter können auch die verschiedenen Lesarten ambiger Wörter einander konzeptuell über- (Hyperonymie) oder untergeordnet (Hyponymie) sein, sowie eine gemeinsame, übergeordnete Bedeutung haben (Kohyponymie). So ist etwa in WordNet\footnote{WordNet kann unter \url{http://wordnetweb.princeton.edu/perl/webwn} online abgefragt werden.} \emph{bet} mit der Lesart `the money risked in a gamble' den weiteren Lesarten `the initial contribution that each player makes to the pot' und `the combined stakes of the betters' übergeordnet (vgl. Abb. \ref{grained}).

Bei einer solchen hierarchischen Strukturierung möglicher Bedeutungen können bei der Evaluierung drei Stufen von Bedeutungsgranularität zugrunde gelegt werden: feine (\emph{fine}), gemischte (\emph{mixed}) und grobe (\emph{coarse}) Körnung \citep[79]{PalmerEtAl2006}. Bei feiner Körnung sind nur exakte Treffer korrekt -- so wäre etwa bei einer korrekten Lesart 7 in Abbildung \ref{grained} die Wahl von 7.1 oder 7.2 nicht korrekt. Bei grober Körnung dagegen wären sowohl die Wahl von 7, von 7.1 als auch von 7.2 korrekt, d.h. Hyponyme, Hyperonyme und Kohyponyme der korrekten Lesart werden auch als korrekt gewertet. Bei gemischter Körnung schließlich zählen Kohyponyme der korrekten Lesart nicht als korrekte Ergebnisse, und für Hyperonyme wird die Wertung abhängig von der Anzahl ihrer Hyponyme proportional verringert. So würde z.B., wenn Lesart 7.1 in Abbildung \ref{grained}, S. \pageref{grained} korrekt wäre, die Wahl von 7.1 mit 1, die von 7 mit 0.5 und die von 7.2 mit 0 bewertet. 

Die drei Granularitäten können als Konfigurationsparameter bei der Modellierung von Experimenten dienen, vergleichbar mit der dargestellten Konfiguration von Merkmalen und Klassifikationsverfahren (s. Abschnitt \ref{appl-senseval-results}, S. \pageref{appl-senseval-results}, vgl. \citealt[16]{AgirreAndEdmonds2006b}; \citealt{MihalceaEtAl2004}).
  
\subsubsection{Agentenmodellierung} \label{agentenmodellierung}

Ausgangspunkt der Modellierung ist das Korpus, dessen getrennte Aspekte (Kontexte und korrekte Lesarten) durch getrennte Agenten implementiert\footnote{Für eine konsistente Darstellung werden hier nur Scala-Signaturen verwendet, tatsächlich wurden die Agenten teils in Scala und Teils in Java implementiert. Die Verwendung von Java- und Scala-Agenten in einem Experiment zeigt, wie eng in TM2 Komponenten in Scala und Java kombiniert werden können.} werden können. Zum Einen liefern die Korpora die Kontext der Zielwörter:

\begin{lstlisting}
class SensevalData(s:String) extends Agent[String, Context]
\end{lstlisting}

Zum Anderen liefern die Korpora Zugriff auf die Zielwörter der Kontexte, die beim Training disambiguiert werden und bei der Klassifikation ambig sind:

\begin{lstlisting}
class SensevalSense(s:String) extends Agent[Context, Ambiguity]
\end{lstlisting}

Für Training und Klassifikation sollen die Kontexte numerisch repräsentiert werden:

\begin{lstlisting}
class ContextFeatures extends Agent[Context, FeatureVector]
\end{lstlisting}

Der Klassifikator schließlich kann als Agent Lesarten für Merkmale ermitteln, auf Basis eines Modells, das Merkmale auf gelernte Klassen bezieht:

\begin{lstlisting}
class SensevalClassifier 
  extends Agent[FeatureVector, Sense] with Model[FeatureVector, Ambiguity]
\end{lstlisting}

\subsubsection{Native Evaluierung mit dem Senseval-Scorer}

Die Senseval-Workshops verwenden ein eigenes Programm, das die Ergebnisse der verschiedenen WSD-Verfahren auswertet und dabei die reichen semantischen Metadaten zur Struktur der Lesarten des Senseval-Korpus (s.o.) ausnutzt, etwa indem automatisch Ergebnisse für die verschiedenen Bedeutungsgranularitäten berechnet werden. 

Im Sinne nachvollziehbarer und vergleichbarer Ergebnisse ist es in einem solchen Fall sinnvoll, die Ergebnisse tatsächlich von diesem Programm berechnen zu lassen, statt auch hier den generischen Vergleichsmechanismus aus der Pseudowort-Evaluierung zu verwenden (etwa indem der Senseval-Goldstandard in TM2-Annotationen transformiert würde). Die Implementierung eines solchen, quasi nativen Evaluationsagenten erfolgt hier durch den internen Aufruf des Senseval-Bewertungsprogramms, und einer anschließenden Verarbeitung der Programmausgabe mit regulären Ausdrücken (zu Details siehe Hinweise in Anhang \ref{anhang-archiv}, S. \pageref{anhang-archiv}).

\subsubsection{Modellierung der Experimente} \label{appl-wsd-model}

Die Modellierung der Senseval-WSD in TM2 auf Basis der oben dargestellten Agentenmodellierung findet sich in Listing \ref{code-wsd-senseval}, S. \pageref{code-wsd-senseval}\footnote{Im Gegensatz zum Experiment mit Pseudoambiguität weiter oben werden hier Scalas \emph{singleton objects} verwendet, wodurch für Agenten, von denen in allen Experimenten nur eine Instanz benötigt wird, eine Erzeugung mit \emph{new} unnötig wird.}. In Listing \ref{code-wsd-senseval} werden die Typ-Angaben (z.B. \lstinline!Agent[FeatureVector, Sense]!) weggelassen. Der Scala-Compiler überprüft dennoch die Typisierung mithilfe von Typ-Inferenz. Um die Typisierung explizit zu machen, können die Typen wie generell in Scala-Code optional hinzugefügt werden (vgl. Listing \ref{code-wsd-senseval-annotated}, S. \pageref{code-wsd-senseval-annotated} für eine vollständig explizit typisierte Version der Experimentserie).

\begin{figure}
\begin{center}
  \includegraphics[width=8.5cm]{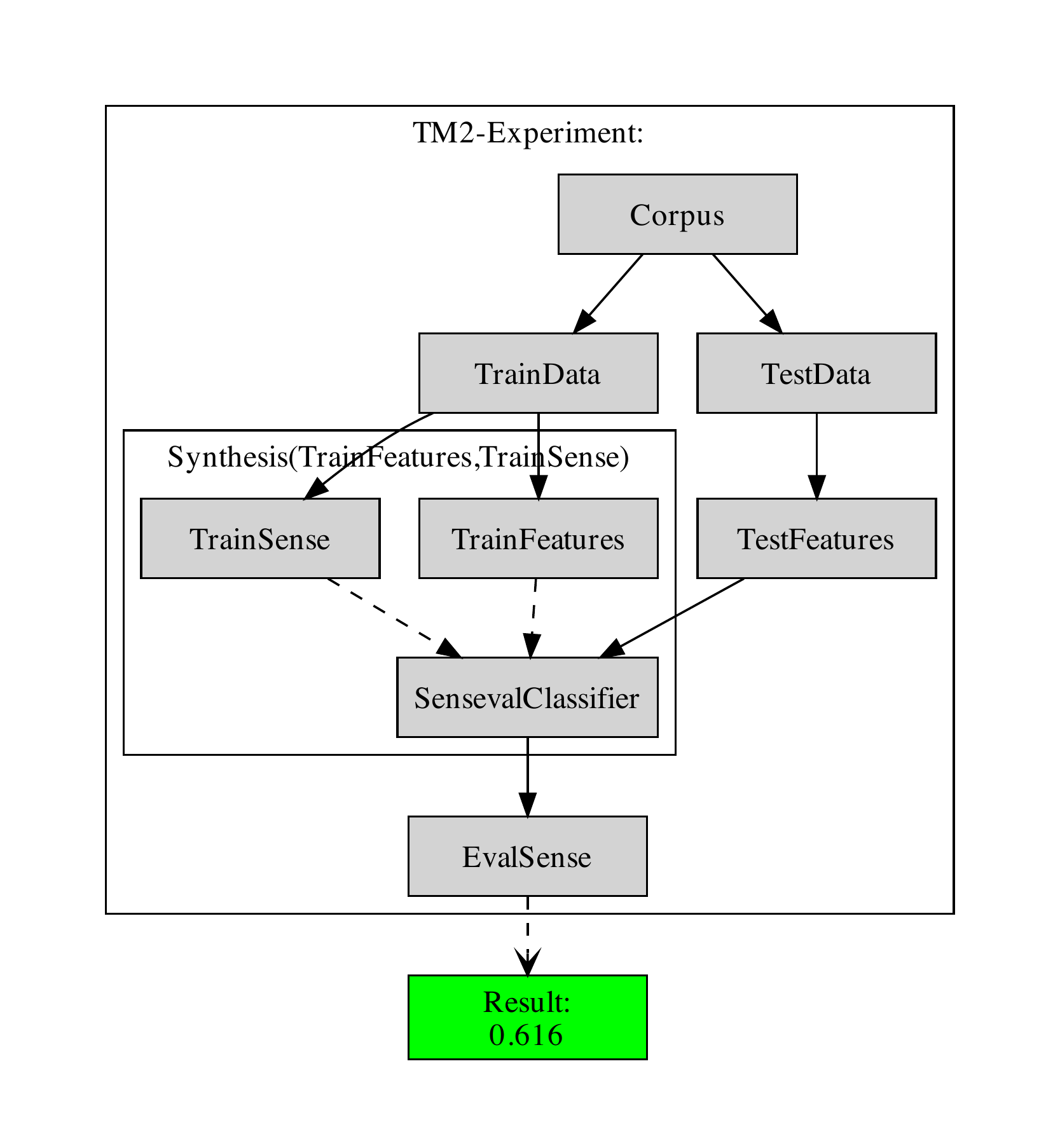}
  \caption{Generiertes Diagramm für WSD-Experiment in Listing \ref{code-wsd-senseval}, S. \pageref{code-wsd-senseval}}
  \label{anwendung-wsd-senseval}
\end{center}
\end{figure}

Die Verwendung der optionalen \lstinline!run!-Methode führt zu einer automatischen parallelen Ausführung der Experimente und einer Aufbereitung der unterschiedlichen Ergebnisse (hier für 4*4*3=48 Experimente). Die generierte Dokumentation der einzelnen Experimente enthält eine "Ubersicht über den Aufbau des Experiments, die in Abbildung \ref{anwendung-wsd-senseval}, S. \pageref{anwendung-wsd-senseval} wiedergegeben ist (vgl. Implementierung der WSD-Experimente mit Pseudoambiguität in Abschnitt \ref{appl-pseudoeval}, S. \pageref{appl-pseudoeval}).

\begin{lstlisting}[float, label=code-wsd-senseval, caption={WSD-Experiment mit Senseval-Daten, vgl. Abb. \ref{anwendung-wsd-senseval}, S. \pageref{anwendung-wsd-senseval}}]
object trainData extends SensevalData("EnglishLS.train.xml")
object testData extends SensevalData("EnglishLS.test.xml")
object trainSense extends SensevalSense("EnglishLS.train.xml")
object corpus extends Corpus
class TrainFeatures extends Features; class TestFeatures extends Features

run {
  /* Variable Konfiguration in den unterschiedlichen Durchlaeufen: */
  for {
    algo <- List(new NaiveBayes, new BayesNet, new SMO, new HyperPipes)
    feat <- List("3-gram", "7-gram", "word", "length")
    eval <- List("fine", "mixed", "coarse")
    classifier = new SensevalClassifier(algo)
    trainFeat = new TrainFeatures(feat)
    testFeat = new TestFeatures(feat)
    evaluation = new SensevalEval(eval)
  }
  /* Festes Zusammenspiel der unterschiedlich konfigurierten Agenten: */
  yield {
    /* Vorverarbeitung: */
    corpus -> (trainData, trainSense) | 
    /* Training: */
    trainData -> (trainFeat, trainSense) | 
    (trainFeat, trainSense) -> classifier |
    /* Klassifikation: */
    testData -> testFeat |
    testFeat -> classifier |
    /* Evaluierung: */
    classifier -> evaluation
  }
}
\end{lstlisting}

\begin{lstlisting}[float, label=code-wsd-senseval-annotated, caption={Explizit typisiertes WSD-Experiment, vgl. Listing \ref{code-wsd-senseval}, S. \pageref{code-wsd-senseval}}]
class TrainData extends SensevalData("EnglishLS.train.xml")
class TestData extends SensevalData("EnglishLS.test.xml")
class TrainSense extends SensevalSense("EnglishLS.train.xml")
class TrainFeatures(s: String) extends ContextFeatures(s)
class TestFeatures(s: String) extends ContextFeatures(s)

run {
  for {
    /* Konfigurationen: */
    algo: weka.classifiers.Classifier <- List(new NaiveBayes, new BayesNet, new SMO)
    feat: String <- List("3-gram", "7-gram", "word", "length")
    grain: String <- List("fine", "mixed", "coarse")
    /* Agenten: */
    corpus: Agent[String, String] = new Corpus
    trainData: Agent[String, Context] = new TrainData
    testData: Agent[String, Context] = new TestData
    trainSense: Agent[Context, Ambiguity] = new TrainSense
    trainFeat: Agent[Context, FeatureVector] = new TrainFeatures(feat)
    testFeat: Agent[Context, FeatureVector] = new TestFeatures(feat)
    classifier = new SensevalClassifier(algo)
    classifierAgent: Agent[FeatureVector, Sense] = classifier
    model: Model[FeatureVector, Ambiguity] = classifier
    evaluation: Agent[Sense, String] = new SensevalEval(grain)
    /* Interaktionen: */
    corpusData: Analysis[String] = corpus -> (trainData, testData)
    corpusContext: Analysis[Context] = trainData -> (trainFeat, trainSense)
    trainClass: Synthesis[FeatureVector, Ambiguity] = (trainFeat, trainSense) -> model
    testContext: Analysis[Context] = testData -> testFeat
    classify: Analysis[FeatureVector] = testFeat -> classifierAgent
    evaluate: Analysis[Sense] = classifierAgent -> evaluation
  } /* Ablauf: */ 
    yield corpusData | corpusContext | trainClass | testContext | classify | evaluate
}
\end{lstlisting}

\subsubsection{Exemplarische Fragestellungen und Ergebnisse} \label{appl-senseval-results}

\begin{figure}
\begin{center}
  \includegraphics[width=12cm]{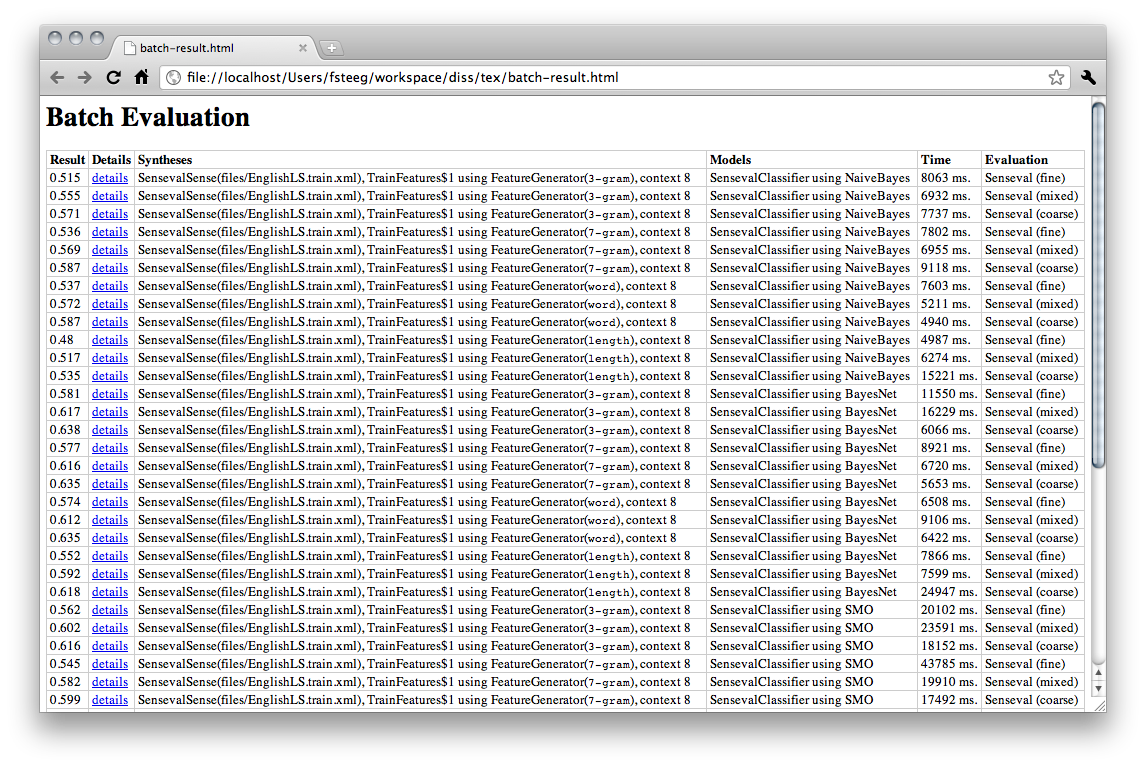}
  \caption{Generierte Übersicht der Ergebnisse für Listing \ref{code-wsd-senseval}, S. \pageref{code-wsd-senseval}}
  \label{generated-table}
\end{center}
\end{figure}

Die Ergebnisse der so modellierten und unterschiedlich konfigurierten Experimente werden automatisch tabellarisch in Form einer HTML-Datei aufbereitet (s. Abb. \ref{generated-table}, vgl. konvertierte\footnote{Die generierte HTML-Dokumentation wurde mithilfe des Programms \emph{html2latex} (\url{http://htmltolatex.sourceforge.net/}) für die Darstellung in dieser Arbeit konvertiert.} Darstellung in Anhang \ref{anhang-erg-wsd}).

\section{Zusammenfassung, Einordnung und Ausblick} \label{chapter-ausblick}

\subsection{Typsichere Modellierung im Text-Mining} \label{tm2-fazit} \index{Kognition!Modellierung durch maschinelles Lernen} \index{Modellierung!typsicher} \index{Typisierung!bei der Modellierung}

In der vorliegenden Arbeit wurden, basierend auf der Zielsetzung einer Vereinfachung der Entwicklung von Software für Text-Mining, Werkzeuge zur typsicheren Modellierung von Experimenten im Text-Mining entwickelt (Kapitel \ref{chapter-tools}, S. \pageref{chapter-tools}). Zur Evaluierung der Werkzeuge durch eine exemplarische Anwendung wurden in Kapitel \ref{chapter-appl}, S. \pageref{chapter-appl} basierend auf einer Vorstellung von Wortsinn als Kontextabstraktion Text-Mining-Experimente umgesetzt. Diese basieren auf Konzepten des korpusbasierten maschinellen Lernens (s. \ref{chapter-wsd-ml}, S. \pageref{chapter-wsd-ml}) und verwenden Daten des BNC zur Evaluierung.

Die Umsetzbarkeit dieser Konzepte mit dem entwickelten Framework zeigt, dass die formale, typsichere Repräsentation von Annotationen und Interaktionen im TM2-Framework zur Modellierung von Text-Mining-Experimenten verwendet werden kann und so ein nützliches Werkzeug für die Entwicklung von Verfahren für besseres Text-Mining ist.

Wie in Kapitel \ref{chapter-appl} dargestellt, lassen sich mit TM2 vergleichsweise einfach modulare sprachverarbeitende Komponenten entwickeln, die in komplexen Experimenten miteinander kombiniert und gegeneinander evaluiert werden können. Diese Modularität, in Verbindung mit der automatisch generierten Dokumentation, maximiert die Wiederverwertbarkeit solcher Komponenten und erleichtert die Integration in andere Umgebungen. So ist die Rolle von TM2 die eines Frameworks zur experimentellen Entwicklung von Komponenten, die evaluiert und optimiert werden können, und dann als wohlgekapselte Elemente in andere Umgebungen integriert werden können. Diese Umgebungen können Aspekte umsetzen, die von TM2 selbst nicht abgedeckt werden, wie verteilte Datenpersistenz oder graphische Modellierungswerkzeuge. Zugleich ist denkbar, dass graphische Werkzeuge oder Persistenzframeworks intern TM2 als Modellierungsformalismus verwenden.

\subsubsection{Generische Annotation für zirkuläre Informationsanreicherung}

\begin{figure}
\begin{center}
  \includegraphics[width=6cm]{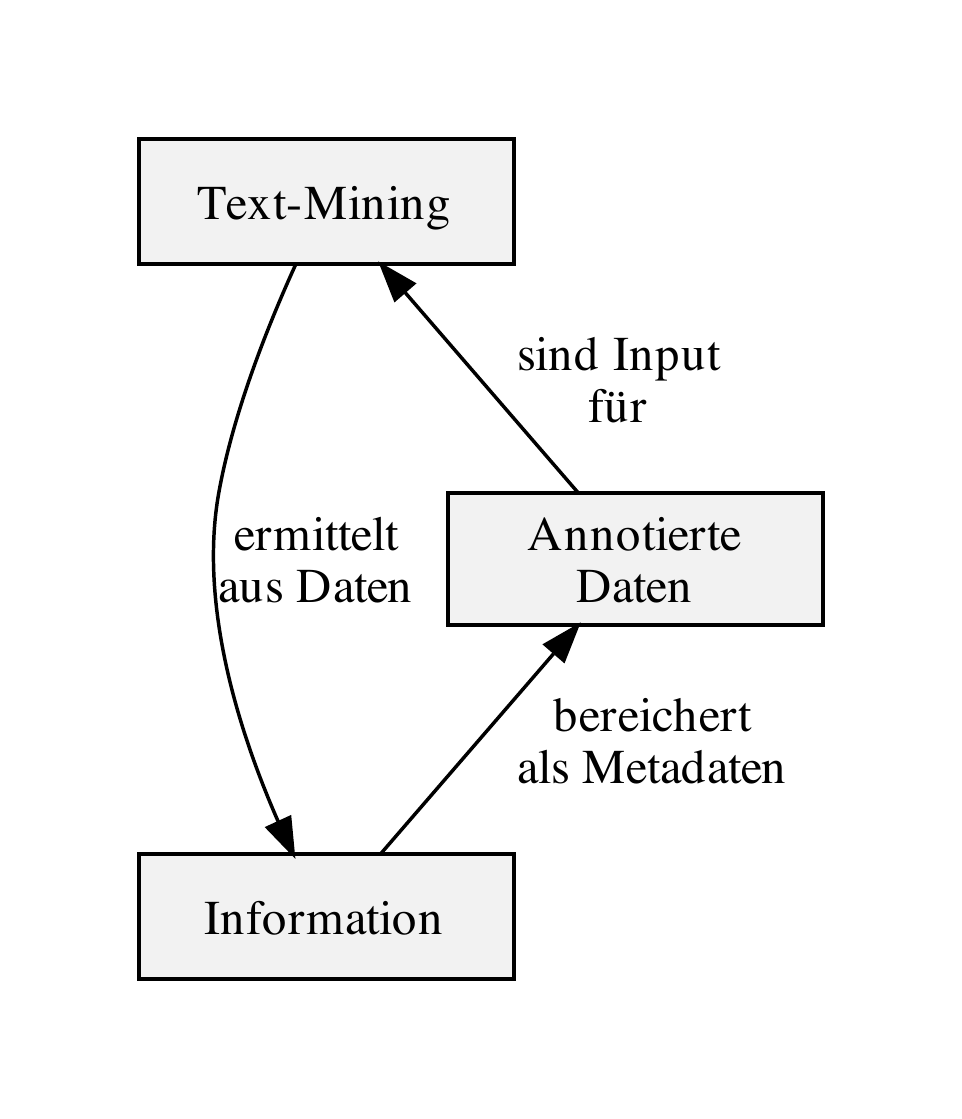}
  \caption{Zirkuläre Natur des Text-Mining} \label{zirkulaer-fazit}
\end{center}
\end{figure}

Ein Grundthema, das sich durch diese Arbeit hindurchzieht, ist die zirkuläre Natur des Text-Mining: Informationen reichern als Metadaten Daten an, und können selbst wieder zur Gewinnung neuer Informationen genutzt werden (s. Abb. \ref{zirkulaer-fazit}). Dieser Prozess, der ein generisches Prinzip der Informationsverarbeitung darstellt, kann mit den beschriebenen Konzepten und Werkzeugen generisch modelliert werden. Dies bietet Anknüpfungspunkte auf verschiedenen Ebenen, etwa unterschiedlichen Anwendungsgebieten (computerlinguistischen und allgemein-informationstechnischen, vgl. Abschnitte \ref{xcl}, S. \pageref{xcl} und \ref{fazit-infomodell}, S. \pageref{fazit-infomodell}), oder deren Teilbereiche (z.B. Vorverarbeitung, Merkmalsberechnung, Klassifikation, etc.). Diese Grundgedanken werden in den folgenden Abschnitten ausgeführt.

\subsubsection{Textuelle Modellierung als adäquater Formalismus} \index{DSL!eingebettet}

\paragraph{Textuelle und graphische Modellierung}

Klassische SALEs enthalten meist graphische Elemente, die eine Modellierung der Experimente vereinfachen sollen. Die grundsätzlichen Ziele einer SALE, wie die Entwicklung von sprachverarbeitetenden Softwarekomponenten und die Modellierung, Durchführung und Evaluierung von Experimenten erfordern jedoch konzeptuell keine GUI. Im Gegenteil kann die GUI die Möglichkeiten beschränken, da kein direkter Zugriff auf flexible formale Modellierungsmechanismen gegeben ist, sondern nur Funktionen für den Nutzer zugänglich sind, die explizit über die GUI verfügbar gemacht werden (z.B. die Durchführung mehrerer, ähnlicher Experimente, etc.). In der Regel erzeugt die GUI auf Basis der zur Verfügung gestellten Möglichkeiten intern die formale Repräsentation, meist in Form einer XML-Datei, vgl. etwa eigenen Vorarbeiten bei der Modellierung der WSD in Tesla \citep[V]{Steeg2007}.

Die textuelle Modellierung von Text-Mining-Experimenten mit einer DSL ermöglicht in TM2 die Entwicklung einfacher und komplexer Experimente und erzeugt aus der textuellen Form graphische Repräsentationen für die Dokumentation, statt diese graphischen Repräsentationen als Modellierungsschicht zu verwenden. Die DSL kann zugleich durch ihre Typsicherheit wie eine spezialisierte GUI Fehler schon beim Modellieren abfangen, jedoch auf Grundlage eines generischen, erweiterbaren Mechanismus: der typsicheren Annotation und Interaktion. Die in Scala eingebettete DSL ist zudem extrem flexibel, da alle Möglichkeiten der Basissprache zur Verfügung stehen. Als flexible Lösung für ein komplexes Problem hat die textuelle Modellierung hier Vorteile gegenüber einer graphischen Modellierung (vgl. \citealt{GreenAndPetre1992}). Beim beschriebenen DSL-Ansatz kann zudem die textuelle Modellierung die Basis für eine graphische Ebene bilden (als API bei der eingebetteten Scala-DSL, vgl. Abschnitt \ref{werk-dsl-intern}, S. \pageref{werk-dsl-intern}) oder dasselbe Schema wie diese verwenden (als EMF-Metamodell bei der eigenständigen Xtext-DSL, vgl. Abschnitt \ref{werk-dsl-extern}, S. \pageref{werk-dsl-extern}). So muss eine textuelle Modellierung die graphische Modellierung nicht ausschließen, sondern sollte je nach Implementierung als eine zugrunde liegende oder komplementäre Ebene betrachtet werden.

\paragraph{Typsichere textuelle Modellierung}

Die Implementierung einer formalen Text-Mining-No\-ta\-tion als eingebettete DSL in einer statisch typisierten Basissprache (vgl. Abschnitt \ref{werk-dsl}, S. \pageref{werk-dsl}) ermöglicht Typsicherheit und die Verwendung von weitergehenden Sprachfeatures der Basissprache. Diese Verbindung macht eingebettete, typsichere DSLs in Scala zu einer sehr interessanten Möglichkeit zur Implementierung von DSLs. Eine Scala-DSL kann zudem auf bestehende Java-APIs aufsetzen und diese mittels \emph{implicits} in veränderter Form zugänglich machen, ohne dass die Java-APIs selbst angepasst werden müssen. Eine solche Lösung ist damit nicht nur theoretisch oder im experimentellen Rahmen interessant, sondern eine realistische, praxis\-taugliche Perspektive für eine Vielzahl von Java-basierten Softwaresystemen.

Die Verwendung von generischen Datentypen in Agenten, Analysen, Synthesen und Modellen (vgl. Abschnitt \ref{werk-api}, S. \pageref{werk-api}) ermöglicht die Entwicklung neuer Abstraktionen, die von den Agenten in den Experimenten verwendet werden\footnote{Dieser Ansatz entspricht so dem generellen Ziel von Scala (\emph{A Scalable Language}, vgl. \citealt[3]{OderskyEtAl2008})}. So kann prinzipiell auf allen konzeptuellen Ebenen von \lstinline!Strings!, über \lstinline!Tokens!, zu \lstinline!Words! oder \lstinline!Senses! modelliert werden. Ein Tokenizer könnte etwa als \lstinline!Agent[String, Token]! modelliert werden, ein Indexer als \lstinline!Agent[Token, Type]!, oder ein POS-Tagger als \lstinline!Agent[Token, POS]!. Die Datenabstraktion ist dabei ein Implementierungsdetail, das in konkreten Agenten ganz unterschiedlich typsiert werden kann, etwa um spezielle Anforderungen oder neue Ideen umzusetzen. Dies macht die Verwendung einer typsicheren DSL zu einem formal strikten, aber zugleich semantisch reichem und flexiblem Werkzeug für die Modellierung von Text-Mining-Experimenten.

\subsection{Ein generelles Modell der Informationsverarbeitung} \label{fazit-infomodell} \index{Information!allg. Modell} \index{Annotation!als allg. Prinzip}

\subsubsection{Konzeptuelle Parallelen zu Binärdaten} \label{xcl}

Die konzeptuelle Einfachheit des Annotationsmodells und die Verwendung von XML als Exportformat für die produzierten Annotationen erlaubt eine vielfältige weitere Nutzung der TM2-Ergebnisse. Als Beispiel soll dazu im Folgenden dargestellt werden, wie in TM2-Experimenten gewonnene Daten in das im Planets-Projekt entwickelte XCDL-Format exportiert und so einer Evaluation mit dem XCL-Comparator zugänglich gemacht werden können.

\paragraph{Datenmigration, Annotation und Evaluierung}

Im Rahmen des EU-geförderten Planets-Pro\-jekts\footnote{PLANETS: Preservation and Long-term Access through NETworked Services, \url{http://www.openplanetsfoundation.org/}} wurden unter anderem Werkzeuge zum Vergleich von Binärdaten entwickelt (XCL: \emph{extensible characterisation languages}, s. \citealt{Thaller2009}).

Binärdaten stellen konzeptuell wie textuelle Annotationen eine inhaltliche Auszeichnung von Daten dar. Ein Datenformat ist dabei die Definition von Bereichen in den Daten, die  unterschiedlichen Zwecken dienen, gewissermaßen ein Tagset oder Schema für die Annotationen. So kann eine konventionelle, nicht-formale Formatspezifikation (z.B. die PNG-Spezifikation\footnote{PNG specification, \url{http://www.w3.org/TR/PNG/}}) in einer formalen XML-Repräsentation wie der XCEL (vgl. \citealt{SchnasseEtAl2009}) beschrieben werden. Instanzen dieses Schemas oder Formats (z.B. konkrete PNG-Dateien) stellen die annotierten Daten dar, und können ebenfalls in einer rein textuellen XML-Repräsentation dargestellt werden (der XCDL, vgl. \citealt{BeylEtAl2009}). Die Erzeugung dieser textuellen Instanz-Repräsentationen von konkreten Dateien (in der XCDL) kann mithilfe des formalen Schemas (der XCEL) automatisiert werden \citep{HeydeggerEtAl2009}. Dateien, für die eine solche XCDL-Repräsentation extrahiert wurde, können mithilfe des XCL-Comparators automatisch und unter Verwendung verschiedenener Metriken miteinander verglichen werden (s. \citealt[174]{HeydeggerEtAl2009}). So kann etwa die Güte einer Datenmigration (z.B. von JPEG zu PNG) evaluiert werden, indem die Ausgangs-XCDL (aus der JPEG-Datei extrahiert) mit der Ergebnis-XCDL (aus der PNG-Datei extrahiert) verglichen wird (vgl. \citealt{SchnasseEtAl2009}).

Aufgrund der konzeptuellen Nähe von TM2 und XCL lassen sich die von den TM2-Werkzeugen erzeugten Daten grundsätzlich auch mit den XCL-Werkzeugen nutzen, die zum Vergleich von Binärdateien entwickelt wurden. Dazu müssen die TM2-Exportdaten in das benötigte XCDL-Format transformiert werden.

\paragraph{XCDL-Transformation}

\begin{lstlisting}[float, label=code-xslt, caption={Haupt-Template einer XSL-Transformation der TM2-Exportdaten in die XCDL}]
<xsl:template match="experiment">
    <xcdl id="0">
        <object id="o1">
            <normData type="text" id="nd1"><xsl:value-of select="@data" /></normData>
            <property id="p1" source="raw" cat="descr">
                <name id="id58">textualAnnotation</name>
                <xsl:apply-templates mode="valueSets" />
            </property>
            <xsl:apply-templates mode="propertySets" />
        </object>
    </xcdl>
</xsl:template>
\end{lstlisting}

Da Eingabe- und Ausgabeformat der benötigten Transformation XML-For\-ma\-te sind, bietet sich für eine Transformation der TM2-Annotationen in die XCDL die XML-basierte Transformationssprache XSLT an. Für die Umwandlung reichen im Wesentlichen drei XSLT-Templates: Das Haupt-Template transformiert die Ergebnisse eines TM2-Ex\-pe\-ri\-ments in eine XCDL mit \emph{properties} vom Typ \emph{textualAnnotation} (s. Listing \ref{code-xslt}, S. \pageref{code-xslt}\footnote{Für diese Experimente enthalten die TM2-Exportdaten keine Referenz auf die annotierten Daten in Form einer URL im \emph{data}-Attribut, sondern die Daten selbst.}). Ein zweites Template transformiert die \emph{label}-Attribute von TM2-Annotationen in \emph{value sets} der Ziel-XCDL, ein drittes Template die Positionsinformationen der TM2-Annotationen in \emph{property sets} der Ziel-XCDL. Die vollständige Transformation der TM2-Annotationen mithilfe von XSLT findet sich in Anhang \ref{anhang-xcl}, S. \pageref{anhang-xcl}.

\paragraph{XCDL-Vergleich}

Die so generierten XCDL-Dateien (etwa für das Ergebnis einer Informationsextraktion mit TM2 und für einen Goldstandard) können dann mithilfe des XCL-Comparators miteinander verglichen werden. Dabei müssen Metriken des XCL-Comparators gewählt werden, die mehrere \emph{value sets} unterstützen (die textuellen Annotationen werden hier als ein \emph{property} mit unterschiedlichen \emph{value sets} betrachtet), z.B. wie in Listing \ref{code-xcl} (zu Details der Konfiguration des Vergleichs s. \citealt[197]{HeydeggerEtAl2009}).

\begin{lstlisting}[float, label=code-xcl, caption={Beispielkonfiguration für XCL-Vergleich der exportierten Daten}]
<coco>
    <compSet>
        <property id="58" name="textualAnnotation">
            <metric id="101" name="valueSetStat_1" />
            <metric id="102" name="valueSetStat_2" />
            <metric id="121" name="valueSetMatch_1" />
            <metric id="122" name="valueSetMatch_2" />
            <metric id="253" name="dataRefMatch_3" />
        </property>
    </compSet>
</coco>
\end{lstlisting}

\paragraph{Integration}

Im Rahmen des Planets-Projekts wurde zur Integration der XCL-Werkzeuge in das \emph{Planets Interoperability Framework} \citep{RossEtAl2009} auch eine Java-API für die XCL-Werkzeuge entwickelt \citep{SteegEtAl2009}. Auf Basis dieser Java-API kann das Ergebnis eines TM2-Experiments mithilfe der XCL-Werkzeuge komplett über Java-APIs verarbeitet werden. Zu Details der prototypischen Implementierung s. Hinweise in Anhang \ref{anhang-code}, S. \pageref{anhang-code}.

\paragraph{Generizität des Annotationsmodells}

Diese Nutzung der TM2-Exportdaten zeigt nicht nur wie vielseitig die Ergebnisse im XML-Format verwendet werden können, sondern illustriert zudem wie universell das Grundkonzept der Annotation von Daten ist: es ist das grundsätzliche Mittel, mit dem Daten angereichert werden, seien sie ursprünglich binär (etwa beim Vergleich von Bilddaten mit den XCL-Werkzeugen), oder von Anfang an textuell, wie im Bereich des Text-Mining. Durch die Generizität des Annotationkonzepts können so Werkzeuge und Verfahren für den Vergleich von Binärdaten wie Bild- Ton- oder Videomaterial prinzipiell auch für den Vergleich von linguistischem Wissen etwa zu Syntax, Semantik, etc. verwendet werden. Kapitel \ref{chapter-ausblick} diskutiert im Anschluss an die experimentelle Anwendung von TM2 in Kapitel \ref{chapter-appl} diese Gemeinsamkeit so unterschiedlicher Daten und deren Verarbeitung, vgl. speziell die abschließenden Darstellungen in Abschnitt \ref{fazit-infomodell}, S. \pageref{fazit-infomodell}.

\subsubsection{Das Kölner Informationsmodell als konzeptuelle Grundlage}

In Abschnitt \ref{xcl}, S. \pageref{xcl} wurden Ergebnisse der mit TM2 formulierten Experimente mit Werkzeugen verwendet, die im Rahmen des Planets-Projekts zur digitalen Langzeitarchvierung entwickelt wurden. Dieses praktische Ergebnis deutet auf eine gemeinsame konzeptuelle Basis der scheinbar recht verschiedenen Bereiche der maschinellen Sprachverarbeitung und der Migration und Archivierung von Binärdaten. In der Tat lassen sich weitgehende Parallelen feststellen zwischen den beschriebenen Konzepten der typsicheren Annotation und dem zugrunde liegenden Modell der Planets-Werkzeuge zur Evaluierung von Datenmigrationen (XCL, vgl. \citealt{Thaller2009}).

\paragraph{Eine Formalisierung der rekursiven Natur der Informationsverarbeitung}

Die Grundidee sowohl des beschriebenen Ansatzes wie auch der XCL ist, dass Metadaten auch nur Daten sind \citep[223]{Thaller2009b}. Wie aus Daten und Metadaten dabei Informationen werden, lässt sich auf verschiedenen Ebenen beschreiben: z.B. Bits als Daten (\emph{01100001}), assoziierte Zahlen- ('97') oder Zeichenwerte ('a') als Information, oder Zahlen als Daten (\emph{97}) und ihre Einheit als Information ('97 Grad Celsius'), etc. Diese Bespiele aus \citet{Thaller2009} zeigen, dass Daten und Information keine getrennten Kategorien sind, sondern Idealtypen an entgegengesetzten Enden eines Kontinuums, die durch eine Anreicherung schrittweise ineinander übergehen \citep[226]{Thaller2009b}.

Dies entspricht der oben dargestellten zirkulären Natur des Text-Mining (Abb. \ref{zirkulaer-fazit}, S. \pageref{zirkulaer-fazit}). So entspricht etwa die Interpretation von Bits als Datentypen (Zahlen oder Zeichen, konkret in Programmiersprachen etwa 'int' oder 'char') einer typsicheren Annotation (Datenbereich $x$, z.B. \emph{01100001} ist Information vom Typ $y$, z.B. 'int'). Diese schrittweise Bildung von Information aus Daten wird in \citet[224]{Thaller2009b} aufbauend auf \citet{Langefors1995} folgendermaßen formalisiert: Information (I) ist das Ergebnis einer Interpretation (i) vorhergehender Information (die kontinuierlich aus Daten gebildet wird), unter Anwendung eines Vorwissen (S), innerhalb der verfügbaren Zeit (t):

\begin{equation} 
I_x = i (I_{x-\alpha}, S(I_{x-\beta}, t), t) \label{infomod-1}
\end{equation}

Die Implementierung von TM2 entspricht diesem Konzept nicht nur prinzipiell durch die generelle Natur von Metadaten-Annotationen, sondern auch in der formalen Modellierung von Agenteninteraktionen. So ist das Grundkonzept der TM2-API, dass Annotationen eines bestimmten Typs in Annotationen eines anderen Typs verarbeitet werden und so schrittweise Information gesammelt oder angereichert wird (das Vorwissen S ist dabei im TM2-Kontext in der Signatur nicht explizit erwähnt, da es als Blackboard hinter der API gekapselt ist, vgl. Abschnitte \ref{anno-blackboard}, S. \pageref{anno-blackboard} und \ref{impl-blackboard}, S. \pageref{impl-blackboard}):

\begin{lstlisting}
process(input: List[Annotation[I]]): List[Annotation[O]]
\end{lstlisting}

Besonders deutlich wird die konzeptuelle Ähnlichkeit zu Formel \ref{infomod-1} im Zusammenhang mit Synthesen, die aus Daten und Informationen ein Modell bilden (vgl. Abschnitt \ref{agent-interaction}, S. \pageref{agent-interaction}), und so wie in Formel \ref{infomod-1} aus vorhandenen Informationen neue Informationen synthetisieren. Die Unterscheidung von Daten und Informationen ist hier rein metaphorisch (denn sowohl Daten als auch Informationen werden als Annotationen repräsentiert) und entspricht so dem in Formel \ref{infomod-1} ausgedrückten Kontinuum zwischen Daten und Informationen:

\begin{lstlisting}
build(data: List[Annotation[D]], info: List[Annotation[I]]): Model[D, I]
\end{lstlisting}

\paragraph{Annotationen zur Verortung von Daten in einem Merkmalsraum}

Das Konzept der Annotation kann als Verortung von Daten in einem $n$-dimensionalen Merkmalsraum betrachtet werden. Die $n$ Typen der Annotationen stellen dabei die konzeptuellen Achsen des Raums dar, in dem die annotierten Daten beschrieben werden (z.B. Wortart, Bedeutung, oder Typographie). So kann etwa die Wortform \emph{Biggin} in einem zweidimensionalen Merkmalsraum aus Interpretation (mit den Werten 'Personenname' und 'Ortsname') und Visualisierung (mit den Werten 'fett' oder 'kursiv') an vier Stellen verortet werden: als fett gedruckter Personenname, als kursiv gedruckter Personenname, als fett gedruckter Ortsname und als kursiv gedruckter Ortsname \citep[231]{Thaller2009b}. Dis\-tanz\-en zwischen unterschiedlichen, in diesem Raum gemessenen Positionen können dann mit unterschiedlichen, allgemeinen raumbezogenen Metriken gemessen werden \citep[235]{Thaller2009b}.

In dieser Terminologie sind die TM2-Annotationen auf ihre konzeptuelle Achse im Merkmalsraum typisiert, und Agenten erweitern die Dimensionalität dieses Raumes, z.B.:

\begin{lstlisting}
tokenizer: Agent[String, Token]
\end{lstlisting}

Ein solcher Agent etwa kann aus einer bestehenden Verortung von Daten in einem 1-di\-men\-sion\-alen Raum mit der Achse 'String' eine zusätzliche Verortung auf der konzeptuellen Achse 'Token' vornehmen. So können Agenten schrittweise, entsprechend dem beschriebenen Kontinuum zwischen idealtypischen Daten und idealtypischer Information, Daten zu Information verarbeiten, z.B. in einem nächsten Schritt, aufbauend auf die Token:

\begin{lstlisting}
gazetteer: Agent[Token, Sense]
\end{lstlisting}

Basierend auf dieser Darstellung wird deutlich, dass die in dieser Arbeit beschriebene softwaretechnische Lösung für typsicheres Text-Mining dem sehr allgemeinen Informationsmodell aus \citet[237]{Thaller2009b} entspricht: beliebige Informationsobjekte lassen sich als Anordnungen von $m$-dimensionalen Objekten $T$ (Token) mit ihren (durch Annotationen definierten) Positionen in einem $n$-dimensionalen Merkmalsraum $C$ (Context) beschreiben:

\begin{equation} 
<I> ::= \{ T_m, C_n \}
\end{equation}

Im Text-Mining sind die Token dabei eindimensional und die Verortung im $n$-di\-men\-si\-on\-alen Kontext ist durch die Annotationen gegeben, deren $n$ Typen die Achsen des Merkmalsraums bilden. Das beschriebene Informationsmodell bildet damit ein formales Modell, das in so unterschiedlichen Bereichen wie Text-Mining und Datenarchivierung softwaretechnisch umgesetzt werden kann, und in dessen Zentrum eine iterative Bildung von Information aus Daten, sowie die variable Dimensionalität von Datenobjekten und Merkmalsraum steht.

\subsection{Fazit}

Auf Basis der Konzeption von annotationsbasierten Agenten in Kapitel \ref{chapter-theory} wurden in Kapitel \ref{chapter-tools} Werkzeuge und eine formale Notation zur Beschreibung und Durchführung von Experimenten im Text-Mining entwickelt, in Form einer statisch typisierten, eingebetteten DSL. Am Beispiel von maschinellem Lernen zur Klassifikation wurden in Kapitel \ref{chapter-appl} die Vorteile und Möglichkeiten der Kombination von Werkzeugen und DSL dargestellt, bei der auf einfache Weise komplexe Experimentserien durchgeführt und automatisch dokumentiert werden können. In Kapitel \ref{chapter-ausblick} wurde zusammenfassend dargestellt, wie das konsequent genutzte Konzept der generischen, typsicheren Annotation dabei zur Modellierung von Prozessen der Informationsverarbeitung genutzt werden kann, und inwiefern es einem allgemeinen Informationsmodell entspricht, das über die Textprozessierung hinausgeht. Als generischer Formalismus zur Annotation bildet so die typsichere Modellierung ein zentrales Element nicht nur des Text-Mining, sondern einer allgemeinen Disziplin der Informationsverarbeitung.

\newpage
\appendix

\section{Implementierungsdetails} \label{chapter-impl}

\subsection{Projektstruktur und Komponenten} \label{anhang-code} \label{anhang-archiv} \label{anhang-build}

\begin{figure}[bhtp]
\begin{center}
  \includegraphics[width=5cm]{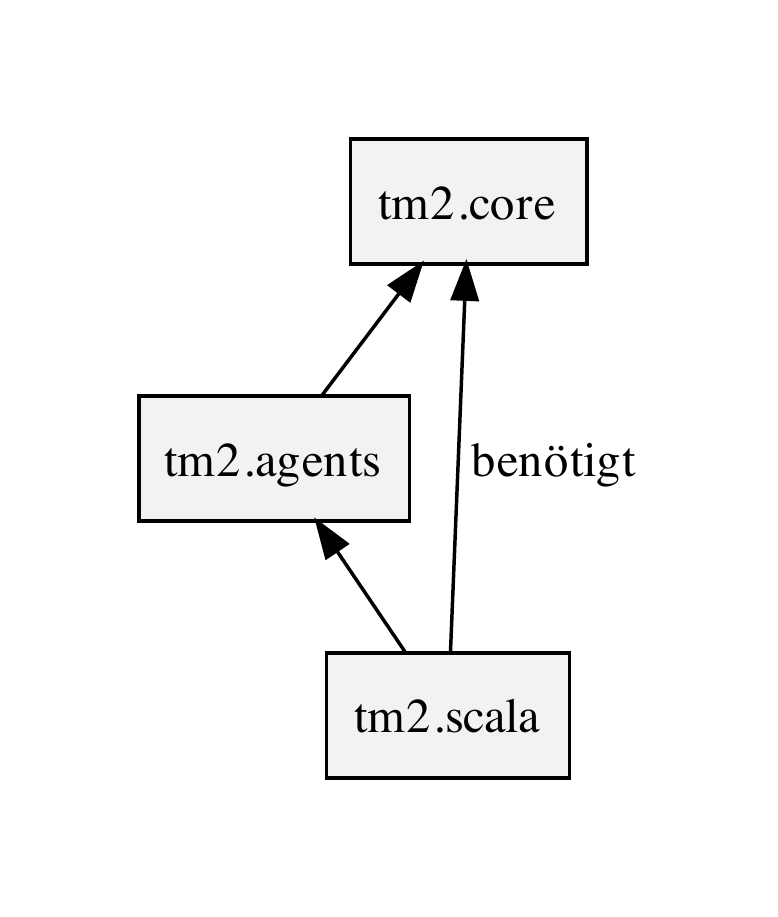}
  \caption[Projektabhängigkeiten]{Projektabhängigkeiten}
  \label{project-dependencies}
\end{center}
\end{figure}

Die Implementierung der in dieser Arbeit beschriebenen Software ist online verfügbar\footnote{TM2: \emph{typesafe modeling in text mining}, \url{http://github.com/fsteeg/tm2}}. Die Software umfasst drei Projekte: tm2.core (Framework und Java-API), tm2.agents (Implementierung von Agenten in Java) und tm2.scala (Scala-DSL und Implementierung von Agenten in Scala). Abbildung \ref{project-dependencies} beschreibt die Abhängigkeiten der drei Projekte.

Für eine Reproduktion der beschriebenen Ergebnisse ist ein plattformunabhängiger, automatisierbarer Erstellungs- und Testprozess von zentraler Bedeutung (vgl. \citealt{Clark2006}). Die beschriebene Software kann daher in einem Schritt mithilfe von Ant\footnote{Apache Ant, \url{http://ant.apache.org/}} gebaut werden. Da das tm2.scala-Projekt von den anderen abhängt (vgl. Abb. \ref{project-dependencies}, S. \pageref{project-dependencies}), enthält dieses das zentrale Ant-Skript, das durch Ausführen von \lstinline!ant! im Verzeichnis \lstinline!tm2.scala/! ausgeführt werden kann. Das zentrale Ant-Skript ruft die Build-Skripte der anderen Projekte auf, erzeugt Jar-Archive, führt die Tests aus und generiert Dokumentation zu API und Testergebnissen.

\subsection{Implementierung der Werkzeuge} \label{anhang-impl}

\subsubsection{API} \label{anhang-api}  \index{Modellierung!typsicher} \index{Typisierung!bei der Modellierung}

In den folgenden Abschnitten finden sich die zentralen Interfaces zur typsicheren Modellierung von Text-Mining-Experimenten mithilfe der Java-API von TM2.

\subsubsection{Agent}

\lstset{language=Java}
\begin{lstlisting}
/**
 * Interface for an agent. This is a Strategy Interface (cf. Bloch 2008, Item 21).
 * @param <I> The type of the input annotations, must be comparable and serializable
 * @param <O> the type of the output annotations, must be comparable and serializable
 */
public interface Agent<I extends Comparable<I> & Serializable, O extends Comparable<O> & Serializable> {
    List<Annotation<O>> process(List<Annotation<I>> input);
}
\end{lstlisting}

\subsubsection{Annotation}

\begin{lstlisting}
/**
 * The interchange format of a text mining application: an annotation of data with meta-data.
 * @param <T> The type of the annotation
 */
public interface Annotation<T> extends Comparable<Annotation<T>>, Serializable {
    Class<? extends Agent<?, ?>> author();
    URL getDataLocation();
    T getValue();
    BigInteger getStart();
    BigInteger getEnd();
}
\end{lstlisting}

\subsubsection{Analysis}

\begin{lstlisting}
/**
 * A simple linear interaction of agents exchanging typed annotations.
 * @param <T> The annotation type exchanged in this analysis
 */
public interface Analysis<T extends Comparable<T> & Serializable> {
    List<Agent<?, T>> sources();
    List<Agent<T, ?>> targets();
    Map<Class<? extends Agent<?, ?>>, List<Annotation<?>>> run(
            final Map<Class<? extends Agent<?, ?>>, List<Annotation<?>>> blackboard);
}
\end{lstlisting}

\subsubsection{Synthesis}

\lstset{language=Java}
\begin{lstlisting}
/**
 * Synthesis of data and info into a model.
 * @param <D> The data type
 * @param <I> The info type
 */
public interface Synthesis<D extends Comparable<D> & Serializable, I extends Comparable<I> & Serializable> {
    Map<Class<? extends Agent<?, ?>>, List<Annotation<?>>> run(
            Map<Class<? extends Agent<?, ?>>, List<Annotation<?>>> blackboard);
    List<Agent<?, D>> data();
    List<Agent<?, I>> info();
    Model<D, I> model();
}
\end{lstlisting}

\subsubsection{Model}

\begin{lstlisting}
/**
 * A model for mapping data of type D to info of type I.
 * @param <D> The data type
 * @param <I> The info type
 */
public interface Model<D extends Comparable<D> & Serializable, I extends Comparable<I> & Serializable> {
    Model<D, I> train(List<Annotation<D>> data, List<Annotation<I>> info);
}
\end{lstlisting}

\subsubsection{Eigenständige DSL} \label{anhang-impl-xpand}

\paragraph{Xtext-Grammatik}

Grammatik der eigenständigen Xtext-DSL (vgl. Abschnitt \ref{werk-dsl-extern}, S. \pageref{werk-dsl-extern}):

\lstset{}
\begin{lstlisting}
/* Xtext experiment grammar */
Experiment :
"Experiment" name=STRING "Data" corpus=STRING "Out" output=STRING
"Import" (imports+=STRING)+
(interactions+=Interaction)+
("Evaluate" (evalAgents+=ID)+ "Against" evalLocation=STRING)?;
Interaction : source=Source  "->" type=ID "->" target=Target "."; 
Source : (sourceAgents+=Agent)+;
Target : (targetAgents+=Agent)+;
Agent : name=ID;
\end{lstlisting}

\paragraph{Textuelles Modell}

Beispiel für ein Experiment mit der eigenständigen DSL (vgl. Abschnitt \ref{werk-dsl-extern}, S. \pageref{werk-dsl-extern}):

\lstset{language={Java}}
\begin{lstlisting}
/* Xtext experiment */
Experiment "NE" Data "data/Corpus.txt" Out "output/Result"
Import "com.quui.amas.agents" "com.quui.amas.types"
Corpus -> String -> Tokenizer. // Read: Corpus via String to Tokenizer
Tokenizer -> String -> Gazetteer Counter.
Evaluate Gazetteer Against "data/Gold_Gazetteer.xml"
\end{lstlisting}

\paragraph{Xpand-Template: Java-Generierung}

Xpand-Template, generiert Code zur Nutzung der TM2-Java-API (s. Abschnitt \ref{werk-api}, S. \pageref{werk-api}) aus der eigenständigen DSL (s. Abschnitt \ref{werk-dsl-extern}, S. \pageref{werk-dsl-extern}). Elemente zwischen \lstinline!<<! und  \lstinline!>>! sind Xpand-Kontrollstrukturen zur Steuerung der Generierung:

\lstset{language={Java}}

\begin{lstlisting}
<<REM>> The generator template for the "TM2-DSL to Java Compiler" <<ENDREM>> <<IMPORT tm2>>
<<DEFINE main FOR Experiment>> <<FILE name+".java">>
import java.util.*; import com.quui.tm2.*; import com.quui.tm2.doc.*; import com.quui.tm2.util.*;
<<FOREACH imports AS i>> import <<i>>.*; <<ENDFOREACH>>
/** Generated TM2 experiment. */
public class <<name>> implements Runnable {
    public static void main(String[] args) { new <<name>>().run(); }
    public void run(){
        /* Instantiate an experiment: */
        String output = "<<output>>";
        Experiment x = Experiments.create("<<name>>", "<<corpus>>", output);
        List<Analysis<?>> interactions = new ArrayList<Analysis<?>>();
        /* Define the interaction of agents in the experiment: */ <<FOREACH interactions AS i>> 
        /* <<i.source.sourceAgents.name>> via <<i.type>> to <<i.target.targetAgents.name>>: */
        Analysis<<<i.type>>> interaction<<interactions.indexOf(i)>> = Analysis.create(); <<FOREACH i.source.sourceAgents AS s>> 
        interaction<<interactions.indexOf(i)>> = interaction<<interactions.indexOf(i)>>.withSource(new <<s.name>>());<<ENDFOREACH>> <<FOREACH i.target.targetAgents AS t>> 
        interaction<<interactions.indexOf(i)>> = interaction<<interactions.indexOf(i)>>.withTarget(new <<t.name>>());<<ENDFOREACH>> 
        interactions.add(interaction<<interactions.indexOf(i)>>);<<ENDFOREACH>>
        /* Run the experiment with all the interactions: */
        for (Analysis<?> i : interactions) { x.addInteraction(i); }
        x.run();
        /* Evaluate against a gold standard: */
        Evaluation evaluation = new Evaluation(output + ".xml");<<FOREACH evalAgents AS e>>
        evaluation.evaluate("<<evalLocation>>", <<e>>.class);<<ENDFOREACH>>
        /* Export both annotations and documentation */
        Documentation.of(experiment);
    }
}
<<ENDFILE>>
<<ENDDEFINE>>
\end{lstlisting}

\paragraph{Generierter Code: Java}

Generierter Java-Code zur Nutzung der TM2-API, auf Basis des in der Xtext-DSL geschriebenen Experiments (s.o.) durch das Xpand-Template (s.o.), vgl. Abschnitt \ref{werk-api}, S. \pageref{werk-api}:

\lstset{language={Java}}
\begin{lstlisting}
import java.util.*; import com.quui.tm2.*; import com.quui.tm2.doc.*; import com.quui.tm2.util.*;
import com.quui.tm2.agents.*; import com.quui.tm2.types.*;
/** Generated TM2 experiment. */
public class NE implements Runnable {
    public static void main(String[] args) { new NE().run(); }
    public void run() {
        /* Instantiate an experiment: */
        String output = "output/Result";
        Experiment x = Experiments.create("NE", "data/Corpus.txt", output);
        List<Analysis<?>> interactions = new ArrayList<Analysis<?>>();
        /* Define the interaction of agents in the experiment: */
        /* [Corpus] via String to [Tokenizer]: */
        Analysis<String> interaction0 = Analyses.create();
        interaction0 = interaction0.withSource(new Corpus());
        interaction0 = interaction0.withTarget(new Tokenizer();
        interactions.add(interaction0);
        /* [Tokenizer] via String to [Gazetteer, Counter]: */
        Analysis<String> interaction1 = Analyses.create();
        interaction1 = interaction1.withSource(new Tokenizer());
        interaction1 = interaction1.withTarget(new Gazetteer());
        interaction1 = interaction1.withTarget(new Counter());
        interactions.add(interaction1);
        /* Run the experiment with all the interactions: */
        for (Analysis<?> i : interactions) { x = x.with(i); }
        x.run();
        /* Evaluate against a gold standard: */
        Evaluation evaluation = new Evaluation(output + ".xml");
        evaluation.evaluate("data/Gold_Gazetteer.xml", Gazetteer.class);
        evaluation.evaluate("data/Gold_Gazetteer.xml", Counter.class);
        /* Export both annotations and documentation */
        Documentation.of(experiment);
    }
}
\end{lstlisting}

\subsubsection{Eingebettete DSL} \label{anhang-embedded-dsl}

Die eingebettete Scala-DSL umfasst 4 Scala-Entitäten, welche die Funktionalität der Java-API anreichern: RichAgent, RichAnalysis, RichSynthesis und RichExperiment. Für diese Entitäten existiert jeweils eine Klasse und ein Singleton-Objekt (vgl. \citealt[59]{OderskyEtAl2008}).

\subsubsection{RichAgent}

\paragraph{Class}

Klasse: \emph{rich wrapper} für \lstinline!Agent!.

\lstset{language={Scala}}

\begin{lstlisting}
/**A rich agent that supports the Scala DSL syntax:*/
abstract class RichAgent[I<:Comparable[I] with Serializable,O<:Comparable[O] with Serializable] 
    extends Agent[I,O]{
  val backingAgent :  Agent[I,O]
  //...
}
\end{lstlisting}

\paragraph{Object}
\enlargethispage{0.5cm}
Singleton-Objekt: \emph{rich wrapper} für \lstinline!Agent!.

\begin{lstlisting}
object RichAgent{
  /** Implicit conversion from Java agents to rich Scala agents: */
  implicit def agentWrapper [I<:Comparable[I] with Serializable,O<:Comparable[O] with Serializable]
     (agent : Agent[I,O]) = { /*...*/ }
  //...
}    
\end{lstlisting}

\subsubsection{RichAnalysis}

\paragraph{Class}

Klasse: \emph{rich wrapper} für \lstinline!Analysis!.

\begin{lstlisting}
/**A rich analysis that supports the Scala DSL syntax:*/
abstract class RichAnalysis[T<:Comparable[T] with Serializable] 
    extends Analysis[T]{
  val backingInteraction :  Analysis[T]
  def | (next: Analysis[_]): RichExperiment = { /*...*/ }
  def | [V<:Comparable[V] with Serializable, C<:Comparable[C] with Serializable] (next: Synthesis[V,C]): RichExperiment = { /*...*/ }
}
\end{lstlisting}

\paragraph{Object}

Singleton-Objekt: \emph{rich wrapper} für \lstinline!Analysis!.

\begin{lstlisting}
object RichAnalysis{
  /** Implicit conversion from Java analyses to rich Scala analyses: */
  implicit def analysisWrapper [T<:Comparable[T] with Serializable] (interaction : Analysis[T]) = 
        new RichAnalysis[T] { /*...*/ } 
  //...
}
\end{lstlisting}

\subsubsection{RichSynthesis}

\paragraph{Class}

Klasse: \emph{rich wrapper} für \lstinline!Synthesis!.

\begin{lstlisting}
/**A rich synthesis that supports the Scala DSL syntax:*/
abstract class RichSynthesis[V<:Comparable[V] with Serializable, C<:Comparable[C] with Serializable] 
    extends Synthesis[V,C]{
  val backingTraining : Synthesis[V,C]
  def | (next: Analysis[_]): RichExperiment = { /*...*/ }
}
\end{lstlisting}

\paragraph{Object}

Singleton-Objekt: \emph{rich wrapper} für \lstinline!Synthesis!.

\begin{lstlisting}
object RichSynthesis{
  /** Implicit conversion from Java syntheses to rich Scala syntheses: */
  implicit def synthesisWrapper [V<:Comparable[V] with Serializable, C<:Comparable[C] with Serializable] 
     (training : Synthesis[V,C]) =
        new RichSynthesis[V,C] { /*...*/ } //...
}    
\end{lstlisting}

\subsubsection{RichExperiment}

\paragraph{Class}

Klasse: \emph{rich wrapper} für \lstinline!Experiment!.

\begin{lstlisting}
/**Wrapper class for Experiment.Builder, builds on-the-fly in the run and eval methods.*/
class RichExperiment(x: Experiment.Builder) {
  //...
  /** Run and export an experiment. */
  def run: RichExperiment = {
    experiment.run; this
  }
  def ! : RichExperiment = { val r = run; documentation = WikitextExport.of(experiment); r }
  def |(next: Analysis[_]): RichExperiment = { x.analysis(next); new RichExperiment(x) }
  def |[V <: Comparable[V] with Serializable, C <: Comparable[C] with Serializable](next: RichSynthesis[V, C]): RichExperiment = {
    x.synthesis(next); new RichExperiment(x)
  }
}
\end{lstlisting}

\paragraph{Object}

Singleton-Objekt: \emph{rich wrapper} für \lstinline!Experiment!:

\begin{lstlisting}
object RichExperiment {
  def apply(interactions: Analysis[_]*)(syntheses: Synthesis[_, _]*): RichExperiment= { /*...*/ }
  // Implicit conversion from tuples of agents to rich analysis and synthesis wrappers
  implicit def expsToExperiments(exps: List[RichExperiment]): List[Experiment] =
    exps.map(_.experiment)
  
  // ALLOWS: a1:Agent[_,D] -> a2:Agent[D,_] : RichAnalysis[D]
  implicit def agentTupleToAnalysis[D <: Comparable[D] with Serializable]
  	(tuple: (Agent[_, D], Agent[D, _])): RichAnalysis[D] = { /*...*/ }
  
  // ALLOWS: a1:Agent[_,D] -> (a2:Agent[D,O], a3:Agent[D,O]) : RichAnalysis[D]
  implicit def agentTupleTupleToAgentList[I <: Comparable[I] with Serializable, D <: Comparable[D] with Serializable, O <: Comparable[O] with Serializable]
  	(tuple: (Agent[I, D], (Agent[D, O], Agent[D, O]))): RichAnalysis[D] = listOfAgentsToAgentList((tuple._1, List(tuple._2._1, tuple._2._2)))
  
  // ALLOWS: a1:Agent[_,D] -> List(a2:Agent[D,O], a3:Agent[D,O]) : RichAnalysis[D]
  implicit def listOfAgentsToAgentList[I <: Comparable[I] with Serializable, D <: Comparable[D] with Serializable, O <: Comparable[O] with Serializable]
  	(tuple: (Agent[I, D], List[Agent[D, O]])): RichAnalysis[D] = agentTupleToAgentList((tuple._1, new AgentList[D, O](tuple._2)))
  
  // ALLOWS: a1:Agent[_,D] -> a2:Agent[D,_] / a3:Agent[D,_]  : RichAnalysis[D]
  implicit def agentTupleToAgentList[O1 <: Comparable[O1] with Serializable, O2 <: Comparable[O2] with Serializable]
  	(tuple: (Agent[_, O1], AgentList[O1, O2])): RichAnalysis[O1] = { /*...*/ }
  
  // ALLOWS: a1:Agent[V,C] + a2:Agent[V,C] -> m:Model[V,C] : RichSynthesis[V,C]
  implicit def agentPairModelTupleToSynthesis[V <: Comparable[V] with Serializable, C <: Comparable[C] with Serializable]
  	(tuple: ((Agent[_, V], Agent[_, C]), Model[V, C])): RichSynthesis[V, C] = { /*...*/ }
}
\end{lstlisting}

\subsubsection{XML-Export} \label{anhang-xml}

XML-Schema-Definition (XSD) des XML-Exportformats für Experimente und Annotationen in TM2:

\lstset{language={XML}}
\begin{lstlisting}
<?xml version="1.0" encoding="UTF-8"?>
<xsd:schema targetNamespace="http://www.quui.com/tm2" elementFormDefault="qualified" xmlns:xsd="http://www.w3.org/2001/XMLSchema" xmlns:tm2="http://www.quui.com/tm2">
    <xsd:complexType name="experiment">
        <xsd:sequence>
        	<xsd:element name="agent" type="tm2:agent" maxOccurs="unbounded" minOccurs="1">
        </xsd:element> </xsd:sequence>
        <xsd:attribute name="data" type="xsd:string" use="required"></xsd:attribute>
    </xsd:complexType>
    <xsd:complexType name="agent">
        <xsd:sequence>
        	<xsd:element name="a" type="tm2:a" maxOccurs="unbounded" minOccurs="1"> </xsd:element> 
        </xsd:sequence>
        <xsd:attribute name="name" type="xsd:string" use="required"></xsd:attribute>
    </xsd:complexType>
    <xsd:complexType name="a">
        <xsd:attribute name="start" type="xsd:string" use="required"></xsd:attribute>
        <xsd:attribute name="end" type="xsd:string" use="required"></xsd:attribute>
        <xsd:attribute name="label" type="xsd:string" use="required"></xsd:attribute>
        <xsd:attribute name="object" type="xsd:string" use="optional"></xsd:attribute>
    </xsd:complexType>
    <xsd:element name="experiment" type="tm2:experiment"></xsd:element>
</xsd:schema>
\end{lstlisting}

\subsubsection{XCDL-Transformation} \label{anhang-xcl}

Transformation von TM2-Annotationen zu XCDL mittels XSLT (vgl. Abschnitt \ref{xcl}, S. \pageref{xcl}):

\lstset{language=XML}
\begin{lstlisting}
<?xml version="1.0" encoding="UTF-8"?>
<xsl:stylesheet version="1.0" xmlns:xsl="http://www.w3.org/1999/XSL/Transform">
    <!-- Transform a TM2 experiment to a Planets XCDL -->
    <xsl:template match="experiment">
        <xcdl xmlns:xsi="http://www.w3.org/2001/XMLSchema-instance" xsi:schemaLocation="http://www.planets-project.eu/xcl/schemas/xcl ../schemas/xcdl/XCDLCore.xsd" xmlns="http://www.planets-project.eu/xcl/schemas/xcl" id="0">
            <object id="o1">
                <normData type="text" id="nd1"> <xsl:value-of select="@data" /> </normData>
                <property id="p1" source="raw" cat="descr">
                    <name id="id58">textualAnnotation</name>
                    <xsl:apply-templates mode="valueSets" />
                </property>
                <xsl:apply-templates mode="propertySets" />
            </object>
        </xcdl>
    </xsl:template>
    <!-- Transform a TM2 annotation label to an XCDL value set -->
    <xsl:template match="a" mode="valueSets">
        <valueSet id="i_i1_i36_i1_i{position()}">
            <labValue> <val> <xsl:value-of select="@label" /> </val> <type>string</type> </labValue>
            <dataRef ind="normSpecific" propertySetId="id_{position()}" />
        </valueSet>
    </xsl:template>
    <!-- Transform TM2 annotation positional information to an XCDL property set -->
    <xsl:template match="a" mode="propertySets">
        <propertySet id="id_{position()}">
            <valueSetRelations> <ref valueSetId="i_i1_i36_i1_i{position()}" name="textualAnnotation" /> </valueSetRelations>
            <dataRef> <ref begin="{@start}" end="{@end}" id="nd1" /> </dataRef>
        </propertySet>
    </xsl:template>
</xsl:stylesheet>
\end{lstlisting}

\subsection{Implementierung der Experimente}

\subsubsection{Informationsextraktion} \label{anhang-ie}

Details zur Implementierung des in Abschnitt \ref{tm2-ie}, S. \pageref{tm2-ie} beschriebenen einführenden Experiments zur Informationsextraktion finden sich im \emph{tm2.agents}-Projekt (vgl. Hinweise in Abschnitt \ref{anhang-code}, S. \pageref{anhang-code}).

\subsection{WSD} \label{anhang-wsd} \label{ref-wsd}

\subsubsection{Text für Beispielexperiment mit Pseudoambiguität}

\emph{Text mining, sometimes alternately referred to as text data mining, refers generally to the process of deriving high quality information from text. High quality information is typically derived through the dividing of patterns and trends through means such as statistical pattern learning. Text mining usually involves the process of structuring the input text (usually parsing, along with the addition of some derived linguistic features and the removal of others, and subsequent insertion into a database), deriving patterns within the structured data, and finally evaluation and interpretation of the output. 'High quality' in text mining usually refers to some combination of relevance, novelty, and interestingness. Typical text mining tasks include text categorization, text clustering, concept/entity extraction, production of granular taxonomies, sentiment analysis, document summarization, and entity relation modeling (i.e., learning relations between named entities).\footnote{Eine Definition des Text-Mining von \url{http://en.wikipedia.org/wiki/Text_mining}}}

\subsubsection{Implementierung der Agenten} \label{anhang-wsd-agent-impl}

Im Folgenden finden sich Details zu den Agenten-Implementierungen für die WSD-Ex\-pe\-ri\-mente in Abschnitt \ref{appl-wsd}, S. \pageref{appl-wsd}, z.T. gekürzt (der komplette Code findet sich in den Projekten tm2.agents und tm2.scala, s. \emph{WsdUsageSenseval.scala} als Einstiegspunkt, vgl. Hinweise in Abschnitt \ref{anhang-code}, S. \pageref{anhang-code}):

\paragraph{SensevalData}

Agent für Zugriff auf Kontexte der Daten (in Scala, \lstinline!Agent[String, Context]!):

\lstset{language={Scala}}
\begin{lstlisting}
abstract class SensevalData(s: String) extends Agent[String, Context] {
  val file = new File(s)
  val trainDataReader = new STAXSensevalDataReader(file);
  val samples: List[Ambiguity] = trainDataReader.getAmbiguities.toList
  val words: java.util.List[String] = trainDataReader.getWords
  def process(input: java.util.List[Annotation[String]]) = samples.map(asAnnotation(_))
  def asAnnotation(amb: Ambiguity): Anno[Context] = {
    Annotations.create[Context](
      classOf[SensevalData], amb.getContext:Context, amb.getContext.targetStart:Int, amb.getContext.targetEnd:Int)
  }
  override def toString = file.toURL.toString
}
\end{lstlisting}

\paragraph{SensevalSense}

Agent für Zugriff auf Zielworte der Daten  (in Scala, \lstinline!Agent[Context, Ambiguity]!):

\begin{lstlisting}
class SensevalSense(s: String) extends Agent[Context, Ambiguity] {
  val words: List[String] = Nil
  val trainDataReader = new STAXSensevalDataReader(new File(s));
  val samples: List[Ambiguity] = trainDataReader.getAmbiguities.toList
  def process(input: java.util.List[Annotation[Context]]) = samples.map(sense(_))
  def sense(amb: Ambiguity): Anno[Ambiguity] = {
    Annotations.create[Ambiguity](
      classOf[SensevalData], amb:Ambiguity, amb.getContext.targetStart:Int, amb.getContext.targetEnd:Int)
  }
  override def toString = "SensevalSense("+s+")"
}
\end{lstlisting}

\paragraph{ContextFeatures}

Agent für die Merkmalsberechnung (in Java, \lstinline!Agent<Context, FeatureVector>!):

\lstset{language={Java}}
\begin{lstlisting}
public abstract class ContextFeatures implements Agent<Context, FeatureVector> {
  /*...*/
  public List<Annotation<FeatureVector>> process(List<Annotation<Context>> input) {
    List<Annotation<FeatureVector>> result = new ArrayList<Annotation<FeatureVector>>();
    for (int i = 0; i < input.size(); i++) {
      Context c = input.get(i).getValue();
      FeatureVector vector = new FeatureVector(generator.getFeatures(c.target, c.all, context * 2), c.lemma, c.id);
      Annotation<FeatureVector> annotation = Annotations.create(getClass(), vector, input.get(i).getStart(), input.get(i).getEnd());
      result.add(annotation);
    }
  return result;
  }
  /*...*/
}
\end{lstlisting}

\paragraph{SensevalClassifier}

Agent für die numerische Klassifikation (in Java, \lstinline!Agent<FeatureVector, Sense>!):

\lstset{language={Java}}
\begin{lstlisting}
public class SensevalClassifier implements Agent<FeatureVector, Sense>, Model<FeatureVector, Ambiguity> {
  /*...*/
  public Model<FeatureVector, Ambiguity> train(List<Annotation<FeatureVector>> value,
    List<Annotation<Ambiguity>> correct) {
    /*...*/
  }
  public List<Annotation<Sense>> process(List<Annotation<FeatureVector>> input) {
    List<Annotation<Sense>> result = new ArrayList<Annotation<Sense>>();
    for (Annotation<FeatureVector> annotation : input) {
      FeatureVector featureVector = annotation.getValue();
      WsdClassifier classifier = lexicon.get(featureVector.lemma);
      String correct = classifier.classify(featureVector.getValues());
      Sense r = new Sense(featureVector.id, featureVector.lemma, correct);
      result.add(Annotations.create(SensevalClassifier.class, r, annotation.getStart(), annotation.getEnd()));
    }
    return result;
  }
  /*...*/
}
\end{lstlisting}

\paragraph{SensevalEval}

Agent f"ur Zugriff auf nativen Senseval-Scorer (in Scala, \lstinline!Agent[Sense, String]!):

\lstset{language={Scala}}
\begin{lstlisting}
class SensevalEval(s: String) extends Agent[Sense, String] with Evaluation {
  //...
  def getResultString = getF.toString
  def process(input: java.util.Set[Annotation[Sense]]) = {
    val sensevalOutput = for (anno <- input; sense = anno.getValue) yield sense.lemma + " " + sense.id + " " + sense.correct
    val res = sensevalOutput.sorted.mkString("\n")
    val file = new File("files/senseval.result")
    val fw = new FileWriter(file)
    fw.write(res); fw.close()
    result = file.getCanonicalPath
    val process = Runtime.getRuntime().exec("files/scorer2 files/senseval.result files/EnglishLS.test.key files/EnglishLS.sensemap -g" + (if (grain=="fine") "" else grain))
    println(Source.fromInputStream(process.getErrorStream).getLines.mkString("\n"))
    response = Source.fromInputStream(process.getInputStream).getLines.mkString("\n")
    Nil
  }
  def getF: Float = {
    val r = """precision: ([^ ]+) """.r
    val Some(res) = r findFirstIn response
    val r(p) = res
    p.toFloat
  }
}
\end{lstlisting}

\subsubsection{Vollständige Ergebnisse} \label{anhang-erg-wsd}

Die folgende Seite enthält die vollständige aus der generierten HTML-Dokumentation mithilfe des Programms \emph{html2latex}\footnote{HTML to LaTeX, \url{http://htmltolatex.sourceforge.net/}} konvertierte Ergebnistabelle für die Experimente in Abschnitt \ref{appl-senseval}, S. \pageref{appl-senseval} (im HTML-Format enthält die Spalte \emph{Details} Hyperlinks zu den Detailseiten eines Experiments). 

\begin{tiny}
\begin{landscape}
\begin{tabular}{l|l|l|l|l|l}

\textbf{Result} & \textbf{Details} & \textbf{Syntheses} & \textbf{Models} & \textbf{Time} & \textbf{Evaluation} \\ 
0.515  & details & SensevalSense(files/EnglishLS.train.xml), TrainFeatures\$1 using FeatureGenerator(
\texttt{3-gram}), context 8  & SensevalClassifier using NaiveBayes  & 8063 ms.  & Senseval (fine)  \\ 
0.555  & details & SensevalSense(files/EnglishLS.train.xml), TrainFeatures\$1 using FeatureGenerator(
\texttt{3-gram}), context 8  & SensevalClassifier using NaiveBayes  & 6932 ms.  & Senseval (mixed)  \\ 
0.571  & details & SensevalSense(files/EnglishLS.train.xml), TrainFeatures\$1 using FeatureGenerator(
\texttt{3-gram}), context 8  & SensevalClassifier using NaiveBayes  & 7737 ms.  & Senseval (coarse)  \\ 
0.536  & details & SensevalSense(files/EnglishLS.train.xml), TrainFeatures\$1 using FeatureGenerator(
\texttt{7-gram}), context 8  & SensevalClassifier using NaiveBayes  & 7802 ms.  & Senseval (fine)  \\ 
0.569  & details & SensevalSense(files/EnglishLS.train.xml), TrainFeatures\$1 using FeatureGenerator(
\texttt{7-gram}), context 8  & SensevalClassifier using NaiveBayes  & 6955 ms.  & Senseval (mixed)  \\ 
0.587  & details & SensevalSense(files/EnglishLS.train.xml), TrainFeatures\$1 using FeatureGenerator(
\texttt{7-gram}), context 8  & SensevalClassifier using NaiveBayes  & 9118 ms.  & Senseval (coarse)  \\ 
0.537  & details & SensevalSense(files/EnglishLS.train.xml), TrainFeatures\$1 using FeatureGenerator(
\texttt{word}), context 8  & SensevalClassifier using NaiveBayes  & 7603 ms.  & Senseval (fine)  \\ 
0.572  & details & SensevalSense(files/EnglishLS.train.xml), TrainFeatures\$1 using FeatureGenerator(
\texttt{word}), context 8  & SensevalClassifier using NaiveBayes  & 5211 ms.  & Senseval (mixed)  \\ 
0.587  & details & SensevalSense(files/EnglishLS.train.xml), TrainFeatures\$1 using FeatureGenerator(
\texttt{word}), context 8  & SensevalClassifier using NaiveBayes  & 4940 ms.  & Senseval (coarse)  \\ 
0.48  & details & SensevalSense(files/EnglishLS.train.xml), TrainFeatures\$1 using FeatureGenerator(
\texttt{length}), context 8  & SensevalClassifier using NaiveBayes  & 4987 ms.  & Senseval (fine)  \\ 
0.517  & details & SensevalSense(files/EnglishLS.train.xml), TrainFeatures\$1 using FeatureGenerator(
\texttt{length}), context 8  & SensevalClassifier using NaiveBayes  & 6274 ms.  & Senseval (mixed)  \\ 
0.535  & details & SensevalSense(files/EnglishLS.train.xml), TrainFeatures\$1 using FeatureGenerator(
\texttt{length}), context 8  & SensevalClassifier using NaiveBayes  & 15221 ms.  & Senseval (coarse)  \\ 
0.581  & details & SensevalSense(files/EnglishLS.train.xml), TrainFeatures\$1 using FeatureGenerator(
\texttt{3-gram}), context 8  & SensevalClassifier using BayesNet  & 11550 ms.  & Senseval (fine)  \\ 
0.617  & details & SensevalSense(files/EnglishLS.train.xml), TrainFeatures\$1 using FeatureGenerator(
\texttt{3-gram}), context 8  & SensevalClassifier using BayesNet  & 16229 ms.  & Senseval (mixed)  \\ 
0.638  & details & SensevalSense(files/EnglishLS.train.xml), TrainFeatures\$1 using FeatureGenerator(
\texttt{3-gram}), context 8  & SensevalClassifier using BayesNet  & 6066 ms.  & Senseval (coarse)  \\ 
0.577  & details & SensevalSense(files/EnglishLS.train.xml), TrainFeatures\$1 using FeatureGenerator(
\texttt{7-gram}), context 8  & SensevalClassifier using BayesNet  & 8921 ms.  & Senseval (fine)  \\ 
0.616  & details & SensevalSense(files/EnglishLS.train.xml), TrainFeatures\$1 using FeatureGenerator(
\texttt{7-gram}), context 8  & SensevalClassifier using BayesNet  & 6720 ms.  & Senseval (mixed)  \\ 
0.635  & details & SensevalSense(files/EnglishLS.train.xml), TrainFeatures\$1 using FeatureGenerator(
\texttt{7-gram}), context 8  & SensevalClassifier using BayesNet  & 5653 ms.  & Senseval (coarse)  \\ 
0.574  & details & SensevalSense(files/EnglishLS.train.xml), TrainFeatures\$1 using FeatureGenerator(
\texttt{word}), context 8  & SensevalClassifier using BayesNet  & 6508 ms.  & Senseval (fine)  \\ 
0.612  & details & SensevalSense(files/EnglishLS.train.xml), TrainFeatures\$1 using FeatureGenerator(
\texttt{word}), context 8  & SensevalClassifier using BayesNet  & 9106 ms.  & Senseval (mixed)  \\ 
0.635  & details & SensevalSense(files/EnglishLS.train.xml), TrainFeatures\$1 using FeatureGenerator(
\texttt{word}), context 8  & SensevalClassifier using BayesNet  & 6422 ms.  & Senseval (coarse)  \\ 
0.552  & details & SensevalSense(files/EnglishLS.train.xml), TrainFeatures\$1 using FeatureGenerator(
\texttt{length}), context 8  & SensevalClassifier using BayesNet  & 7866 ms.  & Senseval (fine)  \\ 
0.592  & details & SensevalSense(files/EnglishLS.train.xml), TrainFeatures\$1 using FeatureGenerator(
\texttt{length}), context 8  & SensevalClassifier using BayesNet  & 7599 ms.  & Senseval (mixed)  \\ 
0.618  & details & SensevalSense(files/EnglishLS.train.xml), TrainFeatures\$1 using FeatureGenerator(
\texttt{length}), context 8  & SensevalClassifier using BayesNet  & 24947 ms.  & Senseval (coarse)  \\ 
0.562  & details & SensevalSense(files/EnglishLS.train.xml), TrainFeatures\$1 using FeatureGenerator(
\texttt{3-gram}), context 8  & SensevalClassifier using SMO  & 20102 ms.  & Senseval (fine)  \\ 
0.602  & details & SensevalSense(files/EnglishLS.train.xml), TrainFeatures\$1 using FeatureGenerator(
\texttt{3-gram}), context 8  & SensevalClassifier using SMO  & 23591 ms.  & Senseval (mixed)  \\ 
0.616  & details & SensevalSense(files/EnglishLS.train.xml), TrainFeatures\$1 using FeatureGenerator(
\texttt{3-gram}), context 8  & SensevalClassifier using SMO  & 18152 ms.  & Senseval (coarse)  \\ 
0.545  & details & SensevalSense(files/EnglishLS.train.xml), TrainFeatures\$1 using FeatureGenerator(
\texttt{7-gram}), context 8  & SensevalClassifier using SMO  & 43785 ms.  & Senseval (fine)  \\ 
0.582  & details & SensevalSense(files/EnglishLS.train.xml), TrainFeatures\$1 using FeatureGenerator(
\texttt{7-gram}), context 8  & SensevalClassifier using SMO  & 19910 ms.  & Senseval (mixed)  \\ 
0.599  & details & SensevalSense(files/EnglishLS.train.xml), TrainFeatures\$1 using FeatureGenerator(
\texttt{7-gram}), context 8  & SensevalClassifier using SMO  & 17492 ms.  & Senseval (coarse)  \\ 
0.563  & details & SensevalSense(files/EnglishLS.train.xml), TrainFeatures\$1 using FeatureGenerator(
\texttt{word}), context 8  & SensevalClassifier using SMO  & 20638 ms.  & Senseval (fine)  \\ 
0.6  & details & SensevalSense(files/EnglishLS.train.xml), TrainFeatures\$1 using FeatureGenerator(
\texttt{word}), context 8  & SensevalClassifier using SMO  & 17404 ms.  & Senseval (mixed)  \\ 
0.614  & details & SensevalSense(files/EnglishLS.train.xml), TrainFeatures\$1 using FeatureGenerator(
\texttt{word}), context 8  & SensevalClassifier using SMO  & 17573 ms.  & Senseval (coarse)  \\ 
0.558  & details & SensevalSense(files/EnglishLS.train.xml), TrainFeatures\$1 using FeatureGenerator(
\texttt{length}), context 8  & SensevalClassifier using SMO  & 19311 ms.  & Senseval (fine)  \\ 
0.595  & details & SensevalSense(files/EnglishLS.train.xml), TrainFeatures\$1 using FeatureGenerator(
\texttt{length}), context 8  & SensevalClassifier using SMO  & 28047 ms.  & Senseval (mixed)  \\ 
0.61  & details & SensevalSense(files/EnglishLS.train.xml), TrainFeatures\$1 using FeatureGenerator(
\texttt{length}), context 8  & SensevalClassifier using SMO  & 17315 ms.  & Senseval (coarse)  \\ 
0.516  & details & SensevalSense(files/EnglishLS.train.xml), TrainFeatures\$1 using FeatureGenerator(
\texttt{3-gram}), context 8  & SensevalClassifier using HyperPipes  & 6839 ms.  & Senseval (fine)  \\ 
0.55  & details & SensevalSense(files/EnglishLS.train.xml), TrainFeatures\$1 using FeatureGenerator(
\texttt{3-gram}), context 8  & SensevalClassifier using HyperPipes  & 5298 ms.  & Senseval (mixed)  \\ 
0.574  & details & SensevalSense(files/EnglishLS.train.xml), TrainFeatures\$1 using FeatureGenerator(
\texttt{3-gram}), context 8  & SensevalClassifier using HyperPipes  & 6056 ms.  & Senseval (coarse)  \\ 
0.487  & details & SensevalSense(files/EnglishLS.train.xml), TrainFeatures\$1 using FeatureGenerator(
\texttt{7-gram}), context 8  & SensevalClassifier using HyperPipes  & 5542 ms.  & Senseval (fine)  \\ 
0.523  & details & SensevalSense(files/EnglishLS.train.xml), TrainFeatures\$1 using FeatureGenerator(
\texttt{7-gram}), context 8  & SensevalClassifier using HyperPipes  & 4910 ms.  & Senseval (mixed)  \\ 
0.552  & details & SensevalSense(files/EnglishLS.train.xml), TrainFeatures\$1 using FeatureGenerator(
\texttt{7-gram}), context 8  & SensevalClassifier using HyperPipes  & 6804 ms.  & Senseval (coarse)  \\ 
0.501  & details & SensevalSense(files/EnglishLS.train.xml), TrainFeatures\$1 using FeatureGenerator(
\texttt{word}), context 8  & SensevalClassifier using HyperPipes  & 4551 ms.  & Senseval (fine)  \\ 
0.536  & details & SensevalSense(files/EnglishLS.train.xml), TrainFeatures\$1 using FeatureGenerator(
\texttt{word}), context 8  & SensevalClassifier using HyperPipes  & 5202 ms.  & Senseval (mixed)  \\ 
0.558  & details & SensevalSense(files/EnglishLS.train.xml), TrainFeatures\$1 using FeatureGenerator(
\texttt{word}), context 8  & SensevalClassifier using HyperPipes  & 4673 ms.  & Senseval (coarse)  \\ 
0.506  & details & SensevalSense(files/EnglishLS.train.xml), TrainFeatures\$1 using FeatureGenerator(
\texttt{length}), context 8  & SensevalClassifier using HyperPipes  & 4492 ms.  & Senseval (fine)  \\ 
0.544  & details & SensevalSense(files/EnglishLS.train.xml), TrainFeatures\$1 using FeatureGenerator(
\texttt{length}), context 8  & SensevalClassifier using HyperPipes  & 16147 ms.  & Senseval (mixed)  \\ 
0.566  & details & SensevalSense(files/EnglishLS.train.xml), TrainFeatures\$1 using FeatureGenerator(
\texttt{length}), context 8  & SensevalClassifier using HyperPipes  & 9908 ms.  & Senseval (coarse) 

\end{tabular}
\end{landscape}
\end{tiny}

\pagestyle{empty}
\bibliography{tm2}

\end{document}